\documentclass[prd,10pt,twocolumn,nofootinbib]{revtex4} 
\usepackage{sistyle}

\usepackage{hyperref}

\usepackage{multirow}

\usepackage[pdftex]{graphicx} 
\usepackage{type1cm}
\usepackage{eso-pic}
\usepackage{color}

\begin{document}

\title{The performance of arm locking in LISA}
\author{Kirk McKenzie, Robert~E.~Spero, and Daniel~A.~Shaddock\footnote{Also at The Centre for Gravitational Physics, Australian National University, ACT 0200, Australia}}
\affiliation{Jet Propulsion Laboratory, California Institute of Technology, Pasadena, California 91109, USA }
\date{\today}

\begin{abstract}
For the laser interferometer space antenna (LISA) to reach it's design sensitivity, the coupling of the free running laser frequency noise to the signal readout must be reduced by more than 14 orders of magnitude. One technique employed to reduce the laser frequency noise will be arm locking, where the laser frequency is locked to the LISA arm length. In this paper we detail an implementation of arm locking. We investigate orbital effects (changing arm lengths and Doppler frequencies), the impact of errors in the Doppler knowledge that can cause pulling of the laser frequency, and the noise limit of arm locking. Laser frequency pulling is examined in two regimes: at lock acquisition and in steady state. The noise performance of arm locking is calculated  with the inclusion of the dominant expected noise sources: ultra stable oscillator (clock) noise, spacecraft motion, and shot noise. We find that clock noise and spacecraft motion limit the performance of dual arm locking in the LISA science band. Studying these issues reveals that although dual arm locking [A. Sutton~\&~D. A Shaddock, Phys. Rev. D \textbf{78}, {082001} (2008).] has advantages over single (or common) arm locking in terms of allowing high gain, it has disadvantages in both laser frequency pulling and noise performance. We address this by proposing a modification to the dual arm locking sensor,  a hybrid of common and dual arm locking sensors. This modified dual arm locking sensor has the laser frequency pulling characteristics and low-frequency noise coupling of common arm locking, but retains the control system advantages of dual arm locking. We present a detailed design of an arm locking controller and perform an analysis of the expected performance when used with and without laser pre-stabilization. We observe that the sensor phase changes beneficially near unity-gain frequencies of the arm-locking controller, allowing a factor of 10 more gain than previously believed, without degrading stability.  With a time delay error of 3~ns (equivalent of 1~m inter-spacecraft ranging error), Time-Delay Interferometry (TDI) is capable of suppressing 300~Hz/$\sqrt{\rm Hz}$ of laser frequency noise to the required level. We show that if no inter-spacecraft laser links fail, arm locking alone surpasses this noise performance for the entire mission. If one inter-spacecraft laser link fails, arm locking alone will achieve this performance for all but approximately 1~hour per year, when the arm length mismatch of the two remaining arms passes through zero. Therefore, the LISA sensitivity can be realized with arm locking and TDI only, without any form of pre-stabilization. 
\end{abstract}

\maketitle

\section{Introduction}

The Laser Interferometer Space Antenna (LISA)~\cite{LISAPPA} is a joint NASA-ESA mission to observe gravitational wave signals of astronomical origin. LISA will consist of three spacecraft arranged in a triangular formation with $5\times10^9$~m sides (arms). Laser interferometry will be used to sense the spacecraft separation to a precision of $\delta x$ = 40~pm$/\sqrt{\textrm{Hz}}$ increasing as $f^{-2}$ below 3mHz, giving a strain sensitivity of $h\sim10^{-20}/\sqrt{\textrm{Hz}}$. Unlike ground based gravitational wave detectors, which are based on Michelson interferometers with equal arm lengths,  the LISA arm lengths vary over time and will be unequal for the majority of the mission.  Over the period of a year, the arm lengths will vary by approximately 3\% of the total arm length,  or by $\pm76,500$~km~\cite{AstriumDoc1}. Given this level of arm length mismatch, there are stringent laser frequency noise requirements: coupling of the free running frequency noise of a Nd:YAG laser to the signal output must be reduced by more than 14 orders of magnitude. 

The frequency noise requirement is expected to be met using three techniques in unison:  (1) Pre-stabilization, where the laser frequency is locked to either the resonance of a Fabry-Perot cavity~\cite{PDH,ThorpeOE2008}, or an mismatched arm length  Mach-Zehnder interferometer~\cite{Heinzel};  (2) arm locking~\cite{SheardPLA,Tinto, Sylvestre}, a technique based on transferring the  stability of the LISA arm length to the laser frequency; and (3) time delay interferometry (TDI)~\cite{TintoPRD1999, TDI1, ShaddockPRD2003}, a post processing technique that synthesizes interferometers with equal arm lengths by combining phase measurements with appropriate delays.

The initial arm locking proposal by Sheard~\textit{et.~al}~\cite{SheardPLA} showed the  round trip propagation delay, $\tau\approx 33$ seconds, of the LISA measurement scheme didn't necessarily limit the arm locking control system to a low bandwidth, low gain system. This is because instantaneous information of the laser phase noise is obtained via the local oscillator field used in each one-way phase measurement. By careful design of the controller, the arm locking control system can encompass many interferometer nulls to deliver a high bandwidth, high gain  system over the LISA measurement band.

High-bandwidth arm locking has been demonstrated in hardware~\cite{Marin, Thorpe, Sheard2} and analyzed theoretically~\cite{Tinto}. Dual arm locking~\cite{SuttonPRD}, built on the proposal of Enhanced Arm Locking by Herz~\cite{Herz},  uses combinations of phase measurements from two arms to increase the frequency of the first null of the sensor from 1/$\tau\approx 30$~mHz to 1/(2$\Delta \tau$) $\gtrsim$ 2~Hz , where $\Delta \tau$ is the 1/2 the difference in light travel round trip times of the two arms used (we define the average round trip time as $\bar{\tau}$). Moving the first null to outside the LISA band allows a more aggressive controller design below 2~Hz and eliminates from the LISA science band noise amplification due to the nulls. 

Although there have been many studies of arm locking, there has been little analysis of how it would be implemented in LISA, the operation of it, the performance limitations due to noise sources,  or the effects associated with the changing arm lengths and Doppler shifts.  This paper is intended to provide a detailed study of these issues and to predict the performance limitations of arm locking in LISA. The analysis is performed for a continuous system, though the implementation on LISA will be digital; we expect the difference in performance will be minor. The following issues are analyzed:
\begin{itemize}
\item The arm locking measurement architecture. That is, which phasemeter signals should be used in the arm locking sensor.
\item {Laser frequency pulling due to an error in the Doppler frequency estimate. The phase measurements used in arm locking require an estimate of the Doppler frequency to be subtracted. Estimation error causes the laser frequency to ramp, pulled from the starting frequency. Laser frequency pulling occurs both at lock acquisition and throughout the orbit. To prevent significant laser frequency pulling, either the Doppler frequency estimate must be updated regularly or low frequency filtering should be implemented.}
\item The significant noise sources in arm locking. These are:
\begin{itemize}
\item Noise of ultra-stable oscillators (USO's), or clocks as we call them henceforth, which enters at each phasemeter measurement. 
\item Spacecraft motion, which is significantly larger than the proof mass motion and is present in most of the phase measurements.
\item Shot noise, due to the quantized nature of the electromagnetic field. 
\end{itemize}
\end{itemize}

These effects are investigated using the common and dual arm-locking sensors~\cite{SuttonPRD}, and are written in a general formalism applicable to any sensor.  We introduce a modification to the dual arm locking sensor that combines the reduced noise and reduced frequency-pulling of common arm locking and the higher gain of dual arm-locking.  We present a controller design that	maximizes frequency noise suppression, including a factor of 10 more gain than previously assumed~\cite{SuttonPRD}, while retaining a	conservative phase margin.  This controller suppresses laser frequency to a level low enough that arm-locking and TDI alone,	without frequency prestabilization, is adequate for the ultimate sensitivity requirement.

The paper is laid out as follows: In section~\ref{section_setup} the notation used throughout the paper is introduced, and the various possible arm locking sensors previously published are listed.  The detailed study begins in section~\ref{section_architecture} by examining which phasemeter signals should be used in the arm locking sensor.  {In section~\ref{section_doppler} laser frequency pulling is examined at lock acquisition and in steady state. The largest noise sources of arm locking are examined in section~\ref{section_noisesources}, and written into a general formalism for the noise floor of arm locking in section~\ref{section_noisebudget}.  In section~\ref{section_modified} the new dual arm locking sensor is presented. In section~\ref{section_controller} we present a design of the arm locking controller.  Finally, in section~\ref{section_performance} we discuss the performance limitations of arm locking to  give a realistic estimate of arm locking performance in LISA.

\section{Notation and spectra}
\label{section_setup}
In this section we introduce notation, write down the arm locking sensors, and the laser frequency noise spectra we shall assume for the rest of the paper. 
\subsection{Notation}

Figure~\ref{dual_arm_configuration_simple} shows a simplified schematic of the LISA measurement scheme and the signal routing for arm locking. For simplicity,  only two of the three spacecraft are drawn and  only laser frequency noise is considered. In section~\ref{section_noisebudget} shot noise, clock noise, and spacecraft motion are added to the calculation. We label the three spacecraft 1,2, and 3, and take spacecraft 1 to be the central spacecraft. The phasemeter output on each spacecraft is given by propagating the laser noise source to the phasemeter. To start with, assume that all control loops are open. In the frequency domain, the phase measured at the output of the phasemeter on spacecraft 3, facing spacecraft 1 (represented by the red circle labelled $A{31}$) is  
\begin{eqnarray}
\phi_{A31}(\omega) &=& \phi_{L3}(\omega)-\phi_{L1}(\omega)e^{-i\omega\tau_{13}},
\end{eqnarray}
where $\phi_{Lj}(\omega)$ is the laser phase noise of the laser on the $jth$ spacecraft pre-arm locking expressed in units of cycles/$\sqrt{\textrm{Hz}}$, and $\tau_{13}$ is the light propagation time between spacecraft~1 and~3. For simplicity we assume\footnote{In general, the light propagation time to and from each spacecraft will not be equal due to the motion of the constellation.} $\tau_{ij}=\tau_{ji}$.  
\begin{figure}[htb]
\centering
\includegraphics[width=.5\textwidth]{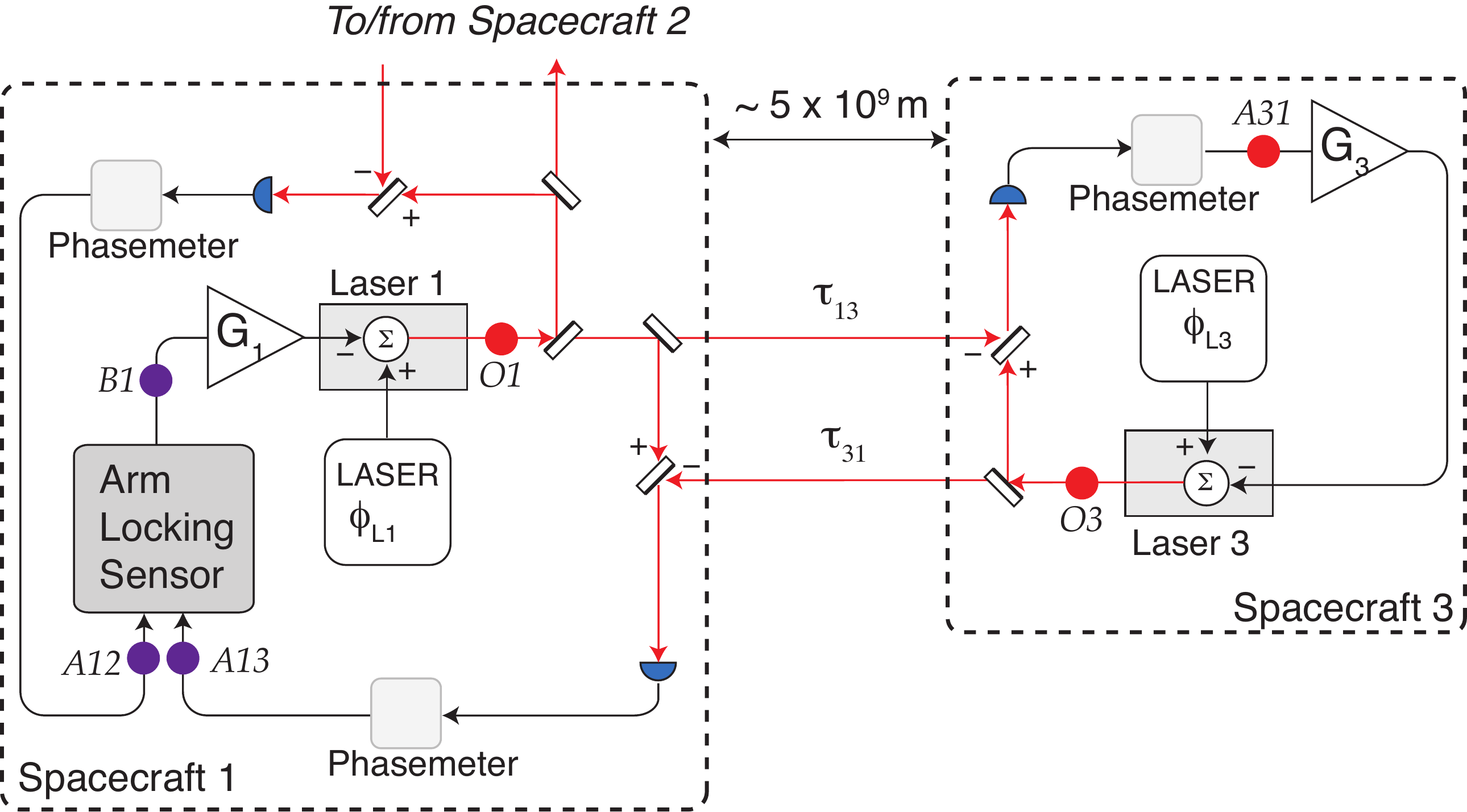}
\caption{\label{dual_arm_configuration_simple} Schematic of arm locking control loop. Laser frequency noise is represented by $\phi_{Lj}$,  with $j$ representing the number of the local spacecraft.}
\end{figure}
If the laser on spacecraft 3 is phaselocked to the incoming light, the closed loop phase noise at the output of laser~3 (represented by the red circle labelled $O3$) is
 \begin{eqnarray}
\phi_{O3}(\omega) &=& \frac{\phi_{L3}(\omega)}{1+G_3(\omega)}+\frac{G_3(\omega)}{1+G_3(\omega)}\phi_{L1}(\omega)e^{-i\omega\tau_{13}},
\end{eqnarray}
where $G_3(\omega)$ is the frequency response of the phase locking controller on spacecraft 3. With the laser on the spacecraft 3 phaselocked to the incoming light, but  the arm locking control loop open ($G_1(\omega)=0$), the phase at the phasemeter output labeled $A13$ on spacecraft 1 is given by
\begin{eqnarray}
\phi_{A13}(\omega) &=&\phi_{L1}(\omega)P_{13}(\omega)-\frac{\phi_{L3}(\omega)}{1+G_3(\omega)}e^{-i\omega\tau_{13}},
 \label{eqn_phi13}
 \end{eqnarray}
where $P_{13}(\omega)$ is the frequency response of the 1-3 arm, including the prompt and delayed signals. The calculation of phase $\phi_{A12}(\omega)$ follows a similar argument.  The frequency responses of the two arms are
\begin{eqnarray}
P_{13}(\omega) &=& 1-\frac{G_3(\omega)}{1+G_3(\omega)}e^{-i2\omega\tau_{13}},\nonumber \\ P_{12}(\omega)& = &1-\frac{G_2(\omega)}{1+G_2(\omega)}e^{-i2\omega\tau_{12}},
\end{eqnarray}
 with $G_2(\omega)$ the controller gain of the phase locking loop on spacecraft 2. 
In the high gain transponder limit\footnote{The phaselocking control loops are expected to have a unity gain frequency $>30~$kHz, yielding high gain across the LISA science band. } $G_j(\omega)/(1+G_j(\omega))\rightarrow1$ and $1/(1+G_j(\omega))\rightarrow0$  for $j = 2,3$, and the frequency responses of the two arms are approximately
\begin{eqnarray}
P_{13}(\omega) \approx 1-e^{-i2\omega\tau_{13}},~~P_{12}(\omega)\approx 1-e^{-i2\omega\tau_{12}}.
\label{p13approx}
\end{eqnarray}
and the phases that enter the arm locking sensor are
\begin{eqnarray}
\phi_{A13}(\omega) &\approx&\phi_{L1}(\omega)P_{13}(\omega),
\label{phiA13_approx}\\
\phi_{A12} (\omega)&\approx&\phi_{L1}(\omega)P_{12}(\omega). \label{phiA12_approx}
 \end{eqnarray}
 We adopt the approximations of (\ref{p13approx}), (\ref{phiA13_approx}), and (\ref{phiA12_approx}) in the sequel.

\subsection{Arm locking sensors}

The phase measurements that enter the arm locking sensor on the central spacecraft are
  \begin{eqnarray}
 \Phi_{A1} = \left[\begin{array}{c}\phi_{A13}(\omega) \\ \phi_{A12}(\omega)\end{array}\right].
 \label{phase_noise_A1}
 \end{eqnarray}

 We are interested in the performance of both common and dual arm locking. This is because the modified dual arm locking sensor, introduced in section~\ref{section_modified}, is a hybrid of the two.  In common arm locking, phase measurements from both arms are simply added.  The mapping of the two input signals for common arm locking is given by the vector
\begin{eqnarray}
\textbf{S}_{+} =\left[\begin{array}{cc}1 & 1\end{array}\right].
\end{eqnarray}
The open loop noise at the output of the arm locking sensor (point $B1$ in figure~\ref{dual_arm_configuration_simple}) is simply
\begin{eqnarray}
\phi_{B1}|_{+}(\omega) = \textbf{S}_{+} \Phi_{A1}.
\end{eqnarray} 
We will be interested in the frequency noise at the laser output with the arm locking control loop closed. This is the noise at the point $O1$ in figure~\ref{dual_arm_configuration_simple}.  With the arm locking control loop closed ($G_1(\omega) \ne 0$) the frequency noise at the laser output is given by 
\begin{eqnarray}
\phi_{O1}|_{+} (\omega) &=&\phi_{L1}(\omega)-\frac{G_1(\omega)P_+(\omega)\phi_{L1}(\omega)}{1+G_1(\omega)P_+(\omega)},  \\
&=&\frac{\phi_{L1}(\omega)}{1+G_1(\omega)P_+(\omega)},
\end{eqnarray}
where $P_+(\omega) = P_{13}(\omega)+P_{12}(\omega)$,  is the frequency response of the common arm locking sensor, also given in table~\ref{table_sensors} \footnote{In defining the sensors in this paper we have assumed the high gain transponder limit, thus we use the approximation of $P_{12}(\omega)$ and $P_{12}(\omega)$ given in equation~\ref{p13approx}.} .

For dual arm locking, the signal mapping vector is~\cite{SuttonPRD}
\begin{eqnarray}
 \textbf{S}_{D} = \left[\begin{array}{cc}1-\frac{E(\omega)}{i\omega\Delta\tau} &1+\frac{E(\omega)}{i\omega\Delta\tau}\end{array}\right].
\end{eqnarray}
The open loop noise at the output of the arm locking sensor (point $B1$ in figure~\ref{dual_arm_configuration_simple}) is 
\begin{eqnarray}
\phi_{B1}|_{D}(\omega) = \textbf{S}_{D} \Phi_{A1}.
\end{eqnarray} 
and the closed loop output is 
\begin{eqnarray}
\phi_{O1}|_{D}  (\omega)
&=& \frac{\phi_{L1}(\omega)}{1+G_1(\omega)P_D(\omega)},
\end{eqnarray}
with $P_D(\omega)$ the frequency response of the dual arm locking sensor. The  signal mapping vectors of, and frequency responses for, single, common, and dual arm locking are contained in table~\ref{table_sensors}. The parameter $E(\omega)$ in the dual arm locking signal mapping vector is a filter designed to ensure stability when the common and difference sensors are combined~\cite{SuttonPRD}. The values used in this paper are given in table~\ref{table_E}.

The sensors in this paper that contain the addition of the two arms (common, dual, and modified dual) have an frequency response magnitude up to twice that of the single arm locking sensor. To take account of this factor or 2 difference, we define the controller for any such sensor as $G_1(\omega)=\frac {G_1^*(\omega)}{2}$. For dual arm locking, which has a frequency response of magnitude 2  at $f<1/{2\Delta \tau}$, this means that we can design a controller $G_1^*(\omega)$ as if we had a sensor with unity magnitude, making the design process more intuitive. 
  \begin{center}
\begin{table*}[t]
\caption{\label{table_sensors}The signal mapping vector and frequency response of different  arm locking configurations. Here $\tau_{ij}$ is the one way light  travel time of the ${ij}$th arm, $\bar{\tau}$ is the average round trip time of the two arms,  and $E(\omega)$ is an filter used to combine the common and difference sensors, given in table~\ref{table_E}. The parameters $H_+(\omega)$ and $H_-(\omega)$ are defined in equations~\ref{Hplus} and~\ref{Hminus}.}
\begin{center}
\begin{tabular}{|l|l|l|}\hline
Configuration & Signal Mapping & Frequency Response \\
 \hline
Single & $\textbf{S}_{S} = \left[\begin{array}{cc}1 & 0\end{array}\right]$ & $P_S(\omega) = 2i\sin\left(\tau_{ij}\omega\right)e^{-i\omega \tau_{ij}}$ \\
Common & $\textbf{S}_{+} =\left[\begin{array}{cc} 1& 1 \end{array}\right]$ & $P_+(\omega) = 2(1- \cos\left(\Delta \tau\omega\right)e^{-i\omega \bar\tau}$) \\
Difference & $\textbf{S}_{-} = \left[\begin{array}{cc}1 & -1\end{array}\right]$ & $P_-(\omega) = -2i\sin\left(\Delta \tau\omega\right)e^{-i\omega \bar\tau}$ \\
Dual & $\textbf{S}_{D} =\left[\begin{array}{cc}1-\frac{E(\omega)}{i\omega\Delta\tau} &1+\frac{E(\omega)}{i\omega\Delta\tau}\end{array}\right]
$ & $P_{D}(\omega) = {P_+(\omega)}-\frac{E(\omega)}{i\omega\Delta\tau}P_{-}(\omega)$ \\
Modified dual & $\textbf{S}_{M} = \left[\begin{array}{cc}H_+(\omega)-H_-(\omega) &H_+(\omega)+H_-(\omega)\end{array}\right]
$ & $P_{M}(\omega)= P_+(\omega)H_+(\omega)-{P_-(\omega)}H_-(\omega)$ \\
 \hline
\end{tabular}
\end{center}
\end{table*}
\end{center}
  
  \begin{table}[htdp]
\caption{Parameters of the filter contained in the dual arm locking sensor. The filter is defined as $E(\omega)=E_1E_2$.}
\begin{center}
\begin{tabular}{|c|c|l|l|}
\hline
Filter & zeros (radians/s) & Poles (radians/s) & Gain \\ \hline 
$E_1$ &	&$p_1 = 2\pi\times 0.1394/\Delta \tau$ & $g_1 =p_1$ \\
 $E_2$ &$z_2 = 2\pi\times10/\Delta {\tau}$	&$p_2=2\pi\times5/(2\Delta{\tau})$& $g_2 = p_2/z_2$\\  \hline \end{tabular}
\end{center}
\label{table_E}
\end{table}%

\subsection{Laser frequency noise spectra}

The level and shape of the laser frequency noise spectrum is required for a number of calculations in this paper. We use three different initial laser noise levels: free running laser noise, and laser noise predicted for two types of pre-stabilization, Fabry-Perot cavity stabilization and Mach-Zehnder stabilization. 
The corresponding laser frequency noise levels are (square root of power spectral density):
\begin{eqnarray}
\Delta \nu_{\rm FR}(f) &=&  30,000\times\textrm{1~Hz}/f~~~\frac{\textrm{Hz}}{\sqrt{\textrm{Hz}}},\label{equation_frln}\\
\Delta \nu_{\rm MZ}(f) &=&  800\times(1+(2.8~\textrm{mHz}/f)^2)~~\frac{\textrm{Hz}}{\sqrt{\textrm{Hz}}},\label{equation_mzn}\\
\Delta \nu_{\rm FP}(f) &=&  30\times(1+(2.8~\textrm{mHz}/f)^2)~~~\frac{\textrm{Hz}}{\sqrt{\textrm{Hz}}}.
\end{eqnarray}
{The free running laser noise level quoted here is conservative. A measurement of the beat between two free running laser Nd:YAG NPRO lasers is shown in figure~\ref{laserfreqnoise}.  These results were taken by interfering two free running lasers  on a beamsplitter (both Lightwave model 126), detecting one output of the beamsplitter on a photodetector and measuring the resultant phase of the beatnote using the LISA phasemeter~\cite{Shaddock2006}. This measurement is always under $25,000 \times 1~\textrm{Hz}/f ~\textrm{Hz}/\sqrt{\textrm{Hz}}$  which, assuming the two lasers have identical noise properties, indicates a free running laser noise of one laser of less than $18,000 \times 1~\textrm{Hz}/f~\textrm{Hz}/\sqrt{\textrm{Hz}}$.}

  \begin{figure}[htb]
\centering
\includegraphics[width=.48\textwidth]{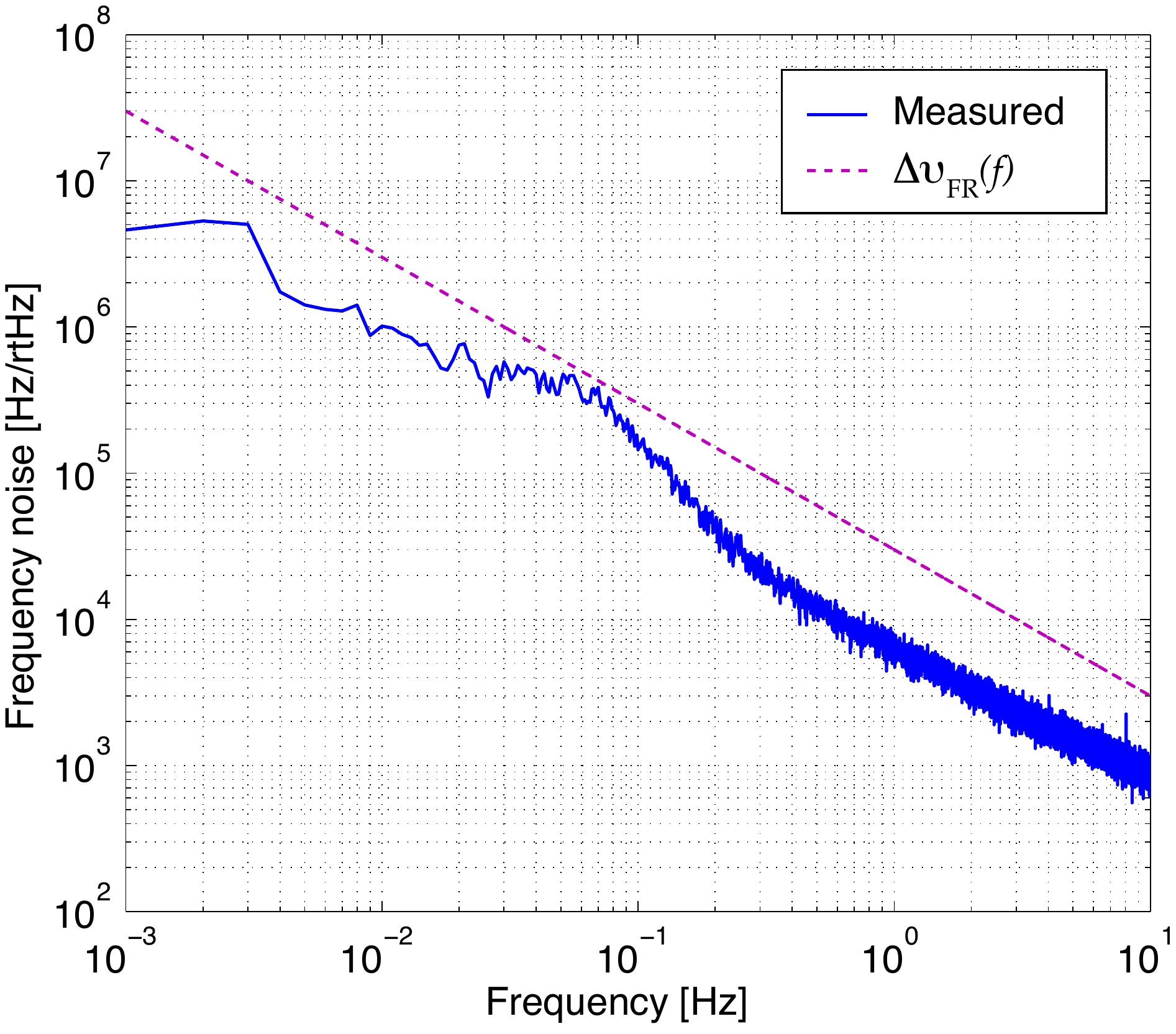}
\caption{\label{laserfreqnoise} A measurement of the frequency noise between two NPRO Nd:YAG lasers (solid trace), and the assumed free running laser noise assumed in this paper (dashed trace).  The phase of the beatnote was measured using the LISA phasemeter. The dashed trace is given by equation~\ref{equation_frln}.}
\end{figure}

\section{Measurement architecture for arm locking}
\label{section_architecture}

The LISA science measurement is performed between proof masses on the local spacecraft and distant spacecraft using the post processing technique, TDI. Time delay interferometry will be implemented on Earth by forming linear combinations of low bandwidth ($\sim$3~Hz) phase measurements with delays determined by inter-spacecraft ranging~\cite{Shaddock2004}. This process requires five phase measurements on each optical bench (with two optical benches per spacecraft). These phase measurements are~\cite{tintoPRD2003}: (1) inter-spacecraft measurement, (2) the backlink measurement, (3) the proof mass to optical bench measurement, (4) and (5), the beatnotes of the upper-upper and lower-lower clock sidebands.  Time delay interferometry forms the displacement measurement using the strap-down architecture~\cite{HeinzelCQG}, which combines the inter-spacecraft measurements, the proof mass to optical bench, and the backlink measurement to remove spacecraft motion. Clock noise in the measurement can be removed using the beatnote of clock sidebands~\cite{tinto2002, tinto, cn1, cn2, deVinePRL}.
 
Unlike TDI, arm locking requires high bandwidth signals ($\sim20$kHz), in real time, and has significantly less stringent noise requirements than the LISA science measurement. Given the relaxed noise requirements,  we assume that arm locking will operate with the most simple measurement architecture - \textit{using only the inter-spacecraft phase measurements}. A consequence  of choosing this measurement architecture is that both clock noise and spacecraft motion will be present in the phase measurements used for arm locking. In section~\ref{section_noisesources} we see that clock noise and spacecraft motion represent the largest noise sources for arm locking in the science band. 
\section{Limiting laser frequency pulling}
\label{section_doppler}

The heterodyne frequency of the inter-spacecraft measurement is set by the Doppler shift of the laser light, due to the relative motion of the spacecraft. The relative velocities in each arm will be up to $18$~m/s, corresponding to a Doppler shift of up to 18~MHz~\cite{Gath1}.  For arm locking to operate stably, this round trip Doppler frequency must be estimated and subtracted in the phase measurements used in the arm locking sensor~\cite{Wand}. In the limit of a high gain DC coupled arm locking control system, an error in the estimated Doppler frequency is compensated for by changing the local laser frequency to maintain the desired beat note frequency. In single arm locking, this frequency change will appear on the light returning from the distant spacecraft 33~s later, necessitating a further change by the local laser frequency to maintain the desired beat note frequency. The closed loop master laser frequency, $\nu_{\rm CL}$, will be changed by the error in the Doppler frequency, $\nu_{\rm DE}$, each round trip, or an average rate of
\begin{eqnarray}
\frac{\delta \nu_{\rm CL}}{\delta t}&=&\nu_{\rm  DE}\frac{c}{2L}\hspace{2cm}[\rm{Hz/s}]
\label{pulling}
\end{eqnarray}
For example, if the Doppler frequency can be estimated to 100~kHz, the laser frequency will be forced to change by 1~GHz in 4 days. Such large pulling of the laser frequency is undesirable, as it could drive the master laser through a mode-hop region, compromising instrument sensitivity. The other lasers in the constellation are also at risk of being pulled into a  mode-hop region as they will be  locked to the master laser frequency.  Additionally, ramping of laser frequency combined with scattered light sources  can couple noise into the science band~\cite{Whitepaper}.

We study frequency pulling in two regimes: 1) in steady state operation, and 2) at lock acquisition. At lock acquisition, the laser frequency can be pulled significantly by an error in the initial Doppler frequency estimate and also in the time derivatives of the Doppler frequency.

{
The solution to this problem explored in detail here is to add high pass filters to the arm locking control loop at a frequency below the LISA science band. This low frequency modification to the controller limits the laser frequency pulling at lock acquisition and allows indefinite operation of the arm locking control system with an acceptable level of pulling.  An alternate and closely related solution presented by Gath~\cite{Gath2}, is based on a DC coupled controller with an additional control loop operating at low frequencies to limit the amplitude of the controller signal at these frequencies.  This active solution may have precision advantages in implementation.}

The arm locking control system operates as follows: before arm locking is engaged,  measurements of the Doppler frequency and the Doppler rate (the first time derivative of the Doppler frequency) will be made (Appendix~\ref{appendix_DE}) and subtracted from the phasemeter measurement. After the control loop is closed the error in the Doppler frequency measurement will cause the laser frequency to ramp at a rate proportional to the product of the error and the step response of the controller. Whilst locked, the Doppler frequency estimate will not need to be updated. The arm locking control loop will be unlocked and re-locked periodically to perform mission tasks, such as to change the heterodyne frequencies~\cite{Gath1}. At these times the Doppler frequency and its time derivatives will be known very accurately (as many weeks or months of data can be averaged to measure it) and the impulse to laser frequency will be much smaller than in the first time arm locking is engaged. }

{In the following sections we describe the extent of the frequency pulling and how to limit it. We shall examine the lock acquisition and steady state operation regimes separately. We start by calculating the frequency response of Doppler frequency error into the stabilized laser frequency.  These results are considered in the controller design process presented in section~\ref{section_controller}, where we show that with the frequency pulling limited to an acceptable level, high gain can be achieved across the LISA science band. }

 \subsection{Frequency response to Doppler frequency error}

\label{section_frequency_response_Doppler_error}

\subsubsection{Single arm locking}
\begin{figure}
\begin{center}
\includegraphics[width = 0.35\textwidth]{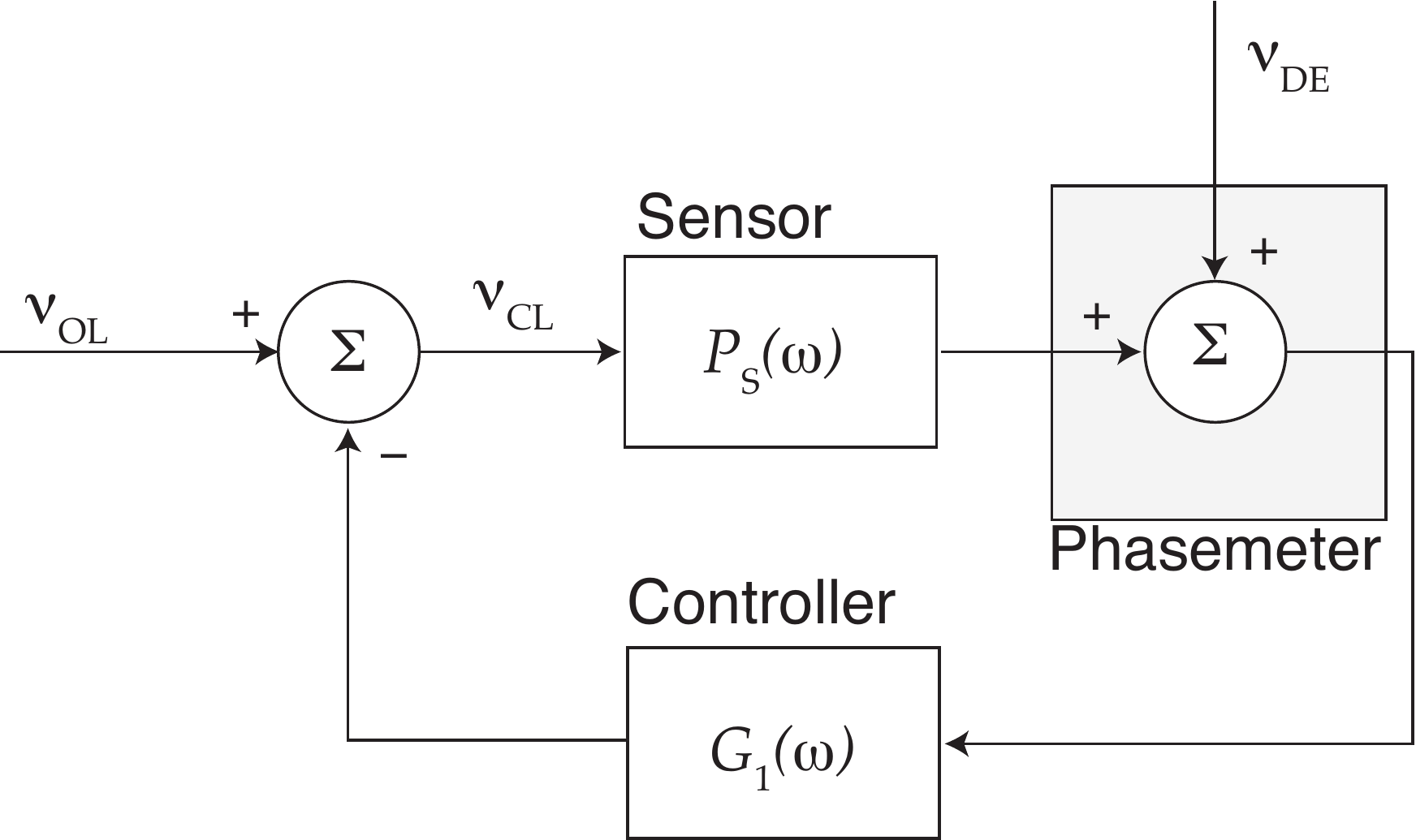}
\caption{Block diagram of the single arm locking control system showing where the Doppler frequency error, $\nu_{\rm DE}$ enters the control loop. $P_S(\omega)$ and $G_1(\omega)$ are frequency responses of the sensor and controller, respectively. The laser frequency with the control loop open and closed are $\nu_{\rm OL}$ and $\nu_{\rm CL}$. \label{Doppler_error}}
\end{center}
\end{figure}

The block diagram in figure~\ref{Doppler_error} shows where the Doppler frequency error enters the single arm locking control loop. Here $\nu_{\rm OL}$ is the initial (open loop) laser frequency, $\nu_{\rm CL}$ the stabilized (closed loop) laser frequency, and $P_S(\omega)$ and $G_1(\omega)$ are the frequency responses of the single arm sensor and controller, respectively. The Doppler frequency error enters the phase measurement in the phasemeter, where an estimate of the Doppler frequency is subtracted from the phase measurement. The closed loop frequency will be pulled by 
 \begin{eqnarray}
 \nu_{\rm CL}|_S &=& \frac{-G_1(\omega){\nu_{\rm DE}}}{1+G_1(\omega)P_S(\omega)}.
 \label{DE_reponse}
 \end{eqnarray}

The frequency response of the Doppler frequency error to the stabilized laser frequency for single arm locking is simply  \begin{eqnarray}
Y_S(\omega) = \frac{ \nu_{\rm CL} |_S}{\nu_{\rm DE}} = \frac{-G_1(\omega)}{1+G_1(\omega)P_S(\omega)}.
 \label{DE_reponse}
 \end{eqnarray}

The frequency pulling due to error in the Doppler frequency can be examined in the extreme cases of low and high loop gain. At high gain,
  \begin{eqnarray}
Y_S(\omega) &\sim& \frac{1}{P_S(\omega)}~~~~~~~(\textrm{for}~~G_1(\omega)P_S(\omega)\gg1),~~~~~~
   \end{eqnarray}
  which, as $P_S(\omega)\approx i\omega \tau$ at low frequencies ($f\lesssim1$~mHz), gives the constant-rate frequency pulling described in equation~\ref{pulling}. At low gain,
  \begin{eqnarray}
Y_S(\omega)   &\sim&~G_1(\omega)~~~~~~~(\textrm{for}~~G_1(\omega)P_S(\omega)\ll1).
 \label{DE_response}
 \end{eqnarray}
This equation shows that for an  AC coupled control loop the Doppler error is multiplied by the low-frequency gain.
 \subsubsection{Common and dual arm locking}

\begin{figure}[h!]
\begin{center}
\includegraphics[width = 0.50\textwidth]{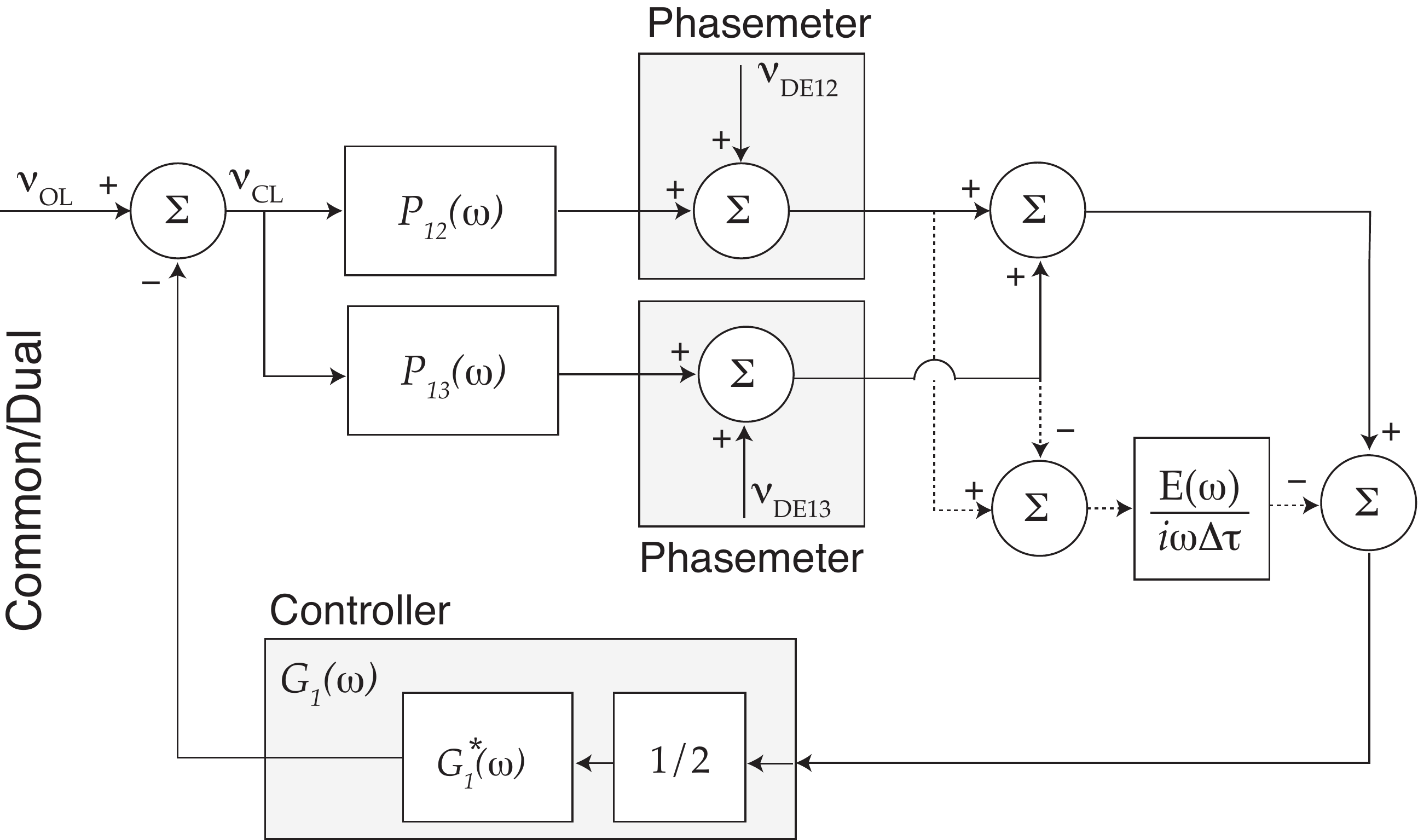}
\caption{Block diagram of the common or dual arm locking control system. Common arm locking would use only the signal paths with solid lines, whereas dual arm locking uses both solid and dashed lines. The Doppler frequency errors that enter at the two phasemeters are labelled $\nu_{\rm DE12}$ and $\nu_{\rm DE13}$. $P_{12}(\omega)$ and $P_{13}(\omega)$ are the single arm frequency responses. \label{Doppler_error_d2}}
\end{center}
\end{figure}

When phase measurements from two arms are used to create the arm locking sensor the analysis becomes  more complex. Figure~\ref{Doppler_error_d2} shows the block diagram for a two signal arm locking control loop, such as used in common or dual arm locking. Common arm locking would use only the signal paths with solid lines, whereas dual arm locking uses both solid and dashed lines. There are two errors in the Doppler frequency, one enters at each phasemeter:  $\nu_{\rm DE12}$ and $\nu_{\rm DE13}$.
The way the Doppler frequency errors enter the control loop with these sensors is less intuitive than in the single arm locking case. The Doppler frequency error is added after the individual arm frequency responses and before the combination of the phase measurements to form the common or dual arm locking sensors.

For common arm locking, the closed loop laser frequency driven by Doppler frequency error is 
\begin{eqnarray}
\nu_{\rm CL}|_+ &=& \frac{-G_1(\omega)(\nu_{\rm DE12}+\nu_{\rm DE13})}{1+G_1(\omega)(P_{12}(\omega)+P_{13}(\omega))}.
\label{common_FR_doppler} \\ 
&=& \frac{-G_1(\omega)\textbf{S}_+\textbf{V}_{\rm DE}}{1+G_1(\omega)P_+(\omega)},
\label{common_DE_fr}
\end{eqnarray}
where 
\begin{eqnarray}
\textbf{V}_{\rm DE} = \left[\begin{array}{cc}\nu_{\rm DE12} & \nu_{\rm DE13}\end{array}\right].
\end{eqnarray}
For common arm locking, the frequency pulling is caused by the sum of the Doppler frequency errors, $\nu_{\rm DE+}= \nu_{\rm DE12}+\nu_{\rm DE13}$. The relevant frequency response is 
\begin{eqnarray}
Y_+(\omega) =  \frac{\nu_{\rm CL}|_+}{ \nu_{\rm DE+}}= \frac{-G_1(\omega)}{1+G_1(\omega)P_+(\omega)}.
\label{Doppler_common_fr}
\end{eqnarray}

 The closed loop laser frequency driven by Doppler frequency error in dual arm locking can be given in a similar form to equation~\ref{common_DE_fr} by replacing the signal mapping vector and frequency response
\begin{eqnarray} 
\nu_{\rm CL}|_D = \frac{-G_1(\omega)\textbf{S}_D\textbf{V}_{\rm DE}}{1+G_1(\omega)P_D(\omega)}.\label{compact_dual_doppler}
\end{eqnarray}

There are two relevant frequency responses to Doppler frequency errors in dual arm locking: due the sum and the difference in the Doppler frequency errors
\begin{eqnarray}
Y_{D}^{(+)} (\omega)&= &\frac{\partial \nu_{\rm CL}|_D}{\partial \nu_{\rm DE+}}   = \frac{-G_1(\omega)}{1+G_1(\omega)P_D(\omega)},\label{Y_D_p}\\ 
Y_{D}^{(-)} (\omega)&=& \frac{\partial \nu_{\rm CL}|_D}{\partial \nu_{\rm DE-}}  = \frac{-G_1(\omega)E(\omega)}{{i\omega\Delta \tau}(1+G_1(\omega)P_D(\omega))},\label{Y_D_m}
\end{eqnarray}
where $\nu_{\rm DE-} = \nu_{\rm DE12}-\nu_{\rm DE13}$. Due to the additional filtering in the difference path of the dual arm locking sensor, $Y_{D}^{(-)} (\omega)$ is inversely proportional to $\omega \Delta \tau$, causing it  to dominate the frequency pulling in dual arm locking.

\subsection{Doppler frequency error }
We  now consider what the Doppler frequency error will be. There will be an error associated with the initial estimate of the Doppler frequency. Also, because the Doppler frequency estimate will not be updated whilst arm locking is operating, (to prevent adding noise in the science band) there will be an error due to the Doppler frequency changing throughout the year. Thus,  the Doppler frequency error must be written as a function of time. The Doppler frequency error in the $ij$ arm in the time domain is
\begin{eqnarray}
\nu_{{\rm DE}ij}(t) &= &\nu_{0 ij}+\int_0^t{\gamma}_{ij}(t)dt+ \int_0^t\int_0^{t^{\prime}} {\alpha}_{ij}(t)dt ^{\prime}dt + \nonumber \\&&  ~(\textrm{Higher~order~terms}),
\end{eqnarray}
where $\nu_{0 ij}$ is the initial error in Doppler frequency, ${\gamma}_{ij}(t)$ is the error in the Doppler rate  (the  first time derivative of the Doppler frequency), and $ {\alpha}_{ij}(t)$ is the error in the second time derivative the Doppler frequency. We neglect higher order time derivatives of the Doppler frequency because they are sufficiently small.

\subsection{Doppler frequency error at lock acquisition}

 The analysis in this section is restricted to the study of transients which occur at lock acquisition. We shall design the controller such that the transients will occur over a period of a few days. Over this period the terms ${\gamma}_{ij}(t),{\alpha}_{ij}(t)$ will change little and for simplicity we shall approximate these terms as constants equal to their initial errors. The Doppler error as a function of time is then 
\begin{eqnarray}
\nu_{{\rm DE}ij}(t)\approx \nu_{0ij}+{\gamma}_{0ij}t+ \frac{{\alpha}_{0ij}t^2}{2},\end{eqnarray}
 with the initial errors in the Doppler rate, change in the Doppler rate given by  ${\gamma}_{0ij}$  and ${\alpha}_{0ij}$.

 In the frequency domain, the Doppler error at lock acquisition is
\begin{eqnarray}
\nu_{{\rm DE}ij} \approx \nu_{0ij} +\frac{{\gamma}_{0ij} }{i\omega}-\frac{{\alpha}_{0ij} }{2\omega^2}.
\label{Doppler_omega}
\end{eqnarray}
Because we are interested in common and dual arm locking configurations we write
\begin{eqnarray}
\nu_{{\rm DE}+}&=& \nu_{0+}+\frac{\gamma_{0+}}{i\omega}-\frac{\alpha_{0+}}{2\omega^2}, \label{Doppler_omega_plus}\\ 
\nu_{{\rm DE}-} &=& \nu_{0-}+\frac{\gamma_{0-}}{i\omega}-\frac{\alpha_{0-}}{2\omega^2}.
\label{Doppler_omega_minus}
\end{eqnarray}
where  $\nu_{0\pm} = \nu_{012}\pm\nu_{013}$, and similarly $\gamma_{0\pm}=\gamma_{012}\pm\gamma_{013}$ and $\alpha_{0\pm} = \alpha_{012}\pm\alpha_{013}$.

At lock acquisition, the closed loop laser frequency driven by  Doppler error in common arm locking can be found using equations~\ref{Doppler_common_fr} and~\ref{Doppler_omega_plus}.
 \begin{eqnarray}
{\nu_{\rm CL }}|_{0+}= \frac{-G_1(\omega)}{1+G_1(\omega)P_+(\omega)}\left(\nu_{0+}+\frac{\gamma_{0+}}{i\omega}-\frac{\alpha_{0+}}{2\omega^2}\right). \label{common_cl_total}
 \end{eqnarray}

The frequency responses to the error in each term in equation~\ref{common_cl_total} written separately are
\begin{eqnarray}
\mathcal{V}_{+}(\omega)  &=& \frac{-G_1(\omega)}{1+G_1(\omega)P_+(\omega)}, \\
\mathcal{G}_{+}(\omega) &=& \frac{-G_1(\omega)}{i\omega(1+G_1(\omega)P_+(\omega))}, \\
\mathcal{A}_{+}(\omega)  &=& \frac{G_1(\omega)}{2\omega^2(1+G_1(\omega)P_+(\omega))}.
 \end{eqnarray}
These equations show the error in the initial Doppler rate, $\gamma_{0+}$  and change in Doppler rate, 
 $\alpha_{0+}$ are integrated over time and as such they can cause significant pulling.

To limit pulling due to $\nu_{0+}$, the control loop must be AC coupled. The lower unity gain frequency can be determined from the maximum allowable pulling and the error in the Doppler frequency. At the unity gain frequency,
\begin{eqnarray}
|G_1(\omega_{\rm ac})P_+(\omega_{\rm ac})|=1.
\label{equation_ac}
\end{eqnarray}
  At frequencies well below $1/\bar{\tau}$  the common arm sensor can be well approximated by
\begin{eqnarray}
P_+(\omega)|_{f<0.1\textrm{mHz}} \approx i2\omega\bar{\tau}.
\label{approx_fr}
\end{eqnarray}
If the controller gain is flat at low frequencies, the response is rolled off purely by the arm response, and the unity gain frequency and the low frequency gain are by, 
    \begin{eqnarray}
|G_1^*(\omega_{\rm ac})|&=&|G_{1\nu}|=\frac{1}{\omega_{\rm ac}\bar{\tau}},
\label{G_lf}
  \end{eqnarray}
  where we have substituted $G_1^*(\omega)=2G_1(\omega)$. This gain gives the factor of amplification of the Doppler knowledge error. Equation~\ref{G_lf} can be rearranged to obtain the AC coupling frequency for a given low frequency gain 
   \begin{eqnarray}
f_{\rm ac}|_\nu&\ge&\frac{1}{2\pi\bar{\tau} |G_{1\nu}|}.
\label{equation_accoupling}
  \end{eqnarray}
This result sets the lower bound on the unity gain frequency based on the allowable gain of the Doppler frequency error.  To give maximum gain, we set the lower unity gain frequency to as low as possible, for example if we allow $|G_{1\nu}|=1000$ then $f_{\rm ac}\ge4.8\times10^{-6}~~\textrm{Hz}$. To limit pulling due to $\gamma_{0+}$ and $\alpha_{0+}$, further constraints are required.

At lock acquisition, the frequency responses to Doppler errors in dual arm locking can be calculated following a similar method. Using equations~\ref{Y_D_p} and~\ref{Doppler_omega_plus} the frequency responses for the common path of dual arm locking are
\begin{eqnarray}
\mathcal{V}_{D}^{(+)}(\omega) &=& \frac{-G_1(\omega)}{1+G_1(\omega)P_D(\omega)}, \label{V_D_p} \\
\mathcal{G}_{D}^{(+)}(\omega) &=& \frac{-G_1(\omega)}{i\omega(1+G_1(\omega)P_D(\omega))}, \\
\mathcal{A}_{D}^{(+)}(\omega) &=& \frac{G_1(\omega)}{2\omega^2(1+G_1(\omega)P_D(\omega))}.
 \end{eqnarray}
The frequency responses for the difference path can be found using equations~\ref{Y_D_m} and~\ref{Doppler_omega_minus}.
\begin{eqnarray}
\mathcal{V}_{D}^{(-)}(\omega)  &=& \frac{-G_1(\omega)E(\omega)}{i\omega\Delta \tau(1+G_1(\omega)P_D(\omega))}, \\
\mathcal{G}_{D}^{(-)}(\omega) &=& \frac{G_1(\omega)E(\omega)}{\omega^2\Delta \tau(1+G_1(\omega)P_D(\omega))}, \\
\mathcal{A}_{D}^{(-)}(\omega) &=& \frac{G_1(\omega)E(\omega)}{i2\omega^3\Delta \tau(1+G_1(\omega)P_D(\omega))}.\label{A_D_m}
 \end{eqnarray}
 
A comparison of the frequency responses to common and differential Doppler frequency errors in dual arm locking again reveals that it is the differential Doppler frequency errors which will dominate the frequency pulling at start up. This is because the differential path has an extra zero at DC and is inversely proportional to $\Delta \tau$.

\subsection{Laser frequency pulling at lock acquisition}

We now have the tools needed to examine the frequency pulling in steady state and at lock acquisition. The magnitude of the pulling at lock acquisition depends on the sensor used, the controller design, and the Doppler frequency estimate. The estimate is made by simply observing the beatnote frequency between the incoming and outgoing fields prior to switching arm locking on (Appendix~\ref{appendix_DE}).  In this section we determine the pulling that would occur for both common and dual arm locking at lock acquisition. We shall use the low frequency part of the controller in section~\ref{section_controller}, which was designed to provide sufficient gain over the LISA science band to suppress free-running laser frequency noise to below the expected TDI capability and to limit laser frequency pulling. This design has gain less than unity at $f< 4.8~\mu$Hz, implemented with a series of  4 high pass filters. Further details can be found in appendix~\ref{appendix_sensors}. Although this controller was not designed specifically for these sensors, the calculation gives both an estimate of  pulling that can be expected with each sensor, and a comparison of the two sensors. We calculate the frequency pulling that would occur with Doppler frequency estimates with $T={200}$~s averaging, as part of arm-locking initialization.  
 \subsubsection{Common arm locking}
 
The accuracy of the Doppler frequency estimation needed for common arm locking depends on $\bar{\tau},$ the average arm length.  Estimation errors  for two assumptions of the laser frequency noise: $-$free-running lasers and Mach-Zehnder prestabilization$-$ are given in table~\ref{table_estimates}, with details in appendix~\ref{appendix_DE}.  In the case of free running laser noise, the estimates of  $\gamma_{0+}$, and $\alpha_{0+}$ have an error larger than  the maximum value  determined by the orbital motion (appendix~\ref{appendix_orbits}).  Thus we do not use the laser measurement of these quantities, rather we assume them to be zero. With prestabilization the error in the measurement of $\alpha_{0+}$ is larger than its maximum value and thus we assume it to be zero.

\begin{figure}[t!]
\begin{center}
\includegraphics[width = 0.48\textwidth]{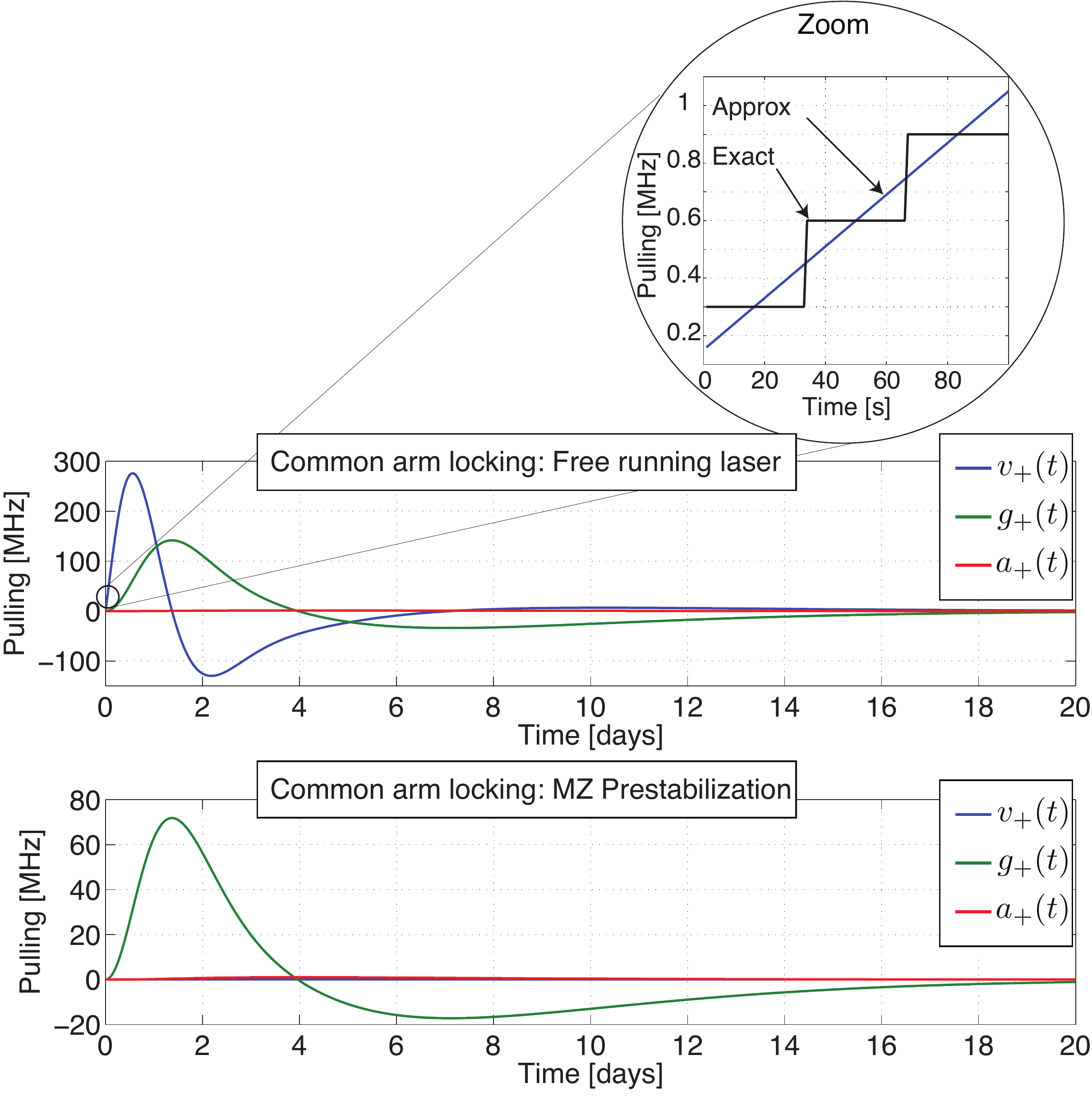}

\caption{Step response of the different drivers of Doppler frequency error for common arm locking with Doppler frequency estimate errors with 200~s averaging made with free running laser noise  (top plot), and prestabilized laser noise (lower plot). \label{CAL_pulling}}
\end{center}
\end{figure}

Frequency pulling during lock acquisition is determined by the step response of the control system. The step response follows from the closed-loop  control system $\nu_{\rm CL}(s)$ according to
\begin{equation}
\nu_{\rm step}(t)={\mathcal L}^{-1}\left(\frac{\nu_{\rm CL}(s)}{s}\right),
\end{equation}
where $s=i\omega$ and ${\mathcal L}^{-1}$ is the inverse Laplace transform operator.
Step responses from the error in the three Doppler derivatives are plotted in figure~\ref{CAL_pulling}, for both free-running and prestabilized lasers.  The small stairstep pattern at the start of acquisition has amplitude  $\nu_{0+}/2$ and period  $\overline{\tau}.$  The maximum pulling from the errors in all the deriviatives of $\nu_{0+}$ is 460 and 90\,MHz, for the respective cases of free-running and prestabilized laser.

  \begin{table}[t!h]
\caption{Initial errors in the common and differential components of Doppler frequency, $\nu_{0+}$,$\nu_{0-}$, Doppler rate, $\gamma_{0+}$,$\gamma_{0-}$, and second time derivative of Doppler frequency, $\alpha_{0+}$,$\alpha_{0-}$, for $T = 200$~s averaging. The columns: free running laser and Mach-Zehnder refer to the initial laser noise spectrum, given by equations~\ref{equation_frln} and~\ref{equation_mzn} respectively. The numbers with $^*$ are the values of the Doppler rate and change in Doppler rate at the nominal start of the mission; the measurement of these parameters with 200~s averaging yields errors larger than the mean values. For the differential Doppler frequency estimates and time derivatives we have taken $\Delta \tau = 0.123$~s.}
\begin{center}
\begin{tabular}{|c|c|c|c|}
\hline
Parameter & Free running laser & Mach-Zehnder & Units\\ \hline 
$\nu_{0+}$ & $6.0\times10^5$ & 45& Hz \\
$\nu_{0-}$ & $2.3\times10^3$ & 0.51 & Hz \\ \hline
$\gamma_{0+}$ & $-4.3^*$ & 2.2 & Hz/s \\
$\gamma_{0-}$ & $1.2^*$ &0.02 & Hz/s \\ \hline
$\alpha_{0+}$ & $-0.37^*$ & $-0.37^*$& $\mu$Hz/s$^2$ \\
$\alpha_{0-}$ & $1.0^*$ & $1.0^*$& $\mu$Hz/s$^2$ \\ \hline
\end{tabular}
\end{center}
\label{table_estimates}
\end{table}%

 \subsubsection{Dual arm locking}

\begin{figure}[t!]
\begin{center}
\includegraphics[width = 0.48\textwidth]{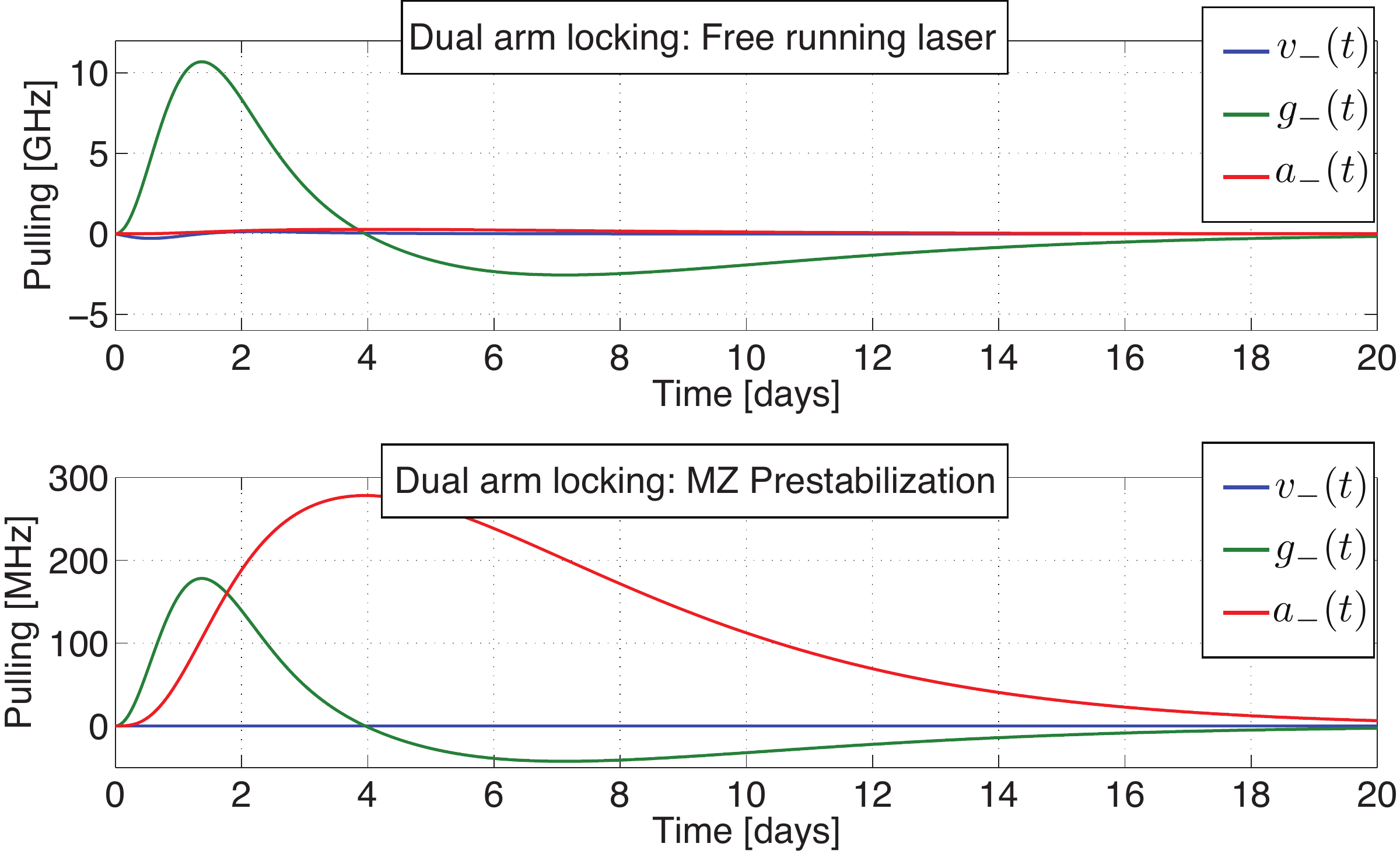}
\caption{Step response of the different drivers of Doppler frequency error for dual arm locking with Doppler frequency estimate errors with 200~s averaging, made with free running laser noise  (top plot) and prestabilized laser noise (lower plot). \label{DAL_pulling}}
\end{center}
\end{figure}

 The error in Doppler estimation for dual arm locking  is computed the same as in common-arm locking, except the relevant delay is  $\Delta \tau = 0.123$~s, corresponding to the  arm length mismatch of the 12-13 arm combination at the start of the mission.  This smaller delay results in smaller error $\nu_{0-}<<\nu_{0+},$ but larger pulling.   As with common arm-locking, when the measurement errors in  $\gamma_{0-} $ and $\alpha_{0-} $ are larger than the maximum values taken from the orbits, they are estimated as zero. The estimation errors are tabulated in table~\ref{table_estimates} and the step responses are plotted in  figure~\ref{DAL_pulling}.
 The maximum pulling with estimation based on free-running lasers is 13\,GHz, unsuitably large for Nd:YAG lasers that have a typical mode-hop free range of approximately 10\,GHz. With prestabilization, the pulling is dominated by the  $\alpha_{0-} $ term, with peak to peak value of 250~MHz. 

 \subsection{Laser frequency pulling in steady state}

We now  consider frequency pulling in steady state, long after lock acquisition. In steady state operation the pulling will be driven by the change in Doppler frequency which occurs as the constellation orbits the Sun. We  determine the pulling in the time domain using a model of the Doppler frequencies over the mission shown in appendix~\ref{appendix_orbits}. 

\subsubsection{Common arm locking}
The laser frequency pulling in steady state for common arm locking is given by the convolution of the $y_+(t)$, the inverse Laplace transform of  $Y_+(\omega)$,  and the common Doppler frequency,
 \begin{eqnarray}
 \nu_{\rm CL} (t) = y_+(t) * \Delta_+(t),
 \end{eqnarray}
where $ *$ is the convolution operator and $\Delta_+(t)$ is the common Doppler shift in the two arms used in common or dual arm locking shown  in figure~\ref{comm_doppler}. The resulting pulling of the laser is shown in the upper plot of figure~\ref{steady_CAL_DAL}.  It can be seen that the laser frequency pulling  with the controller designed in section~\ref{section_controller} is very modest,  less than 8~MHz peak to peak whilst operating in steady state. The pulling is independent of the laser frequency noise as no Doppler frequency estimates are used, and will be an insignificant change compared to the laser frequency drift over this period.

\subsubsection{Dual arm locking}
 
\begin{figure}[h!]
\begin{center}
\includegraphics[width = 0.48\textwidth]{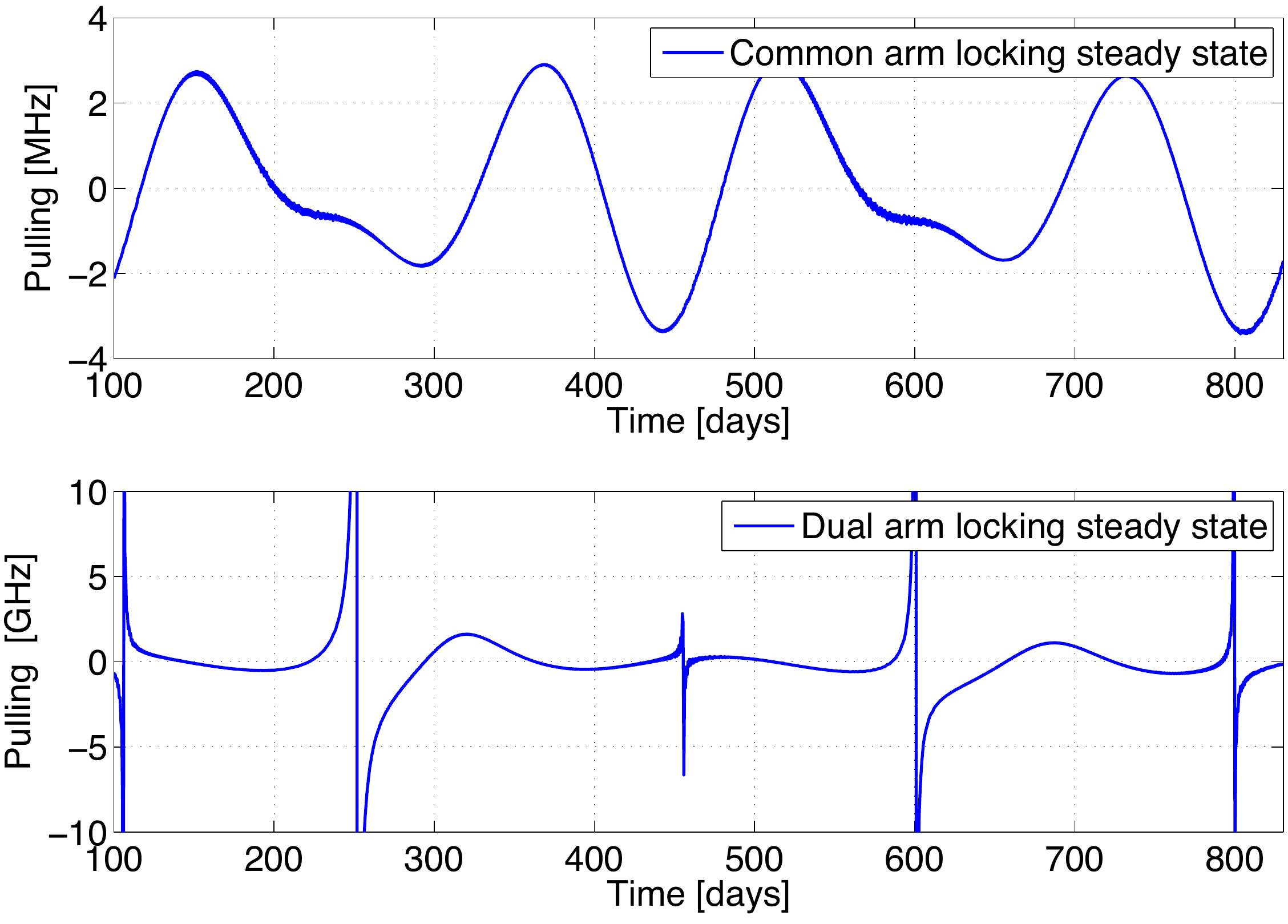}
\caption{The steady state laser frequency pulling of common arm locking (upper plot) and dual arm locking (lower plot).   \label{steady_CAL_DAL}}
\end{center}
\end{figure}

The laser frequency pulling in steady state for dual arm locking is given by  
\begin{eqnarray}
 \nu_{\rm CL} (t) = y_D^{(+)}(t) * \Delta_+(t)+y_D^{(-)}(t) * \Delta_-(t), \label{ss_pulling_dual}
 \end{eqnarray}
 where $\Delta_-(t)$ is the differential Doppler shift of the two arms used by dual arm locking and $ y_D^{(+)}(t) $ and $ y_D^{(-)}(t)$ are the inverse Laplace transforms of  $Y_D^{(+)}(\omega) $ and $Y_D^{(-)}(\omega)$, respectively. The  $ y_D^{(-)}(t)$ term in equation~\ref{ss_pulling_dual} is inversely proportional to the arm length mismatch, which changes significantly throughout the year, dominating the pulling in dual arm locking. The expected pulling for dual arm locking is shown in the lower plot of figure~\ref{steady_CAL_DAL}. This shows unsustainably large pulling of greater than 10~GHz. The pulling which occurs when the arm length mismatch approaches zero becomes rapid and it is unlikely that the lasers could follow this. Also, the time that the frequency pulling is unsustainably large would also be a concern. For example, when the difference in the 12-13 arm passes though zero, the pulling is above 2~GHz for 35 days. If all three arms are available, the central spacecraft in the dual arm locking sensor can be switched around to prevent the arm length mismatch going to zero. In this case the pulling would be reduced.
 
\subsection{Discussion of pulling in common and dual arm locking}

Whilst maintaining large laser frequency noise suppression over the LISA science band (using the controller designed in section~\ref{section_controller}), the laser frequency pulling caused by Doppler frequency errors in common arm locking are relatively modest and do not pose a significant threat of driving the lasers into a mode-hop region. At lock acquisition, with a measurement of the Doppler frequency with only 200~s of averaging, the laser frequency pulling will be less than 460~MHz, if the lasers are free running, or less than 90~MHz if using Mach-Zehnder type prestabilization. This will be further reduced if Fabry-Perot prestabilization is selected. In steady state, the laser frequency pulling is expected to be less than 10~MHz peak to peak, insignificant compared to the drift of a laser frequency over a similar period.  

In contrast, in the form presented in reference~\cite{SuttonPRD}, dual arm locking would have larger pulling,  posing a significant threat of driving the lasers into a mode-hop region, and also of coupling in noise from scattered light sources into the science band. At lock acquisition, with a measurement of the Doppler frequency with only 200~s of averaging, the laser frequency pulling will be approximately 13~GHz, if the lasers are free running, or approximately  250~MHz if using prestabilization. While 250~MHz of pulling at lock acquisition poses only a small threat of pulling the lasers into a mode-hop region, the laser frequency pulling of dual arm locking in steady state is unsustainable. The laser frequency pulling in steady state is dependent on the arm length mismatch which passes through zero twice per year for each pair of arms. Thus the viability of dual arm locking in the form presented in reference~\cite{SuttonPRD} would depend critically on the availability of all inter-spacecraft laser links. If all laser links were available, the pulling could be reduced by switching the central spacecraft when the arm length mismatch becomes small. 

The laser frequency pulling in dual arm locking can be reduced in three ways, 1) reduce the low frequency gain of the controller, which sacrifices the frequency noise suppression in the science band,  2) Implement a scheme to continuously update the Doppler frequency estimate, which would have to be done with a view to not add any noise in the science band or, the solution we explore here,  3) create a modified dual arm locking sensor, a hybrid common- dual arm locking sensor designed in section~\ref{section_modified}. This sensor has the frequency pulling of common arm locking, but maintains other control system advantages of dual arm locking.

\section{Noise sources in Arm Locking}
\label{section_noisesources}

Significant noise sources in the arm locking control system are clock noise, spacecraft motion, and shot noise. These noise sources are common to all arm locking configurations (e.g. single, common, and dual). We develop a general formalism applicable to all arm locking configurations. First though, an overview and approximate amplitude of each of these noise sources is given. 

\subsection{Clock noise}

The  phase of  the photodetector signal is determined by comparing it to an onboard clock (i.e. USO) signal. Consequently, the noise of the clock signal contributes an error in the phase measurement that enters at the phasemeter. The amplitude of the clock noise added to the phase measurement depends on the fractional frequency noise of the clock and scales linearly with heterodyne beatnote frequency, $\Delta_{ij}$ (in units of Hz).  The clock noise added at the $ith$ spacecraft, when measuring the incoming light from the $jth$ spacecraft, is given by 
\begin{eqnarray}
\phi_{Cij}(f) &=& \Delta_{ij}C_{i}(f)~~~~~~\left[\textrm{cycles}/\sqrt{\textrm{Hz}}\right],
\end{eqnarray}
where $C_{i}(f)$ is the clock phase noise normalized to a 1~Hz clock frequency, given by 
\begin{eqnarray}
~~~C_{i}(f) &=&  \frac{y_i(f)}{2\pi f}~~~~~~\left[\textrm{cycles}/\textrm{Hz}\sqrt{\textrm{Hz}}\right],
\end{eqnarray}
and $y(f)$ is the fractional frequency fluctuations of the clock. The amplitude and frequency dependence of $y(f)$ is dependent on the type of clock and operational environment, we use $y(f) = {2.4\times10^{-12}}/{\sqrt{f}}~1/\sqrt{\textrm{Hz}}$, corresponding to  an assumption of the clock stability in LISA~\cite{Gath2}.

Clock noise is correlated for measurements on each spacecraft. Thus, clock noise that enters phasemeter signals on the central spacecraft can add or subtract coherently with common or dual arm locking.

\subsection{Spacecraft motion}

To minimize the disturbances of the space environment on the measurement the spacecraft will fly \textit{drag free} around the local proof masses~\cite{LISAPPA}. This means the location of the spacecraft relative to the proof mass is sensed and controlled to follow the proof mass. Along the axis which the science measurement is made, the spacecraft follows the proof mass with an error of approximately $\Delta \tilde{X} = 2.5\times \sqrt{1+(f/0.3\textrm{Hz})^4}$~nm/$\sqrt{\textrm{Hz}}$~\cite{Gath2}. The inter-spacecraft phase measurements will contain this jitter. Even with this spacecraft motion, the arms represent an excellent frequency reference, as the fractional length stability is still small given the very long arm length. Because dual arm locking utilizes the fractional stability of  the arm length mismatch of two arms, the fractional stability is degraded  accordingly.  Although it is possible to subtract this motion in real time using the same algorithm as used for the LISA science measurement,  this adds complexity and, as we show in section~\ref{section_noisebudget}, the noise performance of arm locking is limited by spacecraft motion for only short periods, twice per year.

If the $ith$ spacecraft moves by an amount of $\Delta \tilde{X}_{ij}(f)$ in the direction of the $jth$ spacecraft, the associated phase shift of the laser field traveling from spacecraft $i$ to spacecraft $j$ is\footnote{The spacecraft motion on the central spacecraft will be correlated in the two arms to some extent. We expect the correlation to be small and neglect it here. }
\begin{eqnarray}
\phi_{Xij}(f) = \frac{\Delta \tilde{X}_{ij}(f)}{\lambda}~~~~~\left[\textrm{cycles}/\sqrt{\textrm{Hz}}\right],
\end{eqnarray}
where $\lambda$ the laser wavelength. Note that in the propagation between the $ith$ and $jth$ spacecraft the laser field will pick up phase noise due to  the motion of both spacecraft.
\subsection{Shot noise}

Shot noise is a phase error  due to the quantization of the electromagnetic field. The phase error of shot noise is inversely proportional to the square-root of the optical power received from the distant spacecraft. Here, we model shot noise as phase error added at the photoreciever. Shot noise is given by~\cite{LISAPPA}
\begin{eqnarray}
\phi_{Sij} = \left( \frac{\hbar c}{2\pi}\frac{1}{\lambda {\rm P}_{d}}\right)^{1/2}~~~~~\left[\textrm{cycles}/\sqrt{\textrm{Hz}}\right],
\end{eqnarray}
where $\lambda$ is the laser wavelength and ${\rm P}_d$ is the optical power received from the distant spacecraft. Shot noise at each $ij$ phase measurement is independent and uncorrelated.  

\begin{figure*}[htb]
\centering
\includegraphics[width=1\textwidth]{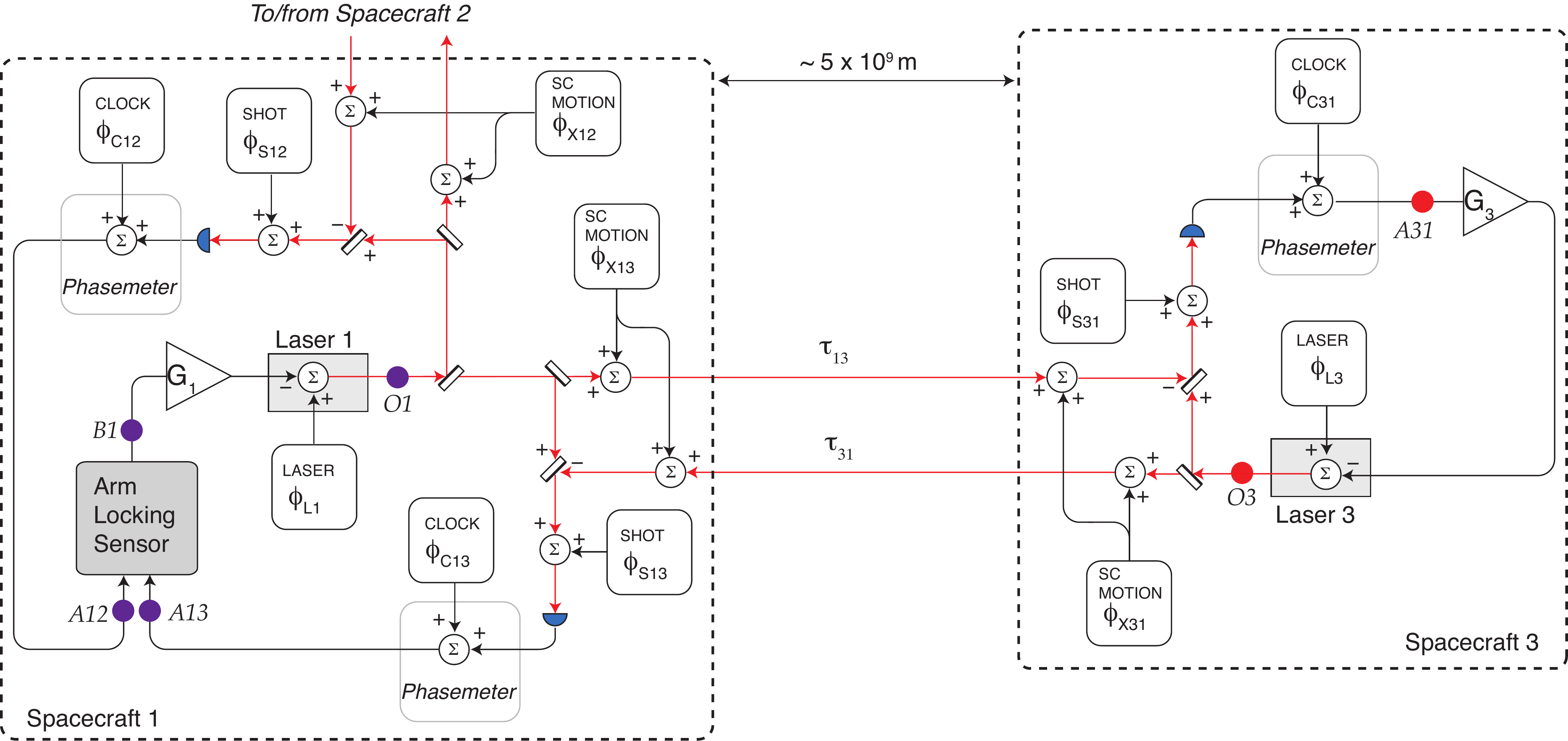}
\caption{\label{dual_arm_configuration_telecon} Schematic of dual arm locking control loop showing where the different noise sources enter the system. These are laser frequency noise, $\phi_{Li}$, clock noise, $\phi_{Cij}$, spacecraft motion, $\phi_{Xij}$, and shot noise, $\phi_{Sij}$  with $i$ and $j$ representing the numbers of the local and distant spacecraft, respectively.}
\end{figure*}
\section{Quantifying the noise performance of Arm Locking}
\label{section_noisebudget}

\subsection{Calculation of noise in arm locking}
In this section we calculate the noise coupling into the arm locking control system building on the calculation in section~\ref{section_setup}. The schematic in figure~\ref{dual_arm_configuration_telecon} indicates where all three noise sources described in section~\ref{section_noisesources} enter the phase measurements. The phasemeter output can be written down in a similar fashion to section~\ref{section_setup}, by propagating each noise source to the phasemeter. Again, to start with, we assume that all control loops are open. The phase measured at the phasemeter on spacecraft 3, facing spacecraft 1 (represented by the red circle labelled $A{31}$) is given by 
\begin{eqnarray}
\phi_{A31}(\omega) &=& \phi_{L3}(\omega)-\phi_{L1}(\omega)e^{-i\omega\tau_{13}}-\phi_{X13}(\omega)e^{-i\omega\tau_{13}}-\nonumber \\ &&\phi_{X31}(\omega)+\phi_{S31}(\omega)+\phi_{C31}(\omega),
\end{eqnarray}
where $\phi_{Lj}$ is the laser phase noise of the laser on the $jth$ spacecraft pre-arm locking. If the laser on spacecraft 3 is phaselocked to the incoming light, the closed loop phase noise at the output of laser~3 (represented by the red circle labelled $O3$) is
 \begin{eqnarray}
\phi_{O3}(\omega) &=& \frac{\phi_{L3}(\omega)}{1+G_3(\omega)}+\nonumber \\ &&\frac{G_3(\omega)}{1+G_3(\omega)}\bigg(\phi_{L1}(\omega)e^{-i\omega\tau_{13}}+\phi_{X13}(\omega)e^{-i\omega\tau_{13}}+\nonumber \\ &&\phi_{X31}(\omega)-\phi_{S31}(\omega)-\phi_{C31}(\omega)\bigg),
\end{eqnarray}
where $G_3(\omega)$ is the frequency response of the phase locking controller on spacecraft 3. With the laser on the spacecraft 3 phaselocked to the incoming light ($G_3(\omega)\gg1$), but  the arm locking control loop open ($G_1(\omega)=0$), the phase at the phasemeter output on spacecraft 1 is given by

 \begin{eqnarray}
\phi_{A13}(\omega) &\approx&  {\phi_{L1}(\omega)P_{13}(\omega)}- \nonumber  \\ &&   \phi_{X13}(\omega)\left(1+e^{-i2\omega\tau_{13}}\right)- 
2\phi_{X31}(\omega)e^{-i\omega\tau_{13}}
 + \nonumber  \\ &&  \phi_{S13}(\omega)+\phi_{S31}(\omega)e^{-i\omega\tau_{13}} + \nonumber  \\ &&  {\phi_{C13}(\omega)+\phi_{C31}(\omega)e^{-i\omega\tau_{13}}}, \end{eqnarray}
 where successive terms represent: laser phase noise, motion of the spacecraft, shot noise, and clock noise. Similarly, the phase at point $A{12}$ is
 \begin{eqnarray}
 \phi_{A12}(\omega) &\approx&  {\phi_{L1}(\omega)P_{12}(\omega)}
- \nonumber  \\ &&  {\phi_{X12}(\omega) \left(1+e^{-i2\omega\tau_{12}}\right)-2\phi_{X21}(\omega) e^{-i\omega\tau_{12}}}
 +\nonumber \\ && {\phi_{S12}(\omega) +\phi_{S21}(\omega) e^{-i\omega\tau_{12}}} +  \nonumber  \\ && {\phi_{C12}(\omega) +\phi_{C21}(\omega) e^{-i\omega\tau_{12}}}.
 \end{eqnarray}
The phase measurements that enter the arm locking sensor on the central spacecraft can be rewritten in the compact form:
  \begin{eqnarray}
 \Phi_{A1} = \left[\begin{array}{c}\phi_{A13}(\omega) \\\phi_{A12}(\omega)\end{array}\right]= \left[ \textbf{N}_L+\textbf{N}_S+\textbf{N}_C+\textbf{N}_X\right],
 \label{phase_noise_A}
 \end{eqnarray}
 where  $\textbf{N}_L$ is the laser noise sensed at the central spacecraft photorecievers, $\textbf{N}_S$ is the shot noise, $\textbf{N}_C$ is the clock noise, and $\textbf{N}_X$ is the optical bench displacement noise. These are given by
 \begin{eqnarray}
\textbf{N}_L&=&\left[\begin{array}{c}P_{12}(\omega)\phi_{L1}(\omega)  \\P_{13}(\omega)\phi_{L1}(\omega) \end{array}\right], \textbf{N}_S = \left[\begin{array}{c}\phi_{S12}(\omega) +\phi_{S21}(\omega) \nonumber \\ \phi_{S13}(\omega) +\phi_{S31}(\omega) \end{array}\right],\\
\textbf{N}_C &=& \left[\begin{array}{c} \Delta_{12}(C_1(\omega) +C_2(\omega) ) \\ \Delta_{13}(C_1(\omega) +C_3(\omega) )\end{array}\right],\nonumber \\ \textbf{N}_X &=& \left[\begin{array}{c}-\phi_{X12}(\omega) \left(1+e^{-i2\omega\tau_{12}}\right)-2\phi_{X21}(\omega)  \\ -\phi_{X13}(\omega) \left(1+e^{-i2\omega\tau_{13}}\right)-2\phi_{X31}(\omega) \end{array}\right].
 \end{eqnarray}
 where we have dropped the time delays in uncorrelated noise terms and assumed the heterodyne frequency on spacecraft 2 and 3 is equal to the one way Doppler shift.

The frequency noise at the laser output with the common arm locking control loop closed  (point $O1$ in figure~\ref{dual_arm_configuration_telecon}) is given by 
\begin{eqnarray}
\phi_{O1}|_{+}(\omega)   &=& \phi_{L1}(\omega) - G_1(\omega)\phi_{B1}|_{+}^{\rm cl}(\omega) , \nonumber \\
&=&\frac{\phi_{L1}(\omega) } {1+G_1(\omega)P_+(\omega)} - \nonumber \\ &&\frac{G_1(\omega)}{1+G_1(\omega)P_+(\omega)} \textbf{S}_{+}\left[ \textbf{N}_S+\textbf{N}_C+\textbf{N}_X\right], \nonumber \\
\label{equation_noise_common}
\end{eqnarray}
where $\phi_{B1}|_{+} = \textbf{S}_{+} \Phi_{A1}$. The control loop acts differently on  laser frequency noise from other noise sources. In the high gain limit, the laser noise is suppressed by the closed loop gain, $P_+(\omega)G_1(\omega)$, whereas the other noise sources are maximumly imposed on the laser light.  

The closed loop laser frequency noise for other arm locking sensors can be calculated in a similar fashion.  The vector representation  and  frequency response  of different arm locking sensors are listed in table~\ref{table_sensors}.
The closed loop noise at the laser output for the dual arm locking is given by
\begin{eqnarray}
\phi_{O1}|_{D}(\omega)   &=& \phi_{L1}(\omega) - G_1(\omega)\phi_{B1}|_{D}^{\rm cl}(\omega) , \nonumber \\
&=& \frac{\phi_{L1}(\omega) } {1+G_1(\omega)P_{D}(\omega)} - \nonumber \\ && \frac{G_1(\omega)}{1+G_1(\omega)P_{D}(\omega)} \textbf{S}_{D}\left[ \textbf{N}_S+\textbf{N}_C+\textbf{N}_X\right], \nonumber \\
\label{equation_noise_dual}
\end{eqnarray}
where $\phi_{B1}|_{D} = \textbf{S}_D\Phi_{A1}$. 
\begin{table*}[htdp]
\caption{LISA parameters and amplitude of noise sources}
\begin{center}
\begin{tabular}{| l | c |l|l|}\hline
Parameter & Symbol & Value & Units\\
 \hline
Average arm length & $\bar{L}$ & $5\times10^9 $&m\\ \hline
Differential arm length & $\Delta L$ & $\leq 76,500,000$ & m\\ \hline
Laser wavelength & $\lambda$ & 1064& nm\\ \hline
Doppler shift arm 13	&$\Delta_{13}$& 15 &MHz \\\hline
Doppler shift arm 12	&$\Delta_{12}$& -14 &MHz \\\hline
Fluctuations of clock & $y(t)$ & $2.4\times10^{-12}/\sqrt{f}$ &$1/\sqrt{\textrm{Hz}}$\\ \hline
Shot noise 			& $\phi_{Sij}$ & 10 &$\mu$cycles/$\sqrt{\textrm{Hz}}$ \\ \hline
Spacecraft motion			& $\phi_{Xij} $ &  $2.5\times \sqrt{1+(f/0.3\textrm{Hz})^4}$ &mcycles/$\sqrt{\textrm{Hz}}$ \\ \hline
TDI capability &$ \nu_{\rm TDI}(f)$&$300\times(1+(3~\textrm{mHz}/f)^2)$&${\textrm{Hz}}/{\sqrt{\textrm{Hz}}}$\\ \hline
 \end{tabular}
\end{center}
\label{table_parameters}
\end{table*}

\subsection{The noise floor of arm locking}
Armed with equations \ref{equation_noise_common} and \ref{equation_noise_dual}, the noise floor of common and dual arm locking can be found.  In this section we neglect laser frequency noise, equivalent to assuming the controller has infinite gain. In section~\ref{section_controller} an arm locking controller is designed and the total noise budget, including laser noise, is plotted in section~\ref{section_performance}. In the limit of high controller gain,  noise sources other than laser frequency noise have little dependence on the controller gain, depending primarily on the sensor and the transfer function into the phase measurement\footnote{The high controller gain assumption mentioned here is valid over the LISA science band as the magnitude of the controller gain is always greater than 100. If the controller gain is low, the noise coupling will be reduced.}. 

\begin{figure}
\begin{center}
\includegraphics[width=0.48\textwidth]{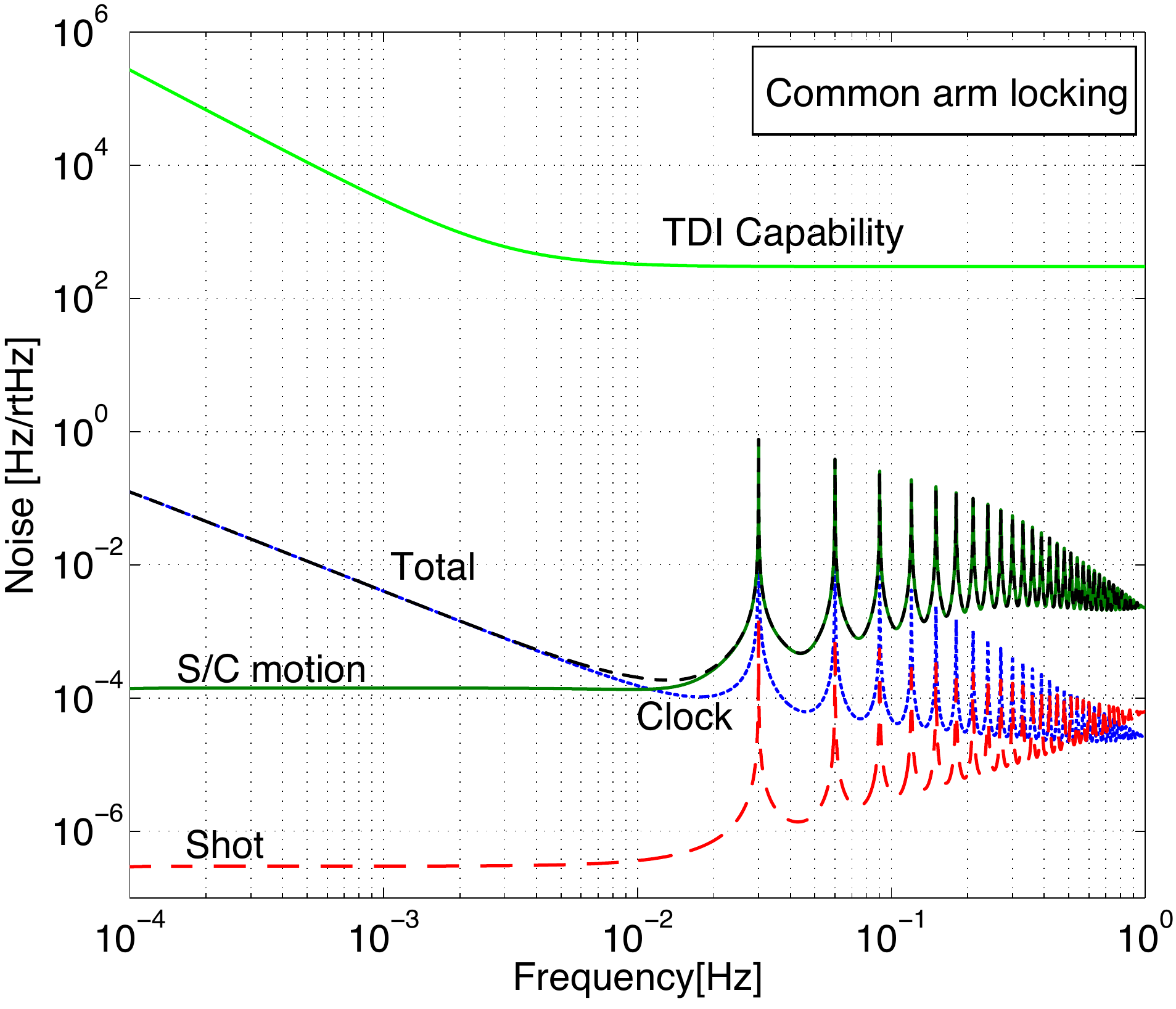}
\caption{\label{common_al_noisefloor} Noise of common arm locking at the laser output. The total comprises clock noise, spacecraft motion, and shot noise. This noise budget is plotted for arm length mismatch $2\Delta \tau = 0.51$s, corresponding to the maximum arm length mismatch. }
\end{center}
\end{figure}

\begin{figure}
\begin{center}
\includegraphics[width=0.48\textwidth]{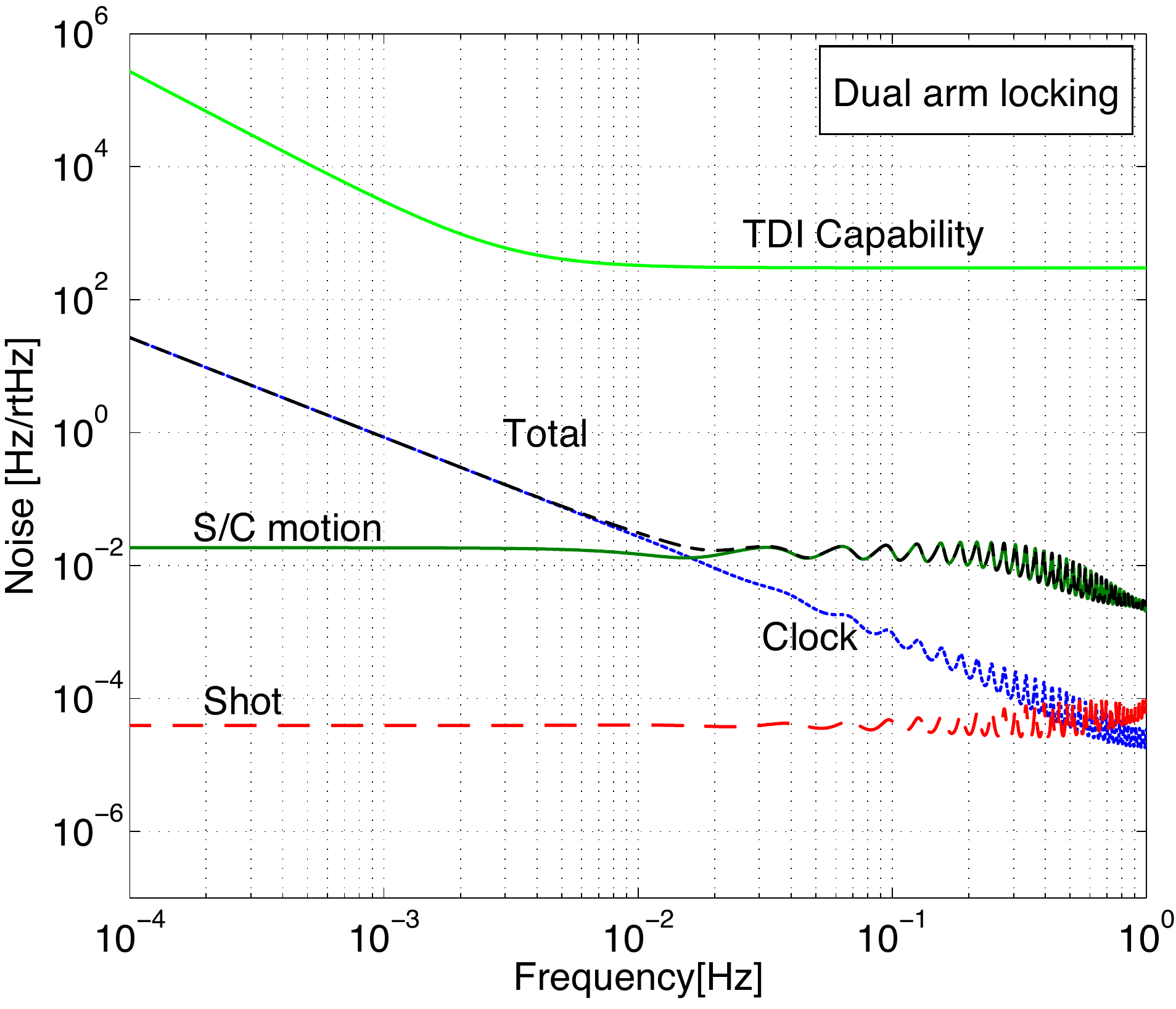}
\caption{\label{dual_al_noisefloor_0_51} Noise of dual arm locking at the laser output. The total comprises clock noise, spacecraft motion, and shot noise. This noise budget is plotted for arm length mismatch $2\Delta \tau = 0.51$s, corresponding to the maximum arm length mismatch.}
\end{center}
\end{figure}

The noise floors of common and dual arm locking are plotted   in figure~\ref{common_al_noisefloor} and~\ref{dual_al_noisefloor_0_51} respectively, in units of frequency noise. These figures were generated assuming high controller gain,  the maximum arm length difference that occurs in the mission, $2\Delta \tau = 0.51$~s,  and parameters listed in table~\ref{table_parameters}.  The total noise (dashed black curve) is a quadrature sum of shot noise (red curve), spacecraft motion (green curve), and the clock noise (blue curve). For the parameters used here, clock noise represents the limiting noise source of dual arm locking below 20~mHz and spacecraft motion represents a noise limit at frequencies above this. Note that clock noise is linearly dependent on the heterodyne frequency at each phasemeter and this noise budget has the worst combination of heterodyne frequencies that can occur for dual arm locking: a maximum difference in Doppler shifts  between the two arms, $\Delta_{13}-\Delta_{12}=29~$MHz. In the science band, shot noise is always smaller than the both clock and spacecraft motion, though  it dominates above band as clock noise and spacecraft motion roll off.

 Also plotted in figures~\ref{common_al_noisefloor} and~\ref{dual_al_noisefloor_0_51} is the expected TDI capability (light green curve). The TDI capability is relaxed below 3~mHz, where acceleration noise of the proof mass becomes significant. The frequency noise curve for the expected TDI capability  is 
 \begin{eqnarray}
\Delta \nu_{\rm TDI}(f)\approx300\times(1+(3~\textrm{mHz}/f)^2)~~\frac{\textrm{Hz}}{\sqrt{\textrm{Hz}}}.
\end{eqnarray}
This capability assumes that TDI will be limited by the time delay error, currently expected to be 3~ns~\cite{FCST}. For  a 3~ns time delay error,  TDI will be capable of suppressing frequency noise of 300~${\textrm{Hz}}{\sqrt{\textrm{Hz}}}$ down to the required value of residual frequency noise ($
 \approx2$~pm/$\sqrt{\rm Hz}$ per inter-spacecraft link)~\cite{McKenzieTDI}. 

\begin{figure}
\begin{center}
\includegraphics[width=.48\textwidth]{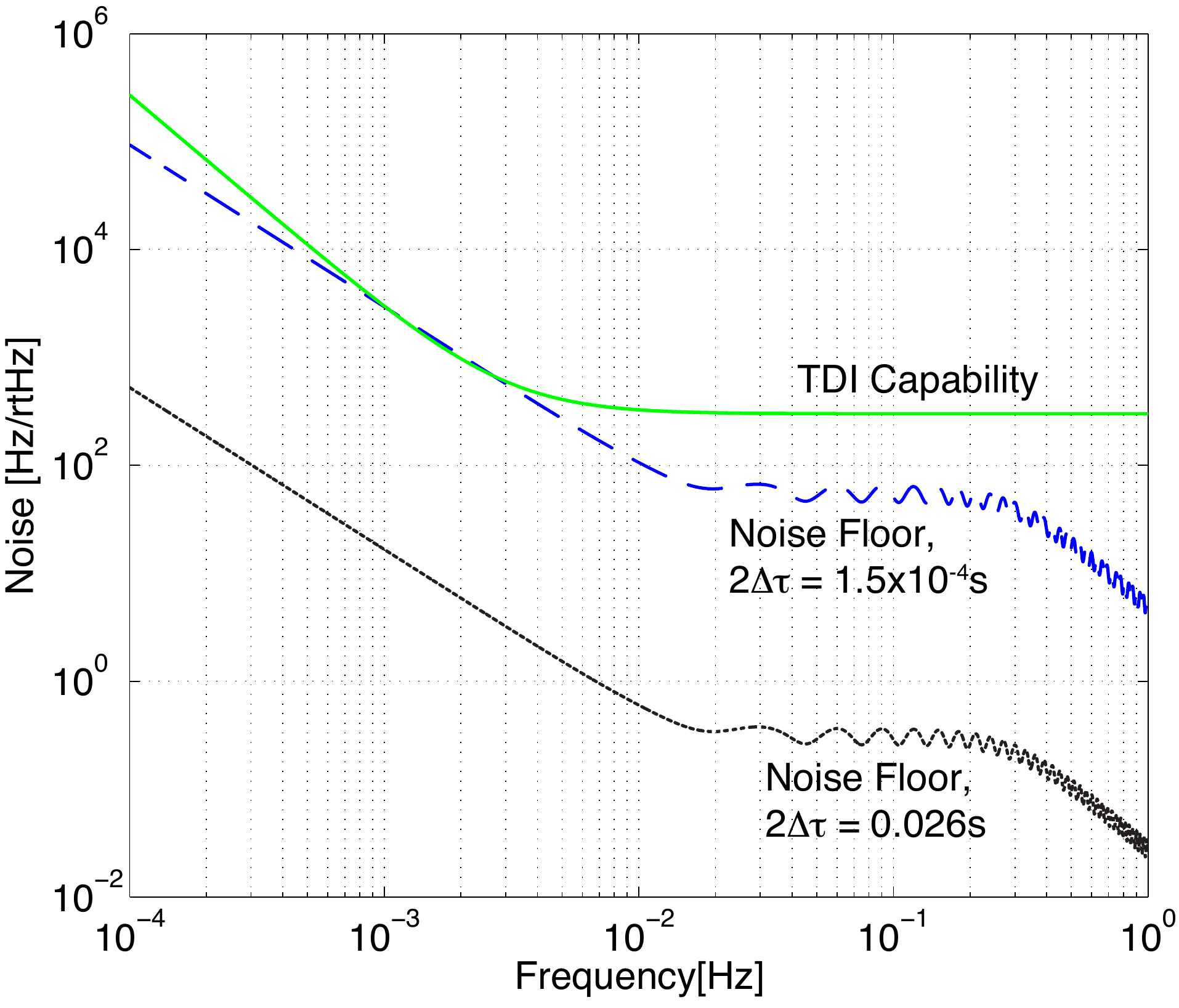}
\caption{\label{dal_t_0_026-t_2_9e-4} Total noise  of dual arm locking at the laser output for different arm length mismatches. The lower   curve has arm length mismatch of $2\Delta \tau = 0.026$~s, corresponding to the minimum arm length mismatch that occurs during the mission if any two of three arms are available to choose from. The upper curve has arm length mismatch of $2\Delta \tau = 1.5\times10^{-4}$~s, which is the smallest arm length mismatch that the total noise is below the TDI capability (green solid curve). }
\end{center}
\end{figure}

Comparison of figures~\ref{common_al_noisefloor} and~\ref{dual_al_noisefloor_0_51}  shows the noise floor of common arm locking is lower than that of dual arm locking over most of the science band, although it peaks higher at frequencies corresponding to nulls in the common arm locking sensor. The noise floors of both common and dual arm locking are substantially lower than the expected TDI capability. The noise floor in figure~\ref{dual_al_noisefloor_0_51} is the optimal noise performance of dual arm locking, as it is plotted with the maximum arm length difference. While the noise floor of dual arm locking varies with inverse proportionality to the arm length mismatch, the noise floor of common arm locking is largely independent of arm length mismatch. Only the magnitude of the peaks at  frequencies of $f = 1/\bar{\tau}$ and integer multiples change with differential arm length change. The dual arm locking noise floor is plotted with smaller arm length mismatches in figure~\ref{dal_t_0_026-t_2_9e-4}. The lower curve (dotted black curve) is plotted with  $2\Delta \tau = 0.026$s,  corresponding to the minimum arm length that arm locking would have to operate with if all inter-spacecraft links are available and the central spacecraft was switched to give the maximum arm length difference~\cite{Gath2}. The noise floor is a factor of 100 below the TDI capability at the closest point showing that if all three inter-spacecraft links are available, the noise floor of dual arm locking is sufficient for the entire mission. 

In the case where an inter-spacecraft laser link fails, only two arms will be available, and therefore the central spacecraft cannot be switched when the arm length mismatch becomes small. In this case, the arm length difference will pass through zero twice per year and at this time the noise floor becomes infinite. The upper  curve (dashed blue) in figure~\ref{dal_t_0_026-t_2_9e-4} shows the smallest arm length mismatch where the total noise floor is below the TDI capability, which occurs for $2\Delta \tau = 150~\mu$s, equivalent to $2\Delta L =44$~km. Here, the clock noise, the dominant effect, becomes equal to the TDI capability near 3~mHz. The spacecraft motion is also less than a factor of 10 away from the TDI capability at this arm length mismatch.  With arm length mismatches smaller than this, the TDI capability cannot be met with dual arm locking.

\begin{figure*}[htb]
\centering
\includegraphics[width=1\textwidth]{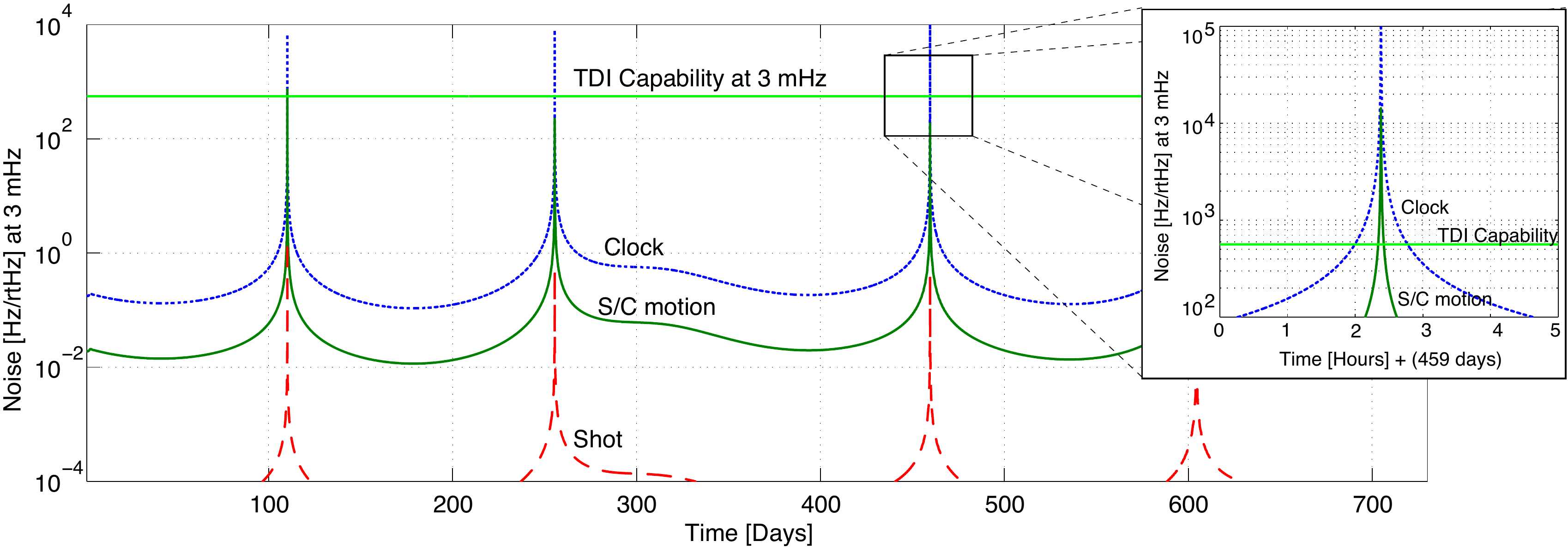}
\caption{\label{dual_noise_at_3mHz} The noise sources of dual arm locking measured at 3~mHz   over the first two years of the LISA mission. This plot  assumes only two of the three LISA arms are available, preventing the central spacecraft from  being switched at small arm length mismatch. The heterodyne frequency assumed for the clock noise curve here is pessimistic, as we have assumed the worst case that occurs in the mission ($\Delta_{12} - \Delta_{13} = 29$~MHz ) for the duration of this plot. The heterodyne frequency over the mission lifetime depends on the Doppler shift, which we have neglected for simplicity.  }
\end{figure*}

The variation of the dual arm locking noise floor  due to the changing arm length mismatch can be seen in figure~\ref{dual_noise_at_3mHz}. This shows the noise sources at 3~mHz over the first two years of the LISA mission assuming only two of the three LISA arms are available. In this case,  the noise floor at 3~mHz is below the TDI capability for the vast majority of the time and breaches the TDI capability for only short periods, twice per year. It also provides some indication of how infrequently and how short a duration dual arm locking can not meet the TDI capability in  case of a critical failure of one arm. The noise performance is insufficient to meet the TDI capability for approximately 1 hour, twice per year.

 \section{Modification of the dual arm locking sensor}
 \label{section_modified}

The benefits of the dual arm locking sensor in terms of achieving high gain across the science band  as well as moving the first null in the sensor to above 2~Hz are significant, but the laser frequency pulling that occurs in steady state is unsustainable. The frequency pulling  at lock acquisition is also larger than in common arm locking.   In terms of noise performance, for the majority of the LISA orbit the noise floor of dual arm locking is well below the TDI capability. However, when the arm length mismatch is small the noise performance of dual arm locking is degraded substantially.  Here we introduce a modification to the dual arm locking sensor to combine the benefits of common and dual arm locking.

The modified dual arm locking sensor comprises the common arm sensor at frequencies below the first null of the arm response and the dual arm locking sensor at frequencies above the first null, with a smooth transition between the two.  This sensor provides the stability and gain advantages of dual arm locking and recovers frequency pulling characteristics and the low frequency noise floor of common arm locking. There are no hardware changes associated with the control system modification. 

\subsection{Design of the modified dual arm locking sensor}

\begin{figure}[htb]
\centering
\includegraphics[width=.32\textwidth]{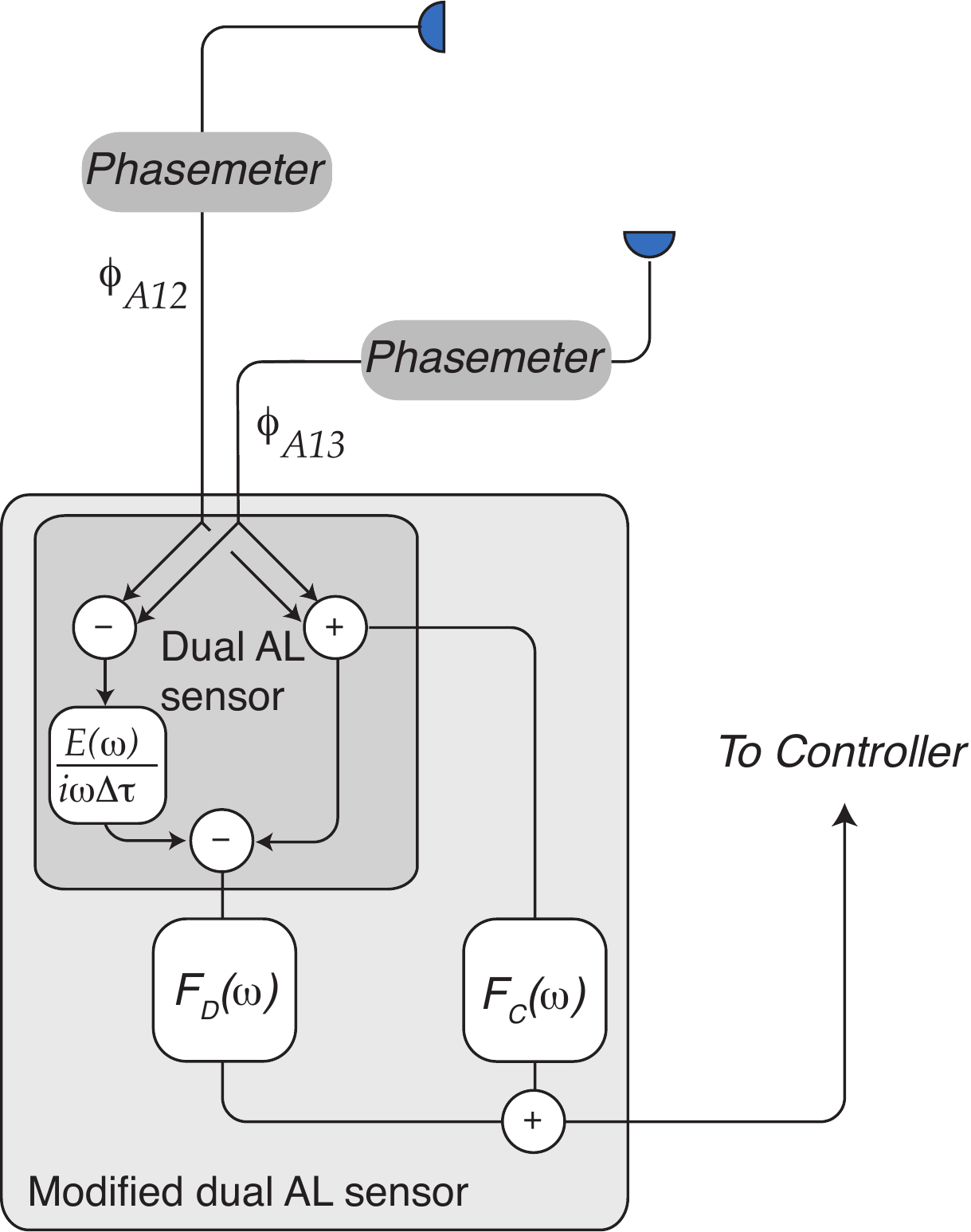}
\caption{\label{modified_dual_sensor} Block diagram of the modified dual arm locking sensor.}
\end{figure}

\begin{figure}
\begin{center}
\includegraphics[width=0.48\textwidth]{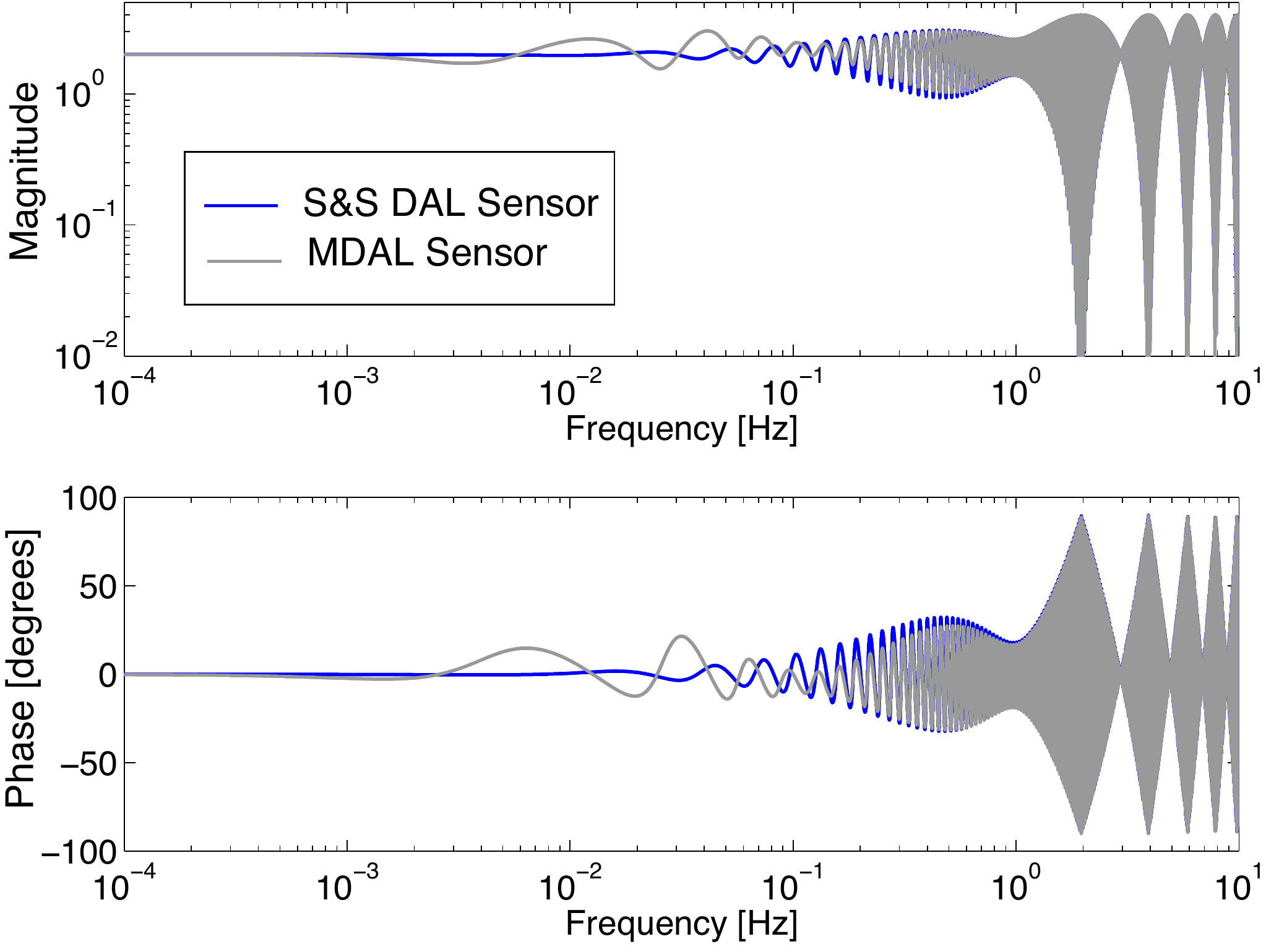}
\caption{\label{new_dual_sensors} Bode plot of the modified dual arm locking sensor (grey curve) and the Sutton and Shaddock~\cite{SuttonPRD} (S\&S) dual arm locking sensor (blue curve). }
\end{center}
\end{figure}

The Bode plot of the modified dual arm locking sensor is plotted in figure~\ref{new_dual_sensors} (grey curve) along with the dual arm locking sensor proposed by Sutton and Shaddock~\cite{SuttonPRD} (blue curve). A block diagram of the modified dual arm locking sensor is shown in figure~\ref{modified_dual_sensor}. The sensor is located on the `central' spacecraft and uses the usual inter-spacecraft phase measurements from two phasemeters which measure the interspacecraft signal, $\phi_{A13}(\omega)$ and $\phi_{A12}(\omega)$ (assuming spacecraft 1 is the central spacecraft). These phase measurements are combined to make the common and difference sensors, by taking the sum and difference, respectively. These signals are then used to construct the modified dual sensor.

\begin{figure}[htb]
\begin{center}
\includegraphics[width=.48\textwidth]{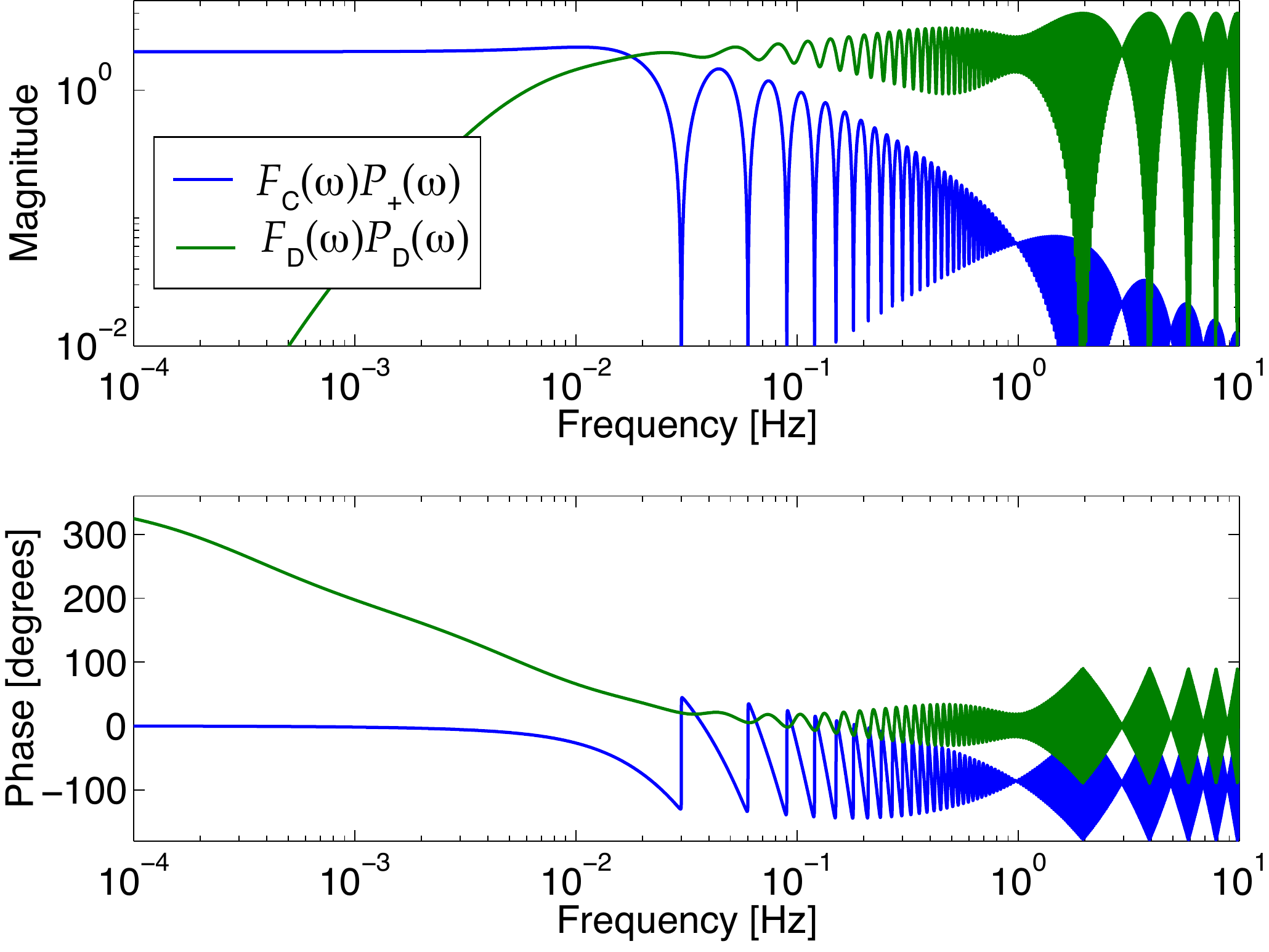}
\caption{\label{common_diff_components} Common and dual components of the modified dual arm locking sensor. The sum of these gives the modified dual arm locking sensor.}
\end{center}
\end{figure}

The modified dual arm locking sensor is designed so that the common arm sensor dominates below the first null of the arm ($f<1/\bar{\tau}$) and the dual sensor dominates above this frequency. The components of the modified sensor are plotted in figure~\ref{common_diff_components}. The frequency response of the sensor is given by
\begin{eqnarray}
P_{M}(\omega) &= &\underbrace{F_C(\omega){P_+(\omega)}}_{\textrm{Common~Part}}+ \underbrace{F_{D}(\omega)P_{D}(\omega)}_{\textrm{Dual~Part}},
\label{P_MD11}
\end{eqnarray}
where the functions $F_C(\omega)$ and $F_{D}(\omega)$ are filters designed to smooth the crossover from the common to dual sensors given by
\begin{eqnarray}
F_C(\omega)& =&\frac{g_a g_b(s+z_b)}{s(s+p_b)}, \\
F_D(\omega) &=&\frac{g_cg_dg_e s^4}{(s+p_c)(s+p_d)(s+p_e)^2},
\end{eqnarray} 
with the parameters given in table~\ref{table_mdal}. 
\begin{table}[htdp]
\caption{Parameters of modified dual arm locking filters}
\begin{center}
\begin{tabular}{|c|c|l|l|}
\hline
Filter & zeros (radians/s) & Poles (radians/s) & Gain \\ \hline 
$F_C(\omega) $&	&$p_a =0$ & $g_a =(\bar{\tau})^{-1}$ \\
 &$z_b = 2\pi\times5/(13\bar{\tau})$	&$p_b =2\pi\times5/(2\bar{\tau})$& $g_b = p_b/z_b$ \\ \hline
 $F_D(\omega) $& 0	&$p_c =7/(5\bar{\tau})$ & $g_c = 1$ \\ 
 &0	&$p_d =11/(20\bar{\tau})$ & $g_d = 1$ \\ &	0 &$p_e =2\pi\times1/(90\bar{\tau})$ & $g_e = 1$ \\  \hline \end{tabular}
\end{center}
\label{table_mdal}
\end{table}%

Equation~\ref{P_MD11} can be rewritten as a function of the common and difference sensors
\begin{eqnarray}
&&P_{M}(\omega)
= P_+(\omega)H_+(\omega)-{P_-(\omega)}H_-(\omega),
\end{eqnarray}
with
\begin{eqnarray}
H_+(\omega)&=& {F_C(\omega)}+F_{D}(\omega),\label{Hplus} \\H_-(\omega)&=& \frac{E(\omega)}{i\omega\Delta{\tau}}F_D(\omega).\label{Hminus}
\end{eqnarray}

 For the sensor in figure~\ref{new_dual_sensors},  $F_C(\omega)$ has a pole at DC along with a lead compensator with a zero at $5/(13\bar{\tau})$~Hz and a pole at $5/(2\bar{\tau})$~Hz. $F_D(\omega)$ is a high pass filter made from four zeros at DC plus poles at $7/(10\pi\bar{\tau})$~Hz, $11/(20\pi\bar{\tau})$~Hz, and two poles at $1/(90\bar{\tau})$~Hz. The frequency response of the modified dual arm locking sensor is similar to the dual arm locking sensor, with an almost flat response below the first null with a magnitude of 2.

\subsection{Frequency pulling in modified dual arm locking}
\label{subsection_mddoppler}
\begin{figure}[h!]
\begin{center}
\includegraphics[width = 0.48\textwidth]{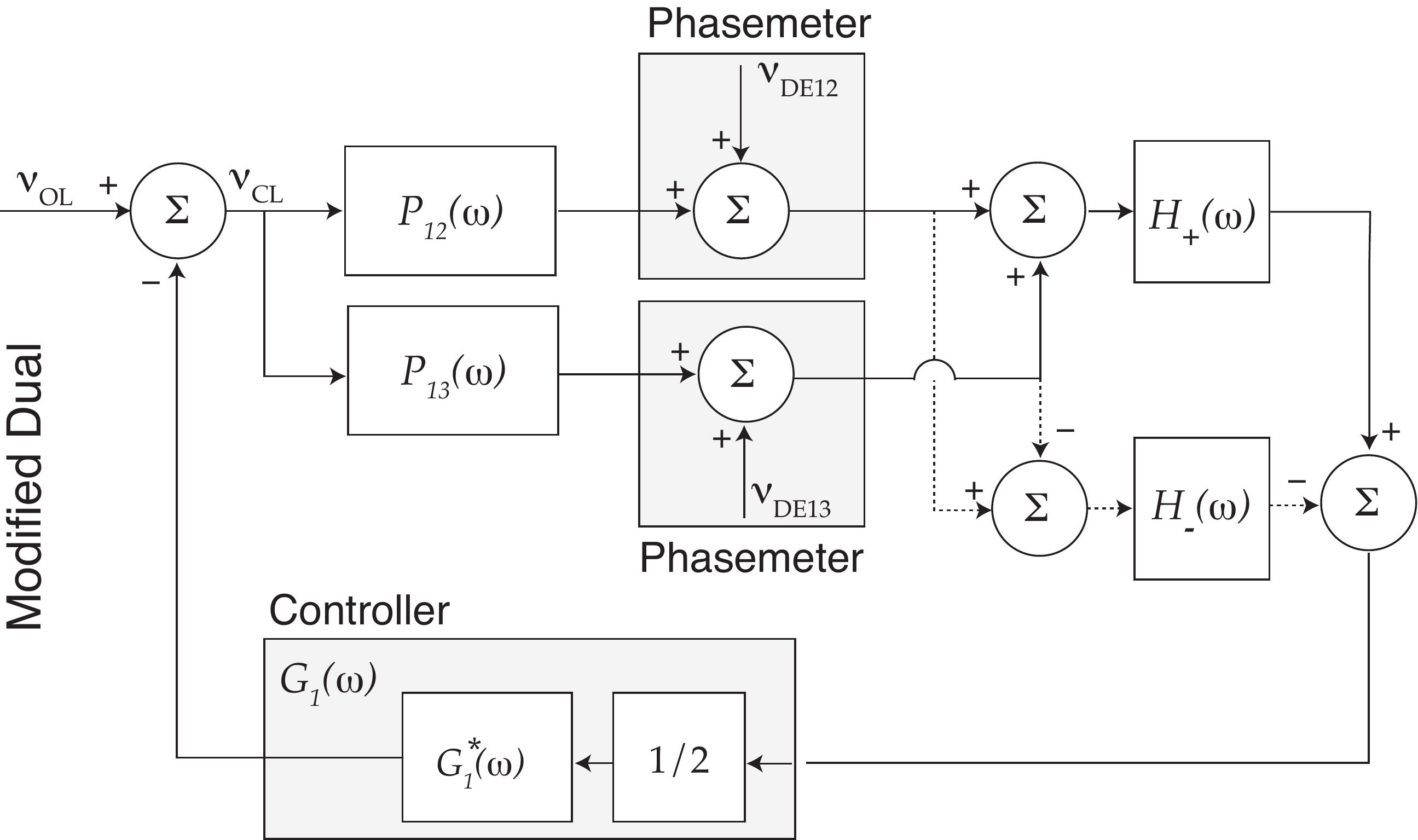}
\caption{Block diagram of the modified arm locking control system showing where the Doppler frequency errors, $\nu_{\rm DE12}$, $\nu_{\rm DE13}$ enter the control loop. \label{Doppler_error_all}}
\end{center}
\end{figure}
 
Following the formalism set out in section~\ref{section_doppler}, the laser frequency pulling in modified dual arm locking is 
\begin{eqnarray}
\nu_{\rm CL}|_{M} = \frac{-G_1(\omega)\textbf{S}_{M}\textbf{V}_{\rm DE}}{1+G_1(\omega)P_{M}(\omega)}.
\end{eqnarray}
where $\textbf{S}_{M}$ is the signal mapping vector given in table~\ref{table_sensors}. Like dual arm locking, there will be pulling due to both the common and differential errors in the Doppler frequency.  The frequency responses  written in terms of equations~\ref{Hplus} and~\ref{Hminus} are
\begin{eqnarray}
Y_{M}^{(\pm)} (\omega)= \frac{\nu_{\rm CL}|_{M}}{\nu_{\rm DE\pm}}   = \frac{-G_1(\omega)H_\pm(\omega)}{1+G_1(\omega)P_{M}(\omega)},  \label{Y_MD_m} 
\end{eqnarray}
where the `$+$' is used for the common path and the `$-$' for the difference path. Figure~\ref{Doppler_error_all} is a block diagram showing where the Doppler error enters the modified dual arm locking sensor.

\subsubsection{Pulling at lock acquisition}

The frequency responses at lock acquisition can be found using equation~\ref{Y_MD_m} with the appropriate $+$ or $-$ sign along with equations \ref{Doppler_omega_plus} and~\ref{Doppler_omega_minus}. The frequency responses are written compactly as
\begin{eqnarray}
\mathcal{V}_{M}^{(\pm)}(\omega)  &=& \frac{-G_1(\omega)H_\pm(\omega)}{1+G_1(\omega)P_{M}(\omega)}, \\
\mathcal{G}_{M}^{(\pm)}(\omega)&=& \frac{-G_1(\omega)H_\pm(\omega)}{i\omega(1+G_1(\omega)P_{M}(\omega))},\\
\mathcal{A}_{M}^{(\pm)}(\omega)&=& \frac{G_1(\omega)H_\pm(\omega)}{2\omega^2(1+G_1(\omega)P_{M}(\omega))}.\nonumber \\
&&
 \end{eqnarray}
 again with the $\pm$ set to a `$+$' for the common path and the `$-$' for the difference path. 
 
  \subsection{Noise performance}
For modified dual arm locking,   the noise at the laser output is given by
\begin{eqnarray}
\phi_{O1}|_{M}(\omega)   &=& \frac{\phi_{L1}(\omega) } {1+P_{M}(\omega)G_1(\omega)} - \nonumber \\ && \frac{G_1(\omega)}{1+P_{M}(\omega)G_1(\omega)} \textbf{S}_{M}\left[ \textbf{N}_S+\textbf{N}_C+\textbf{N}_X\right], \nonumber \\
\label{equation_noise_mdual}
\end{eqnarray}

\begin{figure}[htb]
\centering
\includegraphics[width=.48\textwidth]{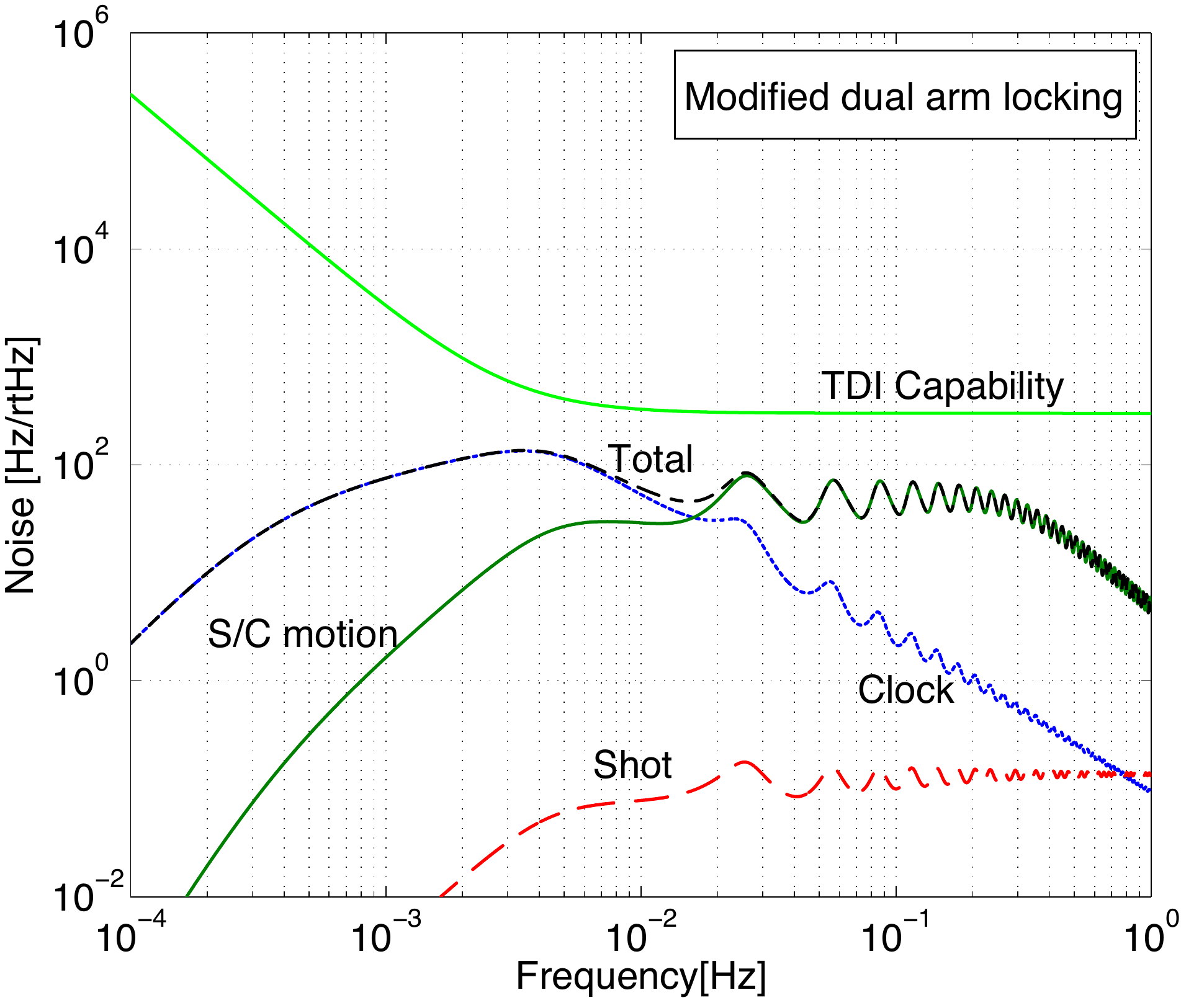}
\caption{\label{mdal_t_5_9e-4} Noise at the laser output for modified dual arm locking with arm length mismatch of $2\Delta \tau = 1.5\times10^{-4}$s. }
\end{figure}

The improved noise performance of modified dual arm locking can be seen by comparing the total noise in figure~\ref{mdal_t_5_9e-4} to the upper curve in figure~\ref{dal_t_0_026-t_2_9e-4}. Figure~\ref{mdal_t_5_9e-4} was plotted assuming the controller has infinite gain (so laser frequency noise is suppressed to zero). The noise floor of modified dual arm locking is similar to that of dual arm locking above $\sim$10~mHz but asymptotes to the noise floor of common arm locking below this. The noise floor of modified dual arm locking at this arm length mismatch no longer breaches the TDI capability.

\section{Design of the Arm Locking Controller }
\label{section_controller}
The arm locking controller is designed for a sensor that has a flat frequency response below 2~Hz with a magnitude of 2, and above 2~Hz has nulls in the response  and a maximum amplitude of 4. Both the dual and modified dual arm locking sensors have this frequency response. This factor of 2 is normalized out of the controller by adding a 1/2 multiplier at the beginning of the controller. Although the controller is designed for the maximum arm length mismatch, it will operate stably for smaller arm length mismatch.

\subsection{Goal of the arm locking control loop}

The goal is to design a controller such that the arm locking control system has sufficient gain across the LISA science band to suppress the laser noise to lower than the TDI capability. The required suppression of laser noise, $S_{\rm req}(f)$, is a ratio of the laser noise before arm locking, $ \nu_{\rm OL}(f)$, and the TDI capability, 
\begin{eqnarray}
S_{\rm req}(f) &=& \frac{\nu_{\rm OL}(f)}{ \nu_{\rm TDI}(f)}.
\end{eqnarray}
If arm locking operates without any pre-stabilization, $\nu_{\rm OL}$ is given by equation~\ref{equation_frln}. If we assume  this level of laser frequency noise and a time delay error of 3~ns, the required suppression is
\begin{eqnarray}
S_{\rm req}(f)&=&\frac{30/{[f/{\rm Hz}]}~~\frac{\textrm{kHz}}{\sqrt{\textrm{Hz}}}}{300\times(1+(2.8~\textrm{mHz}/f)^2)~~\frac{\textrm{Hz}}{\sqrt{\textrm{Hz}}}}, \nonumber \\
&=& \frac{100/[f/{\rm Hz}]}{1+(2.8~\textrm{mHz}/f)^2}~~\textrm{for}~100\mu\textrm{Hz}\le f \le 1\textrm{Hz}.\nonumber \\
\end{eqnarray}

If pre-stabilization is implemented the required gain will be relaxed.

\subsection{Constraints on the controller design}
Compared to a standard phase locking control loop, the arm locking control loop has two additional design constraints. These are: 
\begin{enumerate}
\item {The controller should  have appropriate low frequency filtering  to limit the laser frequency pulling.}
\item The controller must allow for the nulls in the sensor and the additional phase delay associated with them. The nulls in the dual arm locking sensor occur at frequencies above 2~Hz.
\end{enumerate}
These additional constraints limit the achievable gain in the LISA science band and necessitate careful design to ensure loop stability.

In the following sections we design a controller in two parts, treating the low frequency design (at $f<1$~mHz) and high frequency design (at $f>1$~Hz)  separately. Finally, we combine the parts to complete the controller design. 

\subsection{Low frequency filtering}

In section~\ref{section_doppler} the laser frequency pulling that arises at lock acquisition and in steady state operation was studied. At lock acquisition, it was found that low frequency filtering is required to limit frequency pulling due to imperfect knowledge of the Doppler frequency, Doppler rate, and second time derivative of the Doppler frequency. The frequency pulling that occurs for modified dual arm locking, calculated in section~\ref{subsection_mddoppler}, is almost identical to that calculated in section~\ref{section_doppler} for common arm locking.

The lower bound on the lower unity gain frequency for common arm locking (equation~\ref{equation_accoupling}) was designed to be $f_{\rm ac}\ge4.8\times10^{-6}~~\textrm{Hz}$. If the Doppler frequency estimate has an error of $\nu_{0+}=600$~kHz (free running laser for an averaging time of 200~s) the pulling due to the Doppler frequency error will be under 460~MHz, 4.6\% of the mode-hop-free range\footnote{The separation of mode-hops in Nd:YAG non-planar ring oscillator (NPRO) laser are typically $\sim10$~GHz.}.

The frequency pulling  at lock acquisition due to error in the Doppler rate and second time derivative of the Doppler frequency can be limited by further low frequency filtering. We find that three unity gain high pass filters with a corner at 0.8~$\mu$Hz placed in series with the rest of the controller are sufficient to limit this frequency pulling. This low frequency part of the controller limits the frequency pulling in steady state to less than 10~MHz peak to peak for modified dual arm locking.

\subsection{Rolling up the gain at low frequencies}

To ensure a stable control system the controller-sensor system crosses the lower unity gain frequency, $f_{\rm ac}$, with a slope of $f$. To optimize the gain in the LISA science band we consider various slopes that can be used to roll the gain up steeply. To design the roll up we set $>$30 degree phase margin at the unity gain frequency, and aim for the maximum gain at 100$\mu$Hz ($S_{\rm req}(100\mu$Hz) =  $10^3$ is the requirement).

Accounting for the $\sim$30~degrees added by the 0.8~$\mu$Hz high pass filters, a phase margin of 30 degrees and the $f$ slope for crossing the unity gain frequency, leaves an additional 30 degrees of phase at $4.8~\mu$Hz which can be allocated to the increased roll up of the controller. We find that an additional $f^5$ roll up with 5 zeros at 36.6~$\mu$Hz and 5 poles at 185~$\mu$Hz gives a gain at 100~$\mu$Hz of $\approx$1900 and meets the phase requirements.

\subsection{High frequency design of the arm locking controller}

The nulls in the dual arm locking sensor mean careful design of the controller is needed to ensure loop stability and to limit noise amplification.  The nulls in the sensor have additional phase associated with them and mean that there are many unity gain crossings above the first null  (located at 2~Hz for maximal arm length mismatch).

To design the controller slope above 2~Hz we account for all phase delays in the control loop, specify a phase margin, then allocate remaining phase from 180 degrees to slope of the the controller. The delays in the control system are summarized in table~\ref{phase_delay}. A 30 degree phase margin ($\theta_{\rm margin}$) has been chosen at frequencies where there are nulls in the sensor ($f>2$Hz for the dual arm locking sensor). The delays accounted for are: the delay associated with the arm locking sensor, $\theta_{\rm sensor}$, the delays in the PZT actuator, $\tau_{\rm act}$, the phasemeter processing, $\tau_{\rm pm}$,  and in the interaction with the transponder spacecraft and the pre-stabilization loop, denoted $\tau_{\rm trans}$ and $\tau_{\rm ps}$, respectively\footnote{The phase delay of the prestabilization control loop can be removed, as suggested by Sheard~{\it et. al}~\cite{SheardFCST}.}. The remaining phase from 180 degrees to be allocated to the controller is 
\begin{eqnarray}
\theta_{\rm controller} &=&180-\theta_{\rm margin}-\theta_{\rm sensor}-\theta_{\rm delay}, 
\end{eqnarray}
where 
 $\theta_{\rm delay}$ is the total phase due to delays in the control loop given by
 \begin{eqnarray}
\theta_{\rm delay}&=&360 f(\tau_{\rm act}+\tau_{\rm pm}+\tau_{\rm trans}+\tau_{\rm ps}).
\end{eqnarray}
\subsection{Phase of the sensor at unity gain points}

{In arm locking studies published to date~\cite{SheardPLA,SuttonPRD} the phase delay associated with the nulls in the sensor has been set to $\theta_{\rm sensor}=-\pi/2$. This phase delay was allocated because it is the maximum possible
phase delay of the sensor at the nulls~\cite{SheardPLA}. Here, we show that this allowance is overly conservative when the controller gain is low. To determine the sensor phase at frequencies where the total open loop gain crosses unity, $\theta_{\rm sensor}|_{\rm UG}(\omega)$, we  solve for the phase of the sensor when the 
loop gain has a magnitude of 1, $\left|G_{L}(\omega)\right|=1$. The open loop frequency response of the control system is
\begin{eqnarray}
G_{\rm L}(\omega)=G_1(\omega)P_{M}(\omega)e^{-i\theta_{\rm delay}}.
\end{eqnarray} 
{The common arm locking sensor dominates the frequency response of modified dual arm locking above the first null 
\begin{eqnarray}
P_{M}(\omega)|_{f>\frac{1}{2\Delta \tau}} &\approx &P_{\rm D}(\omega) \approx P_+(\omega) ,\\
&=&2(1- \cos\left(\Delta \tau\omega\right)e^{-i\omega \bar\tau}) .
\end{eqnarray}
We are interested in the nulls in the sensor which occur when $\cos\left(\Delta \tau \omega\right) = 1$. Thus let
\begin{eqnarray}
P_{M}(\omega)|_{f>\frac{1}{2\Delta \tau}} \approx2(1- e^{-i\omega \bar\tau}) .
\end{eqnarray}
Define $\theta=\pi -\omega\tau/2,$ giving 
\begin{eqnarray}
P_{M}(\omega)|_{f>\frac{1}{2\Delta \tau}}=2(1+e^{2i\theta}).
\end{eqnarray}
Using the identity $1+e^{2i\theta}=2\cos \theta e^{i\theta}$ 
we see that $\theta$ is the phase of the sensor. At
unity gain,  $|G_1(\omega)P_{M}(\omega)|=|G_1^*(\omega)|\cdot 2\cos\theta_{\rm sensor}|_{\rm UG} =1.$
That gives $\cos\theta_{\rm sensor}|_{\rm UG} = 1/|2G_1^*(\omega)|,$ or for the sensor phase at unity gain 
\begin{eqnarray}
\theta_{\rm sensor}|_{\rm UG}&=& -\arccos\left(|2G_{1}^*(\omega) |^{-1}\right)~~~~\textrm{For~}|2G_{1}^*(\omega)| \ge1, \nonumber\\
\label{sensor_UG}
\end{eqnarray}
We have selected  the negative solution  for $\theta_{\rm sensor}|_{\rm UG}$ because that gives maximum phase delay, which is of interest for stability considerations.}

At high controller gain, say $|2G_{1}^*(\omega)|>100$, the point on the sensor which crosses unity gain is very close to the null. Accordingly, the phase delay of the sensor at the unity gain is $\theta_{\rm sensor}\approx-\pi/2$. For a controller gain  or $|2G_{1}^*(\omega)|=1$ the peak of the sensor response will crosses unity gain. The corresponding phase at this point approaches zero ($\theta_{\rm sensor}\rightarrow0$).  The sensor phase at several unity gain frequencies is illustrated in the Nyquist plot in figure~\ref{nyquist_plot}.

 \begin{figure}
\begin{center}
\includegraphics[width=0.37\textwidth]{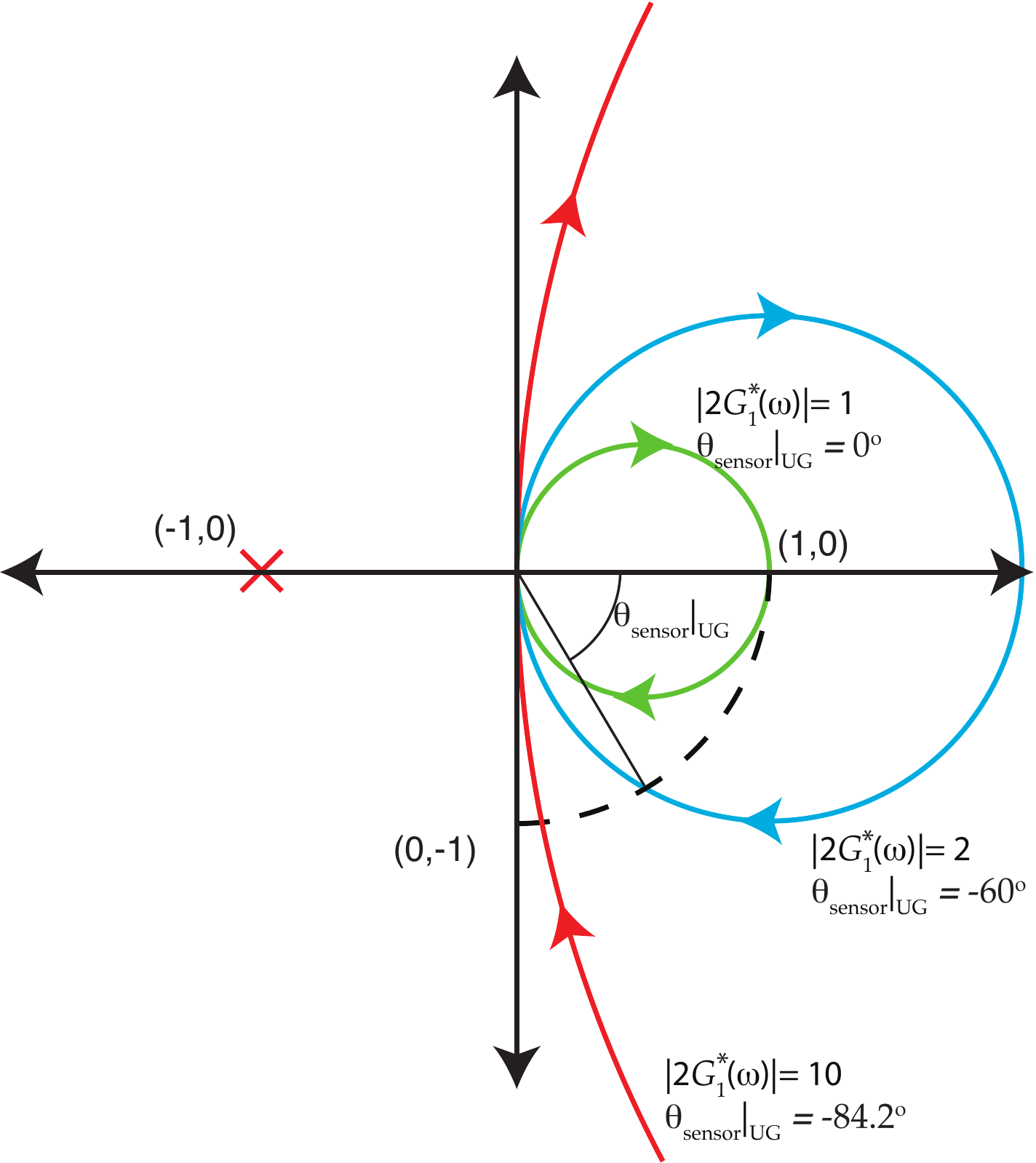}
\caption{Nyquist plot of the sensor for three different controller gains, $|G_1^*(\omega)|$. The phase delay associated with the sensor corresponds to the angle from the x-axis to the point at which the sensor times gain crosses the dashed line which has radius of 1. \label{nyquist_plot}}
\end{center}
\end{figure}

\begin{table*}[htdp]
\caption{Arm locking system delays}
\begin{center}
\begin{tabular}{|l|l|c|l|}
\hline
Type of Delay &Symbol&	Delay  & Notes\\ 
\hline
Phase margin &$\theta_{\rm margin}$& 30$\,^{\circ}$&Control system margin\\
Arm locking sensor&$\theta_{\rm sensor}|_{\rm UG}$ & $-\arccos\left(|2G_{1}^*(\omega)|^{-1}\right)$& Phase of the sensor at unity gain \\
\hline
Actuator delay&$\tau_{\rm act}$ & $5\mu$s & PZT delay ($4\mu$s measured) \\
Phasemeter processing delay&$\tau_{\rm pm}$ & $2\mu$s &DACs have an inherent 1$\mu$s delay \\
Transponding S/C phase delay & $\tau_{\rm trans}$&$({2\pi30\textrm{kHz}})^{-1}$s &30kHz Unity gain frequency assumed\footnote{By approximating the phase accrued in the transponding phase locking system (and that in the pre-stabilization system) by a time delay we underestimate the accrued phase at 10~kHz by 0.3 degrees.}\\
Pre-stabilization phase delay &$\tau_{\rm ps}$& $({2\pi30\textrm{kHz}})^{-1}$s&30kHz Unity gain frequency assumed\\
\hline
\end{tabular}
\end{center}
\label{phase_delay}
\end{table*}%
With the phase associated with the sensor at unity gain and the other delays in table~\ref{phase_delay} we find that approximately 60 degrees of phase can be assigned to the controller,  corresponding to a controller slope of $f^{-0.66}$. With this slope, a unity gain frequency of up to 14.9~kHz can be achieved. A Bode plot of such a controller is shown in figure~\ref{new_controller_bode}. The magnitude is given by $|567/(if)^{0.66}|$.  The total phase of the control system (black curve) is a sum of the system delays (dotted green curve), the phase of the sensor at unity gain (dashed red curve), and phase associated with the controller (blue solid curve). Note the behavior of the system delays and the sensor phase at unity gain is complementary at high frequencies. This effect allows the control bandwidth to be increased and the controller slope to be steeper, giving a factor of 10 higher gain at 1~Hz than without this effect. If  $\theta_{\rm sensor}=-\pi/2$ allocated, the maximal constant controller slope is $f^{-0.58}$ and the control bandwidth is 1.2~kHz, giving a gain at 1~Hz of 61. 

 \begin{figure}
\begin{center}
\includegraphics[width=0.48\textwidth]{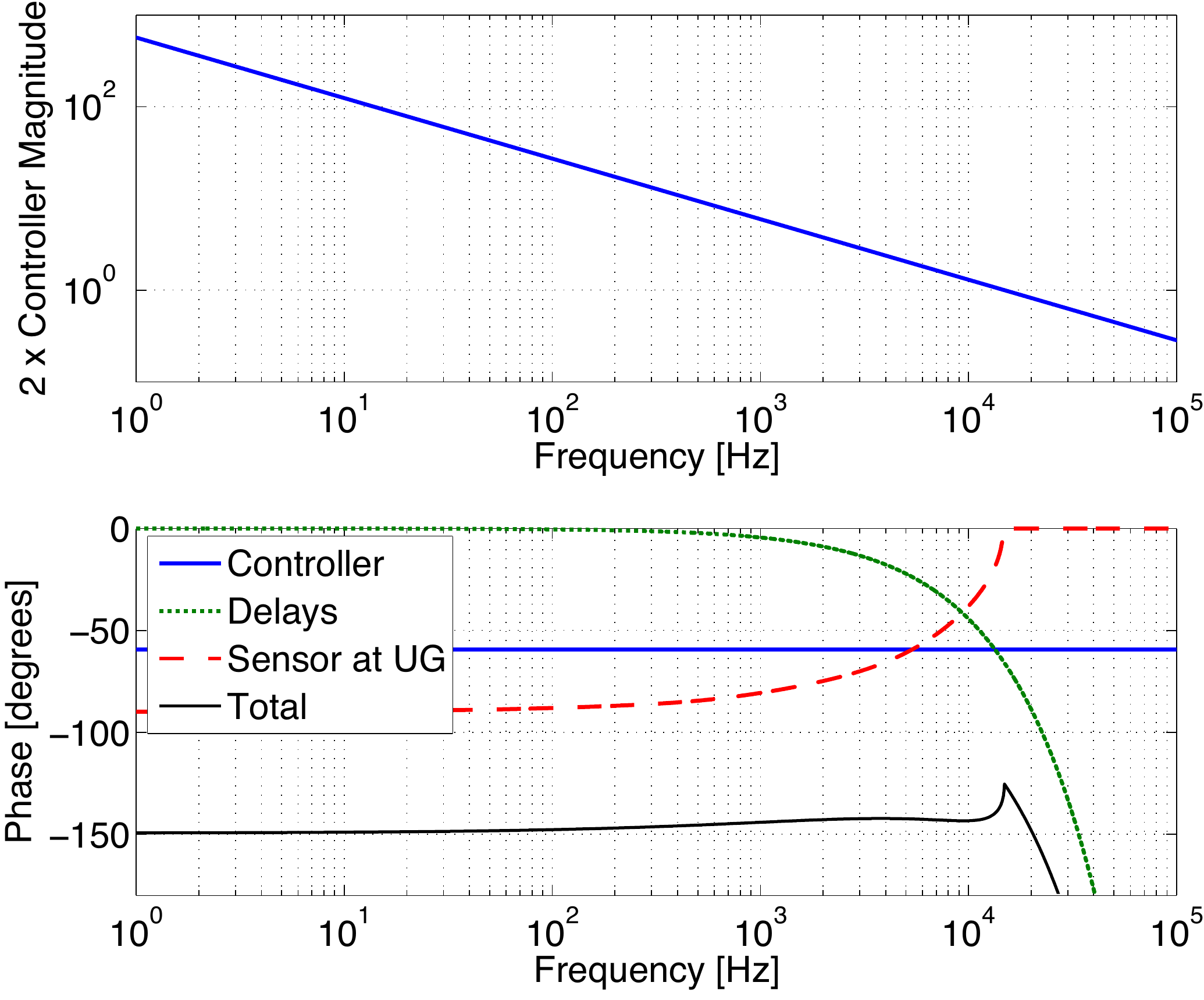}
\caption{Bode plot of the arm locking controller.  The lower plot shows the total open loop phase of the control system and the three components of which it is comprised. These are the controller (-59.4 degrees), the system delays (for parameters used in table~\ref{phase_delay}), and the sensor phase at unity gain, given by equation~\ref{sensor_UG}. \label{new_controller_bode}}
\end{center}
\end{figure}
 
 \subsection{The complete frequency response of the controller}
The low and high frequency components of the controller can be combined to obtain the complete frequency response. Analytically, the controller can be designed with a precise slope at all frequencies.  Here, we present a controller using only poles and zeros. A block diagram of the controller architecture is shown in figure~\ref{controller_block_diagram}. The controller consists of five stages. Stage 1 is the very low frequency part of the controller and comprises three zero-pole pairs to form unity gain high pass filters at 0.8~$\mu$Hz in series. Stage 2 sets the lower unity gain frequency and has a zero at DC and a pole at 210~$\mu$Hz. Stage 3 is a lead stage, which has five zero-pole stages in series to roll up the gain steeply between the lower unity gain frequency and  the low frequency part of the controller.  Stage 4 consists of two poles in series and provides the transition between the low frequency gain and the shallow slope high frequency part of the controller. Stage 5 is the shallow sloped part of the controller. It consists of nine poles in parallel, with the gain for each pole chosen to achieve the required slope of approximately $f^{-0.66}$. The frequency response of the controller is given by
\begin{eqnarray}
G_1^*(\omega) &=& \left(\frac{g_1{s}}{s+p_1}\right)^3\times \left(\frac{g_2{s}}{s+p_2}\right)\times\left(\frac{g_3( s+z_3)}{s+p_3}\right)^5\times\nonumber \\&&\left(\frac{g_4}{(s+p_{41})(s+p_{42})}+\sum_{k=1}^9\frac{g_{5k}}{s+p_{5k}}\right),
\label{equation_controller}
\end{eqnarray}
with  values of the zeros, poles, and gains listed in table~\ref{zpk}\footnote{An additional low pass filter in series with the controller is generally required to roll off the loop gain at the resonance frequency of the laser PZT actuator (near 100~kHz)}.

 \begin{figure}
\begin{center}
\includegraphics[width=0.50\textwidth]{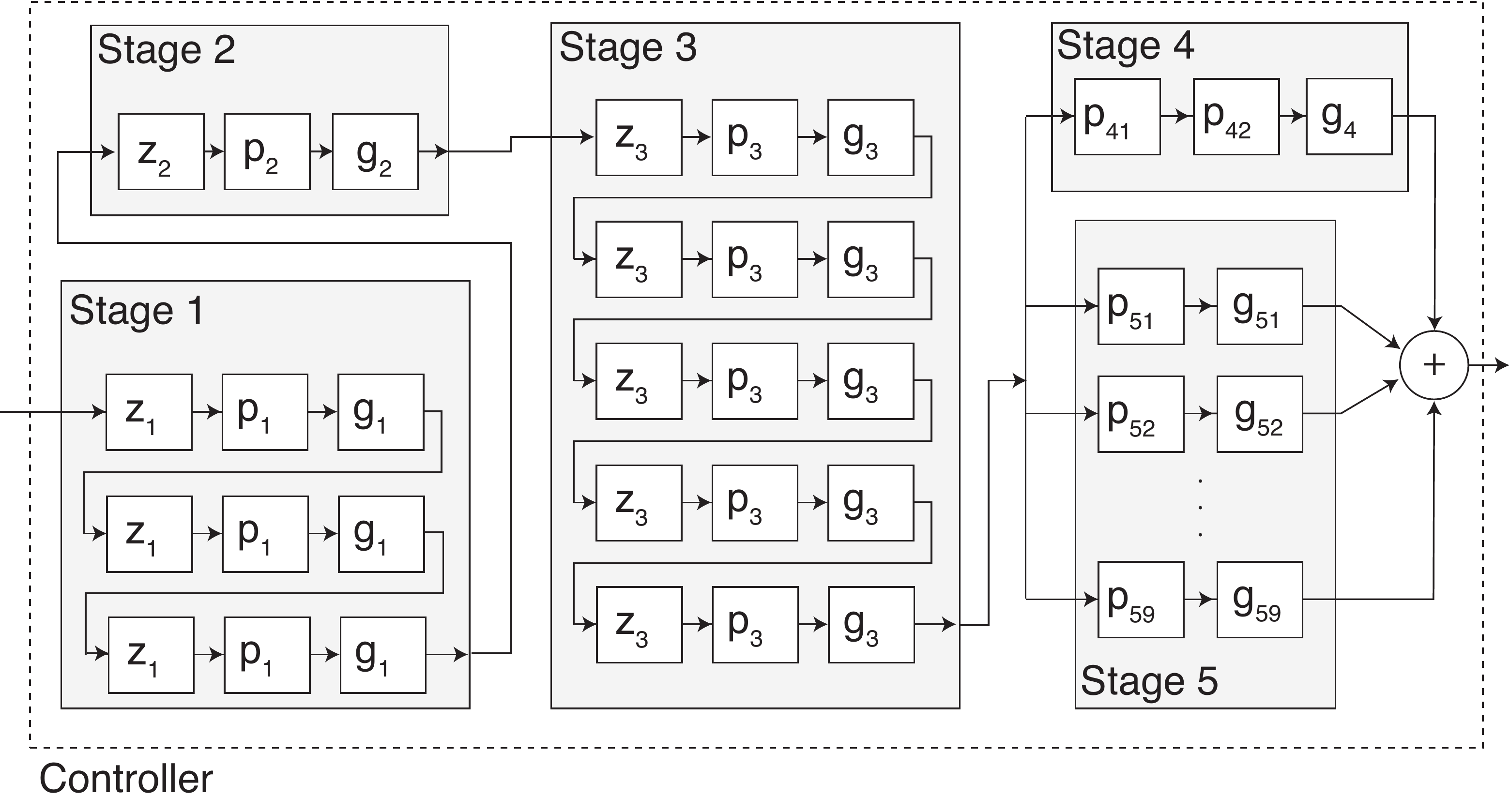}

\caption{Block diagram of the arm locking controller. The controller is built from five stages: stage 1 consist of three very low frequency high pass filters, stage 2, defines the lower unity gain frequency; stage 3, rolls up the gain below the LISA science band; stage 4 has two poles in parallel to effectively transition between stage 3 and stage 5; stage 5 has 9 poles in parallel, with gains individually chosen to generate a slope of approximately $f^{-0.66}$. \label{controller_block_diagram}}
\end{center}
\end{figure}

 \begin{figure}
\begin{center}
\includegraphics[width=.5\textwidth]{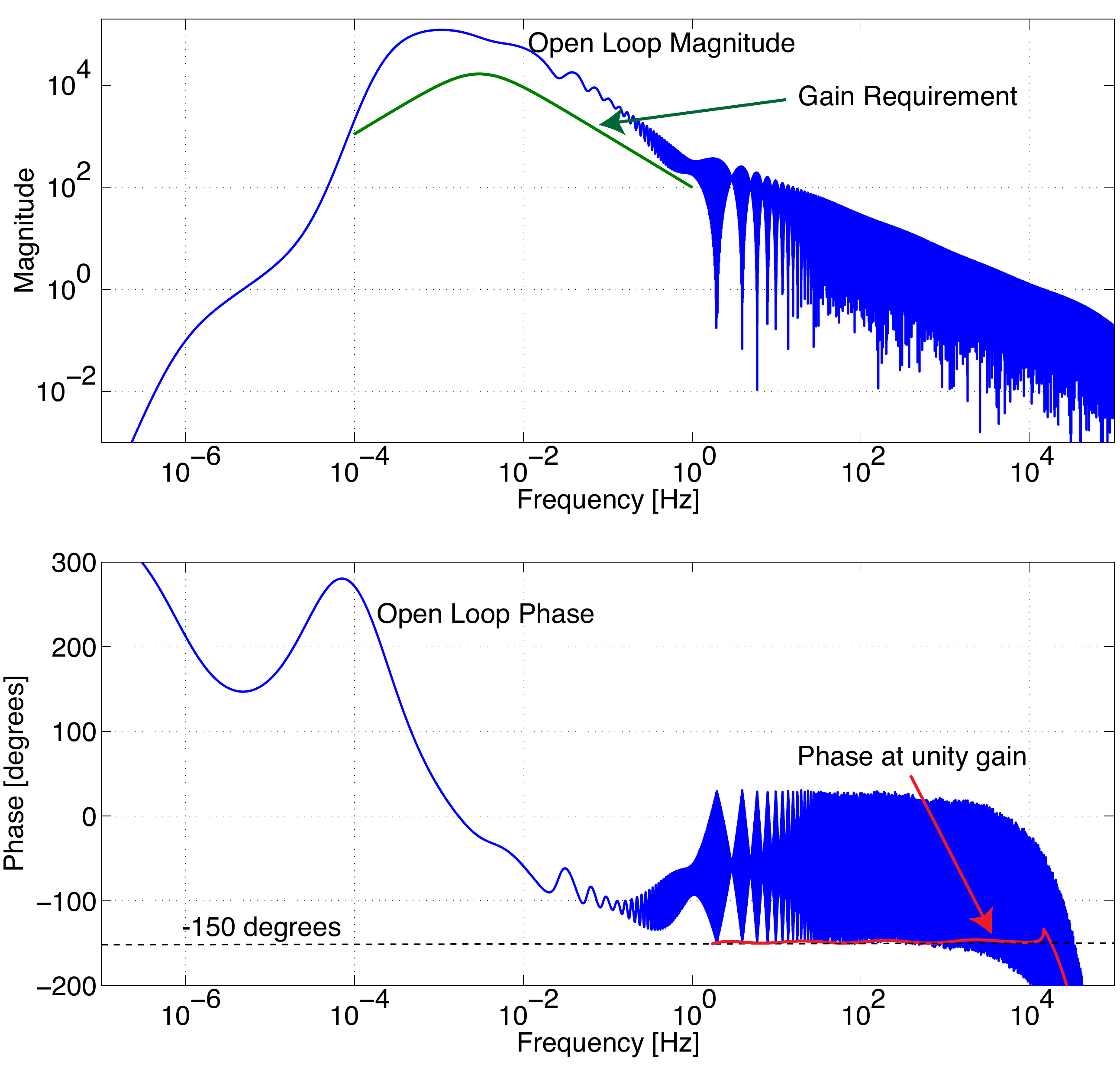}

\caption{Open loop frequency response, $G_{L}(\omega)$, of modified dual arm locking. Also shown in the magnitude plot is the gain required to meet the TDI capability, assuming no pre-stabilization. In the phase plot, the red curve indicates the total phase of the control loop at unity gain (given by equation~\ref{total_UG}). The arm length mismatch is assumed to be 2$\Delta \tau = 0.51$s. \label{open_loop_G}}
\end{center}
\end{figure}

\begin{table*}[htdp]
\caption{Parameters of the arm locking controller}
\begin{center}
\begin{tabular}{|c|l|l|l|} \hline
Stage & zeros (radians/s) & Poles (radians/s) & Gain (radians/radian)\\ \hline 
\hline
1 & $z_1 = 0$ & $p_1=2\pi\times8\times10^{-7}$ & $g_1= 1$ \\ \hline
2 & $z_2 = 0$ & $p_2=2\pi\times210\times10^{-6}$ & $g_2= 0.95/ f_{\rm ac}$ \\ \hline
3 & $z_3=2\pi\times36.6\times10^{-6}$& $p_3=2\pi\times185\times10^{-6}$~\footnote{  If arm locking is implemented with prestabilization, $p_3=2\pi\times178\times10^{-6}$.} & $g_3=$~$p_3$/z$_3$ \\ \hline
4& & $p_{41}=2\pi\times3\times10^{-3}$  & \\
 & & $p_{42}=2\pi\times238\times10^{-3}$   &  $g_4=$~$p_{41}p_{42}$ \\ \hline
 5 
& & $p_{51}=2\pi\times3\times10^{-3}$   &  $g_{51}=1.3\times10^{-3}$ \\
& & $p_{52}=2\pi\times3\times10^{-2}$   &  $g_{52}=3.7\times10^{-3}$ \\
& & $p_{53}=2\pi\times3\times10^{-1}$   &  $g_{53}=4.2\times10^{-3}$\\
& & $p_{54}=2\pi\times3$  		     &  $g_{54}=16\times10^{-3}$ \\
& & $p_{55}=2\pi\times3\times10^{1}$   &  $g_{55}=30\times10^{-3}$ \\
& & $p_{56}=2\pi\times3\times10^{2}$   &  $g_{56}=69\times10^{-3}$ \\
& & $p_{57}=2\pi\times3\times10^{3}$   &  $g_{57}=0.11$\\
& & $p_{58}=2\pi\times3\times10^{4}$   &  $g_{58}=0.33$ \\
& & $p_{59}=2\pi\times3\times10^{5}$   &  $g_{59}=0.70$\\ \hline
\end{tabular}
\end{center}
\label{zpk}
\end{table*}%

The open loop gain of the control system with the modified dual arm locking sensor is given by
\begin{eqnarray}
G_{L}(\omega) = G_1(\omega)P_{M}(\omega)e^{-i\omega(\tau_{\rm act}+\tau_{\rm pm}+\tau_{\rm trans}+\tau_{\rm ps})}.
\end{eqnarray}
 The Bode plot of $G_{L}(\omega)$ is plotted in figure~\ref{open_loop_G}. Note that in this plot we have assumed no prestabilization and thus set $\tau_{\rm ps}=0$. From the magnitude plot it can be seen that the open loop gain is sufficient to meet the required gain, $S_{\rm req}(f)$. Note that looking at the open loop phase is somewhat deceptive. At high frequencies it appears that the open loop phase crosses 180 degrees before 1~kHz, making the 14.9~kHz bandwidth control loop unstable. In fact, the total phase at the unity gain points is always greater than -150 degrees which is indicated by the red solid curve, which is given by
 \begin{eqnarray}
 \theta_{\rm UG} &=& \angle G_1^*(\omega)\times \frac {180} \pi +
\theta_{\rm sensor}|_{\rm UG} + \nonumber \\&&360f(\tau_{\rm act}+\tau_{\rm pm}+\tau_{\rm trans}),
\label{total_UG}
\end{eqnarray}
which is valid for $f\geq1/(2\Delta \tau)$.

\subsection{Closed loop performance of the controller}

The closed loop disturbance suppression function is shown in figure~\ref{suppression_a2}. This function is given by
\begin{eqnarray}
S_{\rm D}(\omega) = \frac{1}{1+G_L(\omega)}.
\end{eqnarray}
 This shows the suppression of the control loop, as well as  amplification of the noise at the nulls in the sensor and above the control loop unity gain frequency. Also plotted is the required suppression to meet the TDI capability. Note that the amplification at the nulls is always less than a factor of 2 except near the final unity gain frequency, near 15~kHz, where the amplitude increases to 5.
  \begin{figure}
\begin{center}
\includegraphics[width=0.48\textwidth]{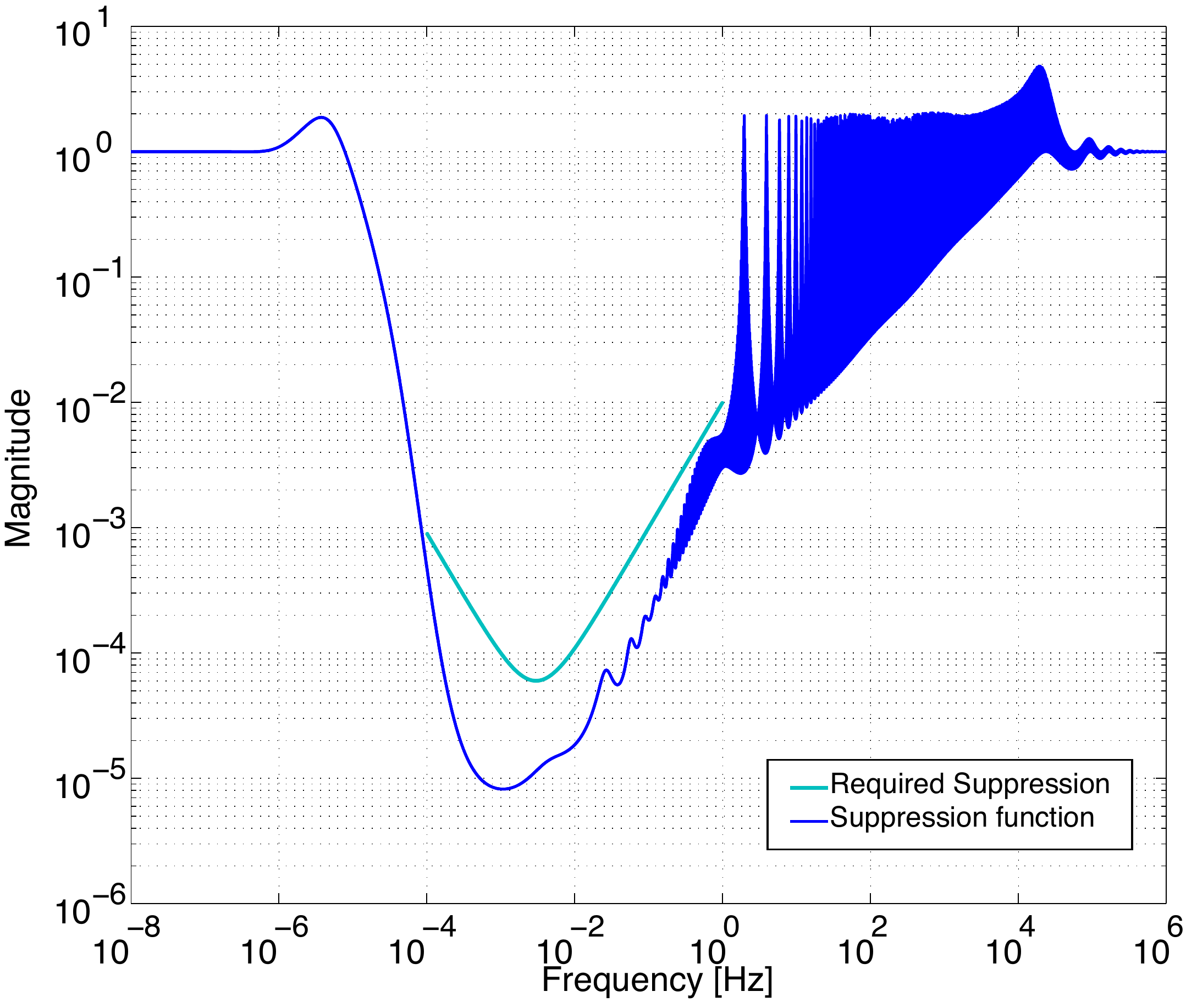}

\caption{Plot of  the disturbance sensitivity, $S_{\rm D}(\omega)$ and the required suppression, $1/S_{\rm req}(\omega)$. Plotted with $2\Delta \tau = 0.55$s. \label{suppression_a2}}
\end{center}
\end{figure}

\section{Expected performance of arm locking}
\label{section_performance}
The performance of arm locking can be predicted using the frequency pulling analysis of section~\ref{section_doppler}, the noise analysis of section~\ref{section_noisebudget}, the controller design given in section~\ref{section_controller}, and an assumption of the laser frequency noise. Here, we calculate noise budgets for three different initial laser noise levels: free running laser noise, and laser noise predicted for two types of pre-stabilization, Fabry-Perot cavity stabilization and Mach-Zehnder stabilization.

 \subsection{Performance assuming free running laser noise}

The expected frequency pulling which would occur at lock acquisition is shown in the upper plot in figure~\ref{Step_fn}.  In this case the Doppler frequency errors are the values of $\nu_{0+}, \gamma_{0+}$, and $\alpha_{0+}$ given in the free running laser column in table~\ref{table_estimates}.   These results are near identical to those for common arm locking shown in section~\ref{section_doppler} because the common arm locking sensor dominates the frequency response at low frequencies.  Likewise, the result for pulling in steady state of modified dual arm locking, shown in figure~\ref{SteadystateMDAL}, is similar to that of common arm locking, with less than 8~MHz peak-peak pulling.

 \begin{figure}
\begin{center}
\includegraphics[width=0.48\textwidth]{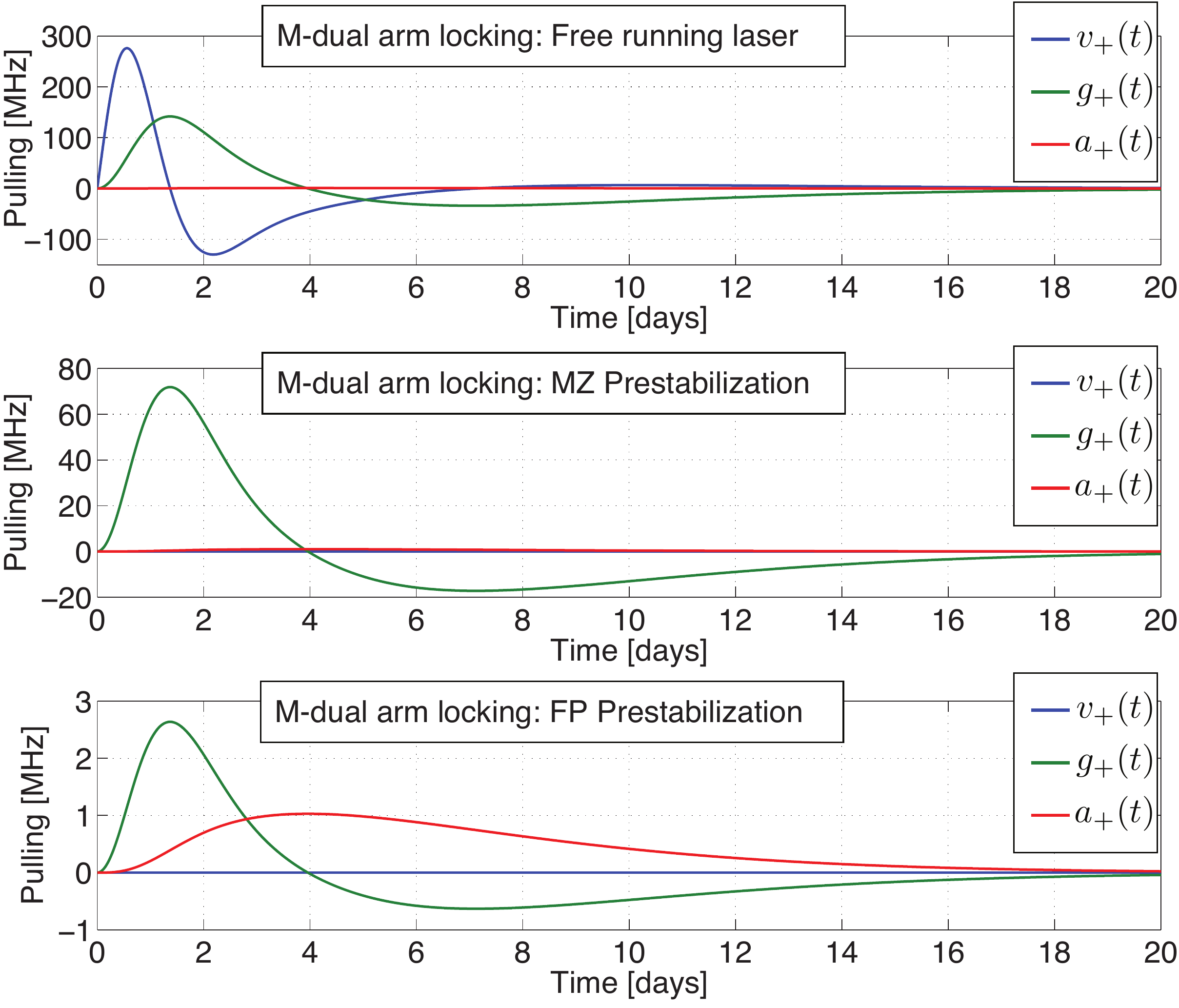}
\caption{The step responses of different drivers of Doppler frequency error for modified dual arm locking. The upper, middle, and lower plots assume free running laser noise, Mach-Zehnder type prestabilization, and Fabry-Perot cavity prestabilzation, respectively, with the Doppler frequency estimates averaged for 200~s.
 \label{Step_fn}}
\end{center}
\end{figure}
 \begin{figure}
\begin{center}
\includegraphics[width=0.48\textwidth]{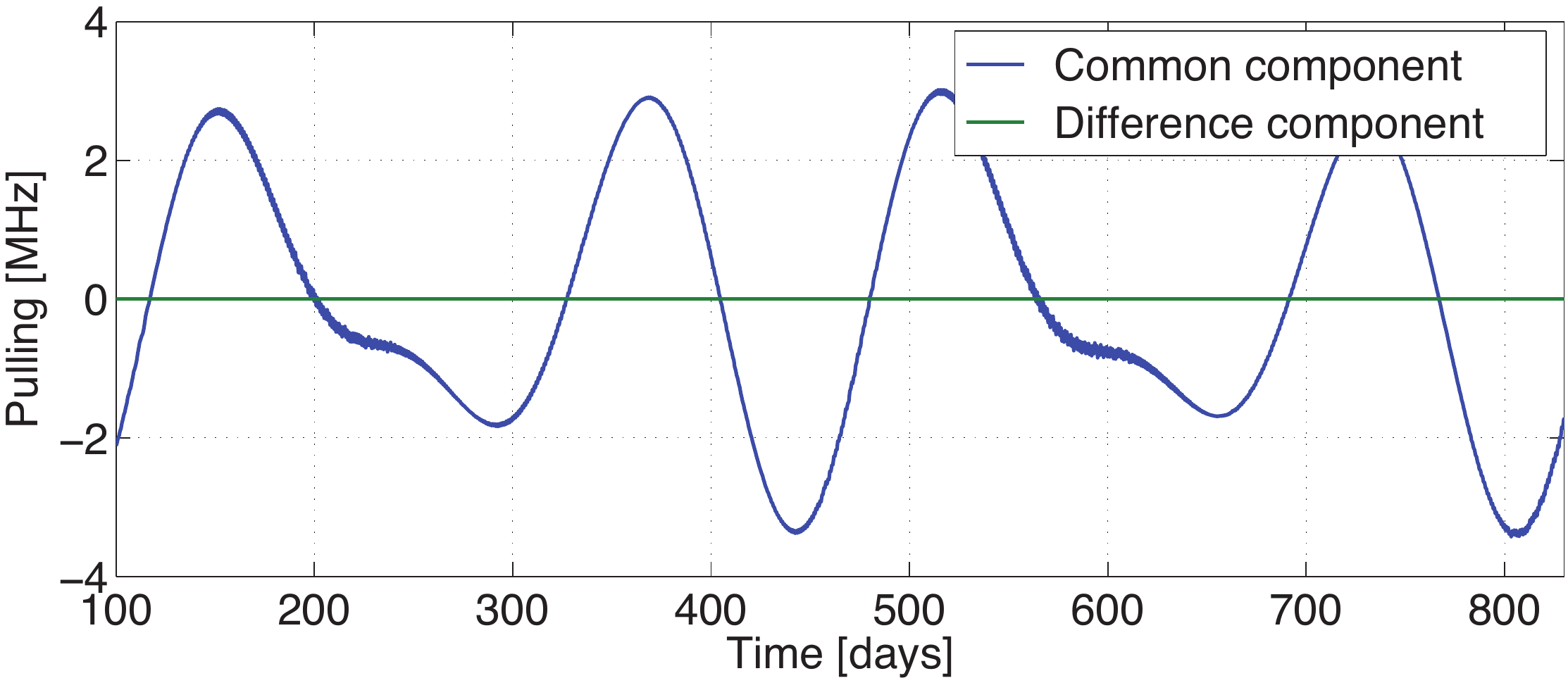}

\caption{ The steady state laser frequency pulling of modified dual arm locking. The pulling due to the common doppler shift is the dominant term.\label{SteadystateMDAL}}
\end{center}
\end{figure}

Figure~\ref{mdual_noise_budget} shows the noise budget of modified dual arm locking with free running laser noise and an arm length mismatch of  $2\Delta \tau =0.51~$s. For this arm length mismatch the laser frequency noise is the limiting noise source (the system is gain limited) with the other system noise sources well below the laser frequency noise. Figure~\ref{mdual_noise_budget}  shows that even without any form of laser pre-stabilization arm locking will meet the TDI capability across the entire LISA science band. At the most sensitive frequency of LISA, 3~mHz the frequency noise is more than a factor of 4 below the TDI capability.  If there is not a  failure of one inter-spacecraft laser link, the dual arm locking central spacecraft can be switched when the arm length mismatch becomes small, and arm locking alone has sufficient performance to meet the TDI capability for the mission.

The performance of arm locking with an inter-spacecraft laser link failure is similar to the no failure case. For the majority of the mission the system will be gain limited as it is in the no link failure scenario,  the difference is for a few hours per year the system will become noise limited, when the arm length mismatch is close to zero. The noise performance will be insufficient to meet the TDI capability for approximately 30 minutes, twice per year. Note that, this noise limited time could be reduced by either flying a clock with better stability or by implementing the clock noise removal algorithm inside the arm- 
locking sensor\footnote{However, to remove clock noise using this method the clocks at the end spacecraft would need to be phase-locked to the central spacecraft clock in an analogous way to the lasers.}.
\subsection{Performance assuming pre-stabilized laser noise}
 The frequency pulling which would occur at lock acquisition assuming Mach-Zehnder and Fabry-Perot   prestabilization  is shown in the middle and   lower plots in figure~\ref{Step_fn}, respectively.  The Doppler frequency errors assumed for $\nu_{0+}, \gamma_{0+}$, and $\alpha_{0+}$ for the former prestabilization are given in the Mach-Zehnder column in table~\ref{table_estimates} and for the  latter are $\nu_{0+}=1.68$~Hz, $\gamma_{0+}=0.08$~Hz/s. This shows less than 90~MHz peak to peak pulling over  a period of 10 days for the Mach-Zehnder prestabilization or less than 4~MHz for Fabry-Perot prestabilization.

\begin{figure}[htb]
\centering
\includegraphics[width=.48\textwidth]{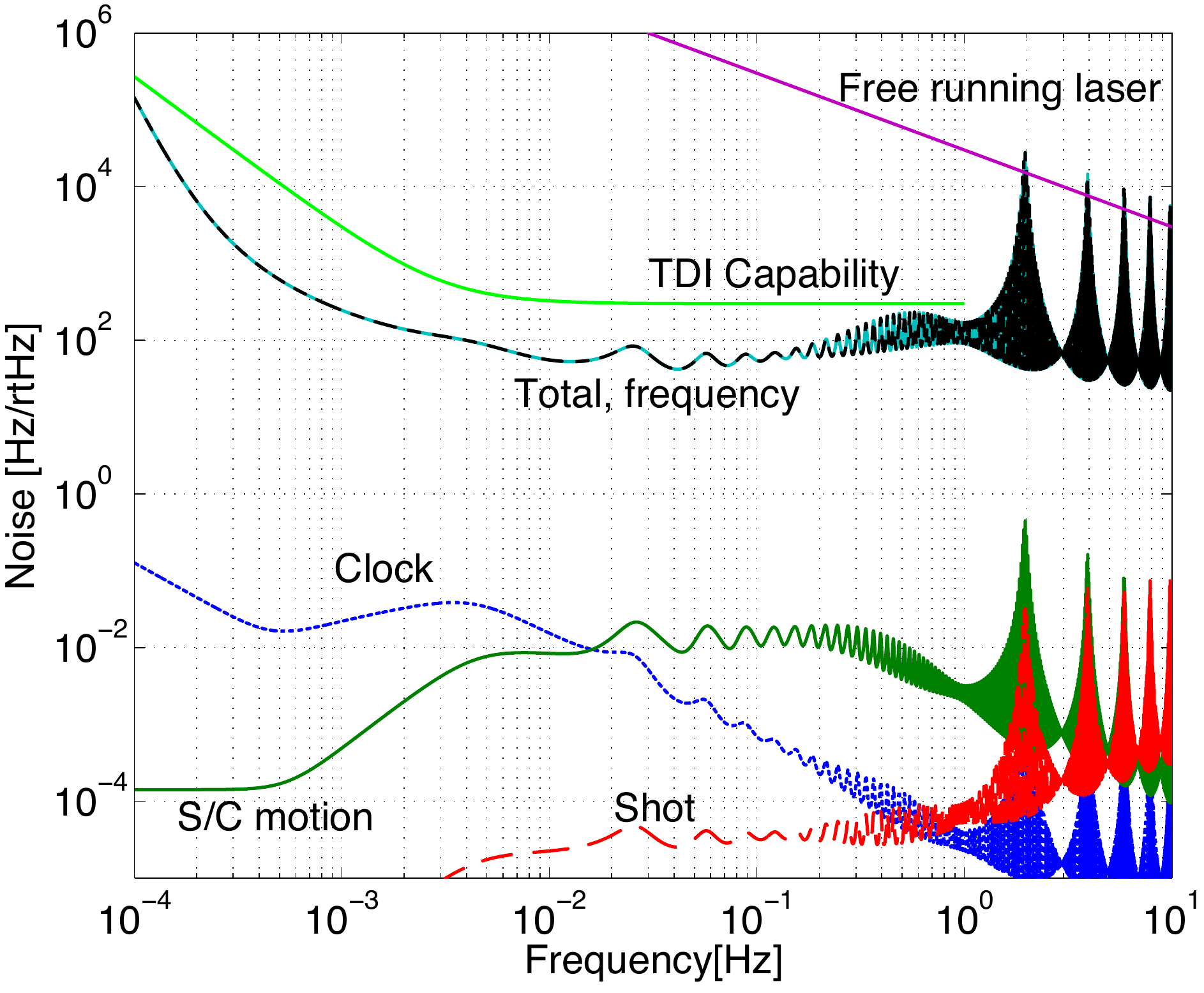}
\caption{\label{mdual_noise_budget} The noise budget of modified dual arm locking with arm length mismatch of $\Delta \tau = 0.51$s. The tperformance was calculated with free running laser noise as an initial condition. }
\end{figure}
 
 \begin{figure}
\begin{center}
\includegraphics[width=0.46\textwidth]{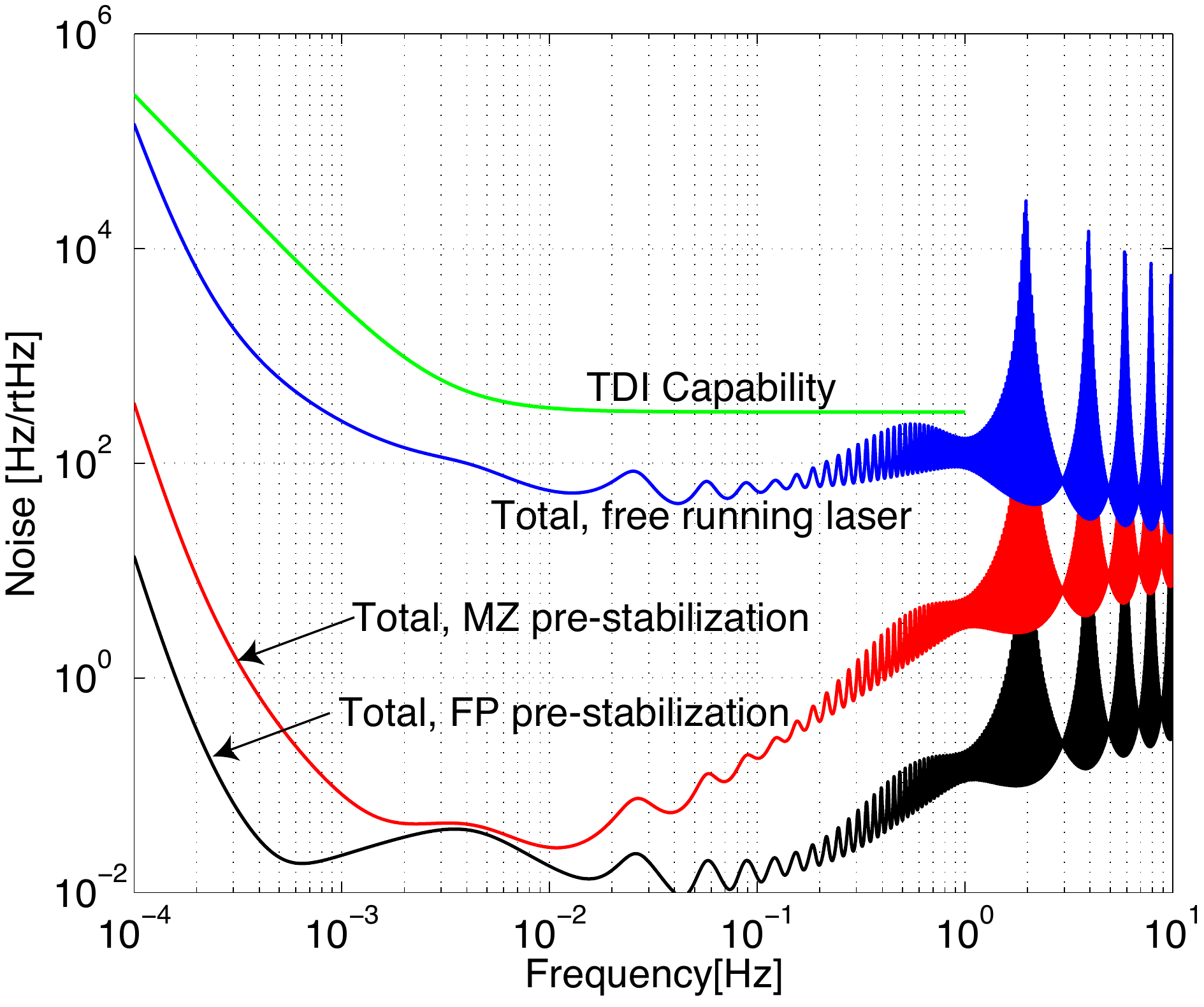}
\caption{Noise floor of modified dual arm locking with different initial laser frequency noise conditions: free running laser noise (blue curve), Mach-Zehnder pre-stabilization (red curve), and Fabry-Perot cavity pre-stabilization (black curve). Arm length mismatch of $\Delta \tau = 0.51$s.\label{Noise_budget_different_prestab}}
\end{center}
\end{figure}

In figures~\ref{Noise_budget_different_prestab} and \ref{mdal_diff_prestab-0_026s} the total noise after arm locking is plotted for no pre-stabilization, Fabry-Perot cavity pre-stabilization, and Mach-Zehnder pre-stabilization.  Figure~\ref{Noise_budget_different_prestab} is plotted for the maximum arm length mismatch, while figure~\ref{mdal_diff_prestab-0_026s}
 is plotted for the minimum arm length mismatch, which will be used if there is no failed inter-spacecraft laser links. It is clear that either pre-stabilization type in combination with arm locking will deliver performance several orders of magnitude better than the TDI capability.  With either pre-stabilization system the performance is limited by clock noise and spacecraft motion in the region of a few mHz. This is the case even with the largest arm length mismatch.
 
Again, the performance of arm locking with a single link failure is similar to the no failure case with pre-stabilization, except at some frequencies the system will be noise limited for the whole mission. The ultimate performance at 3~mHz, for example, will be entirely dictated by the noise sources. The noise floor will exceed the TDI capability as the arm length mismatch passes through zero. The performance is insufficient to meet the TDI capability for about 30 minutes, twice per year.

 \begin{figure}
\begin{center}
\includegraphics[width=0.46\textwidth]{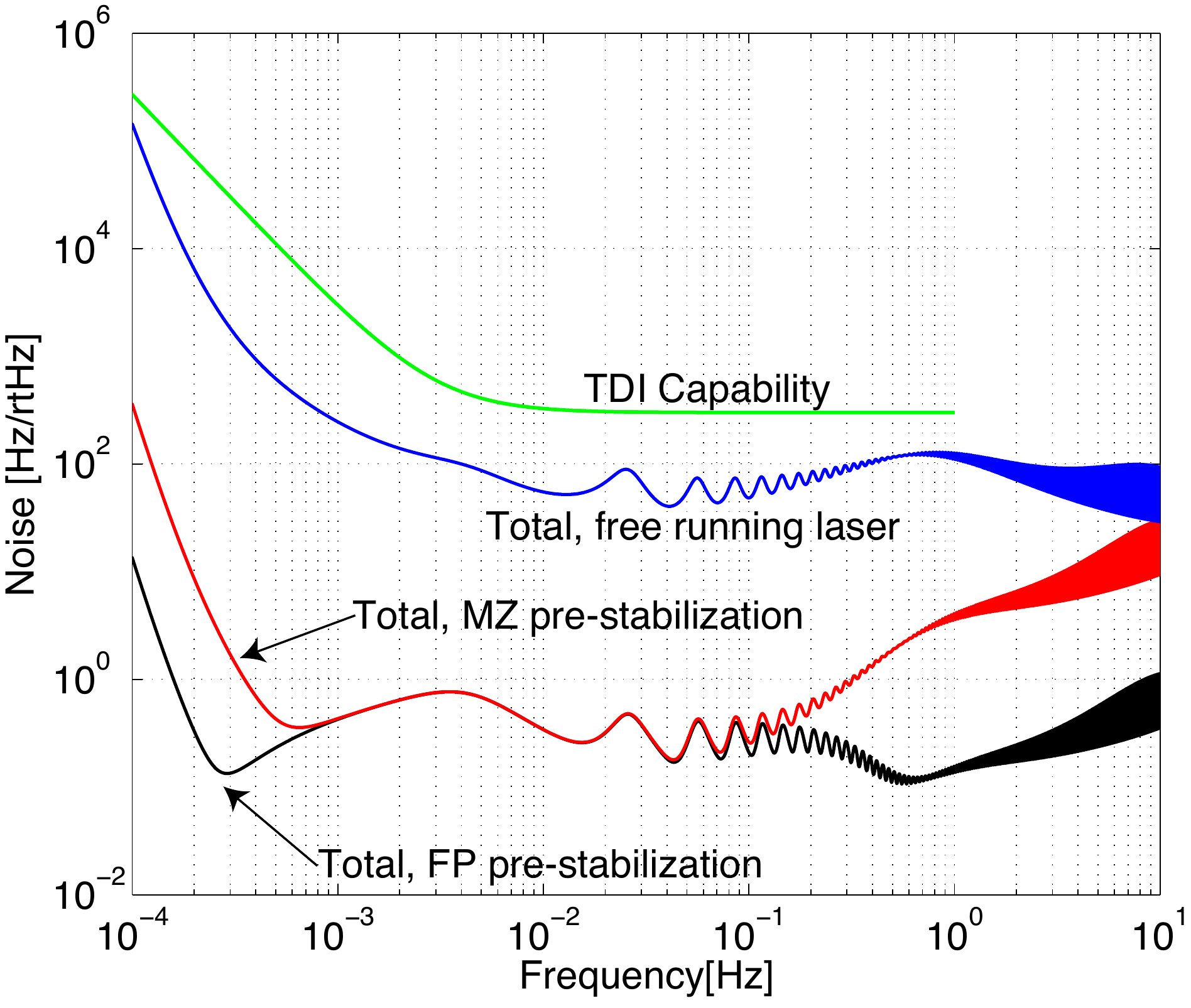}
\caption{Noise floor of modified dual arm locking with different initial laser frequency noise conditions: free running laser noise (blue curve), Mach-Zehnder pre-stabilization (red curve), and Fabry-Perot cavity pre-stabilization (black curve). Arm length mismatch of $\Delta \tau = 0.026$s. \label{mdal_diff_prestab-0_026s}}
\end{center}
\end{figure}
\section{Conclusions}

{We have performed a detailed analysis of the expected performance of arm locking in LISA. This analysis includes many of the orbital effects expected on LISA:  changing arm lengths and Doppler frequencies as well as  laser frequency  pulling due to errors in the Doppler frequency estimate. It was found that laser frequency pulling could be limited to an acceptable level by high pass filtering the control loop. We also noted that not only the Doppler frequency error will cause laser frequency pulling at lock acquisition, but because the Doppler frequencies continuosly change, the error in Doppler rate can too be significant.  In terms of the magnitude of the pulling, if arm locking is used without prestabilization, the Doppler frequency can be estimated to 600~kHz over 200 seconds, giving 460~MHz of pulling at lock acquisition for the controller designed here. This magnitude of pulling would occur only the first time the control loop is engaged, as after LISA has been operating over longer periods the Doppler frequency estimate would be improved.  Less pulling at lock acquisition will be achieved if either a longer estimate of the Doppler frequency was made, or if arm locking was used in combination with prestabilization.

{The noise analysis presented here included the expected dominant noise sources in arm locking; clock noise, spacecraft motion, and shot noise. It was found that clock noise and spacecraft motion are the dominant noise sources in the LISA science band and that they will limit the noise performance when the arm length mismatch is small. }

{We introduced a new sensor design for the dual arm locking sensor, which uses a combination of the common and dual arm  sensor at frequencies below $1/\bar\tau$  and the dual arm locking sensor frequencies above  $1/\bar\tau$. This modified dual sensor has the control system advantages of dual arm locking with the frequency pulling charachtoristics and low frequency noise performance of common arm locking. }

{We designed an arm locking controller which maximizes gain in the science band, minimizes frequency pulling and has a phase margin of greater than 30 degrees to ensure stability. We noted an effect where the arm locking sensor phase changes beneficially near the unity-gain frequencies of the control system. With this effect taken into account,  it was found that the control bandwidth could be up to 14.9~kHz and still maintain a 30 degree phase margin. This control bandwidth is 10 times higher than if this  frequency dependent phase effect is not taken into account. }

{Using the noise analysis of section~\ref{section_noisebudget} and the controller design given in section~\ref{section_controller}, we calculated noise budgets for different initial laser frequency noise levels. The laser frequency noise levels we chose correspond to the noise of lasers free running, pre-stabilization to a  Mach-Zehnder interferometer and to a Fabry-Perot cavity.  A key result is that  arm locking alone can meet the TDI capability. With pre-stabilization, the TDI capability would be met by more than 2 orders of magnitude across the science band.}

{We analyzed the expected performance of arm locking with and without a failure of an inter-spacecraft laser link. If there is no failure, arm locking will have sufficient noise performance to meet the TDI capability for the entire mission, independent of whether pre-stabilization is used or not (though the performance is significantly better with pre-stabilization). If there is a loss of one inter-spacecraft laser link, the noise floor of arm locking will exceed the TDI- capability for approximately 30 minutes, twice per year, when the arm length mismatch is less than 12~km.  This would occur with or without pre-stabilization.}

\acknowledgements 

The authors would like to acknowledge the LISA Frequency Control Study Team. In particular we thank, Guido Muller, Vinzenz Wand, Bill Klipstein, Brent Ware and Glenn de Vine for useful discussions. We also thank Ira Thorpe for useful comments on the manuscript. This research was supported by an appointment to the NASA Postdoctoral Program at the Jet Propulsion Laboratory, California Institute of Technology, administered by Oak Ridge Associated Universities through a contract with NASA. Copyright 2009 California Institute of Technology. Government sponsorship acknowledged.

\appendix
 
\section{Doppler Frequency Estimation}
\label{appendix_DE}

The measurement concept for initial estimation of the Doppler rate of a given arm is illustrated in Figure~\ref{estimate-optics}.
\begin{figure}
\centering
\includegraphics[width=0.48\textwidth]{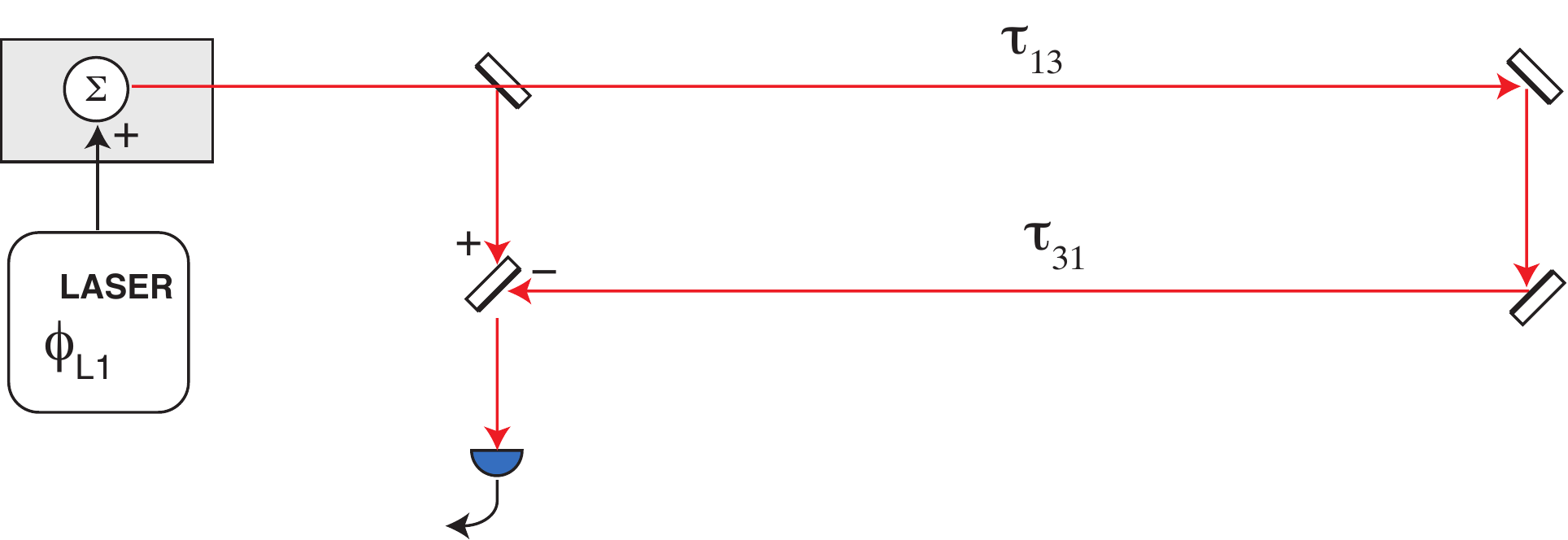}
 \caption{Measurement of Doppler frequency before initialization of arm-locking can be made by comparing the prompt and delayed (and Doppler frequency shifted) laser frequencies. \label{estimate-optics}}
\end{figure}
The transponded, interfered and detected LISA science signal $\phi_M$  ($M$ for measured) contains the doppler signal $\nu_M=v/\lambda,$ where $v$ is the relative velocity between spacecraft.  
The variance of the measured frequency, $\sigma^2,$ is defined as the  integral of the PSD $\tilde{\nu}_M^2(f)$:
\begin{eqnarray}
\label{sigdef}
\sigma^2=\int_0^{\infty}{df\  \tilde{\nu}_M^2(f)}.
\end{eqnarray}
Averaging over a window of duration $T$ introduces the filter
$
H(f)={\rm sinc}(fT)  =\sin(\pi fT)/(\pi fT).
$ 
The phase measurement is
\begin{eqnarray}
\phi_M(t)=\phi(t)-\phi(t-\tau),
\end{eqnarray}
where $\phi(t)$ is the inherent phase of the laser output and  $\tau$ is the round-trip travel time. Differentiating, the corresponding frequency values are
\begin{eqnarray}
\nu_M(t)=\nu(t)-\nu(t-\tau).
\end{eqnarray}
Equation~\ref{sigdef} then becomes
\begin{eqnarray}
\label{general}
\sigma^2=\int_0^{\infty}{df\   \tilde{\nu}^2(f) |H(f)|^2 |L(f)|^2},
\end{eqnarray}
with
\begin{eqnarray}
L(f)=1-\exp(2\pi i f \tau).
\end{eqnarray}

Using
\begin{eqnarray}
|L(f)|^2=2[1-\cos(2\pi f \tau)] =4\sin^2(\pi f \tau),
\end{eqnarray}
\begin{eqnarray}
\label{master}
\sigma^2=4\int_0^{\infty}{df\  \tilde{\nu}^2 {\rm sinc}^2(fT)\sin^2(\pi f \tau)}.
\end{eqnarray}
\begin{figure}
\begin{center}
\includegraphics[width=0.48\textwidth]{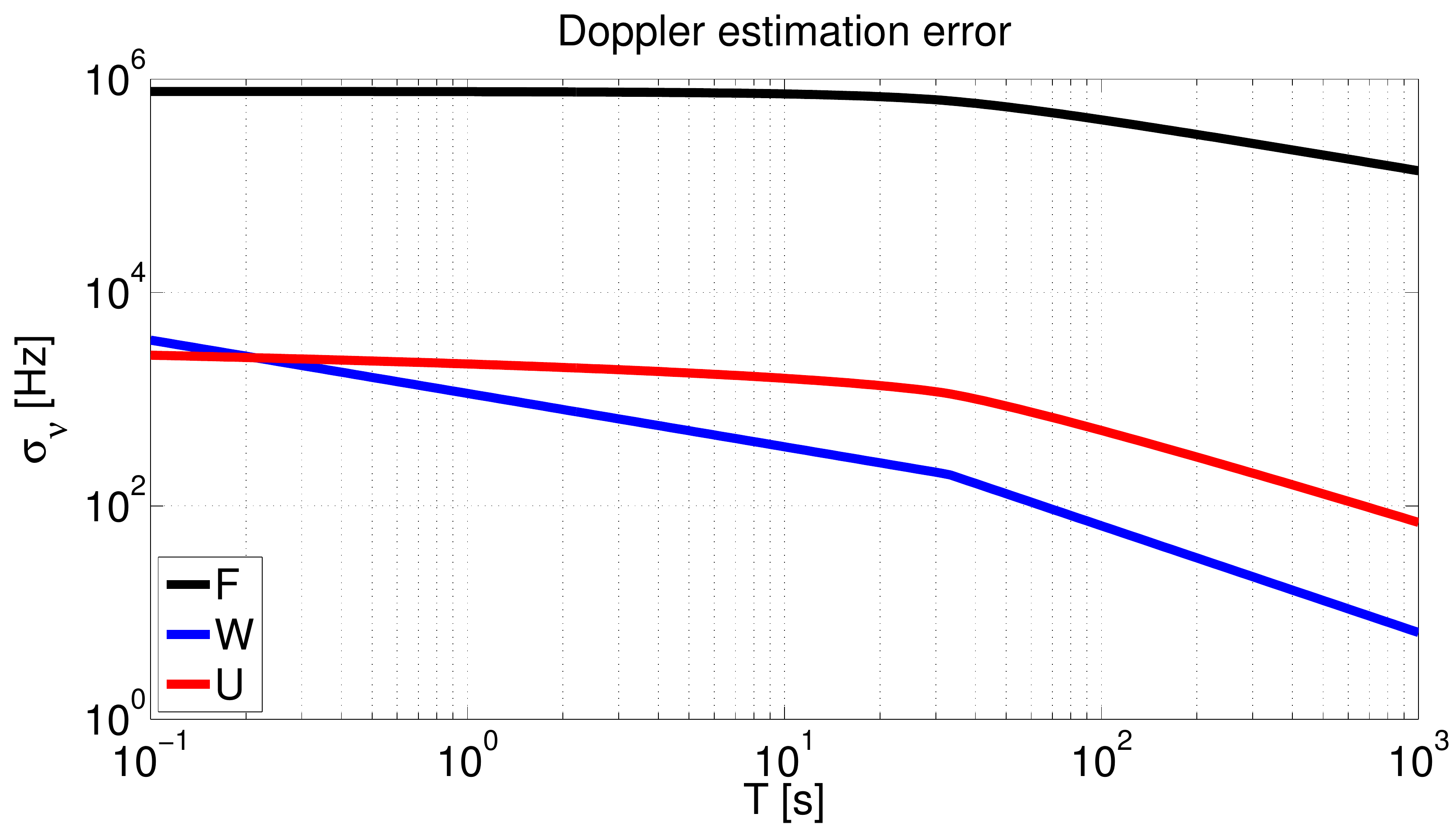}
\caption{Doppler estimation error $\sigma(T)$ from averaging the spectra $\tilde{\nu}(f)$ of Figure~\ref{all-spectra} over period $T.$  
\label{all-sigmas}}
\end{center}
\end{figure}
We consider $\tilde{\nu}(f)=f^{k},\ k=[-1, -\frac 1 2, 0],$ corresponding respectively to random-walk frequency noise, flicker frequency noise, and white frequency noise.  These shapes are typical for the respective sources:  free-running lasers, USO phase noise, and unequal-arm Michelson interferometry limited by white phasemeter noise. Analytic expressions for $\sigma$ are in Table~\ref{sigma-table}, and representative numerical values of $\tilde{\nu}(f)$ and the resulting values of $\sigma$ are plotted in Figures~\ref{all-spectra}~and~\ref{all-sigmas}, respectively. 
\begin{table}
\begin{center}
\begin{tabular}{|c|c|} \hline
$\tilde{\nu}(f)$ & $\sigma$ \\ \hline \hline
$K_Ff^{-1}$ &
  $K_F(\pi/n)\sqrt{\tau/3}\sqrt{
 \left|n-1\right|^3+\left(n+1\right)^3-2\left(n^3+1\right)}$ \\ \hline
$K_Uf^{-1/2}$ & 
$K_U\sqrt{\log{
\left[
 |1-n^{-2}||1-n^2|^{1/n^2}
\left([n+1]/|n-1|\right)^{2/n}
\right]
}}$ \\ \hline
$K_W$ &
$K_W/(\sqrt{2\tau}n)\sqrt{1+n-|1-n|}$ \\ \hline
\end{tabular}
\caption{Error in doppler frequency estimation, $\sigma$, for various spectra of frequency noise $\tilde{\nu}(f).$  Round-trip travel time = $\tau,$ averaging time $=T$, $n=T/\tau.$  Note the distinction between absolute-value brackets $|\ldots|$ and grouping brackets $[\ldots]$ or $(\ldots).$
\label{sigma-table}}
\end{center}
\end{table}

\begin{figure}
\begin{center}
\includegraphics[width=0.48\textwidth]{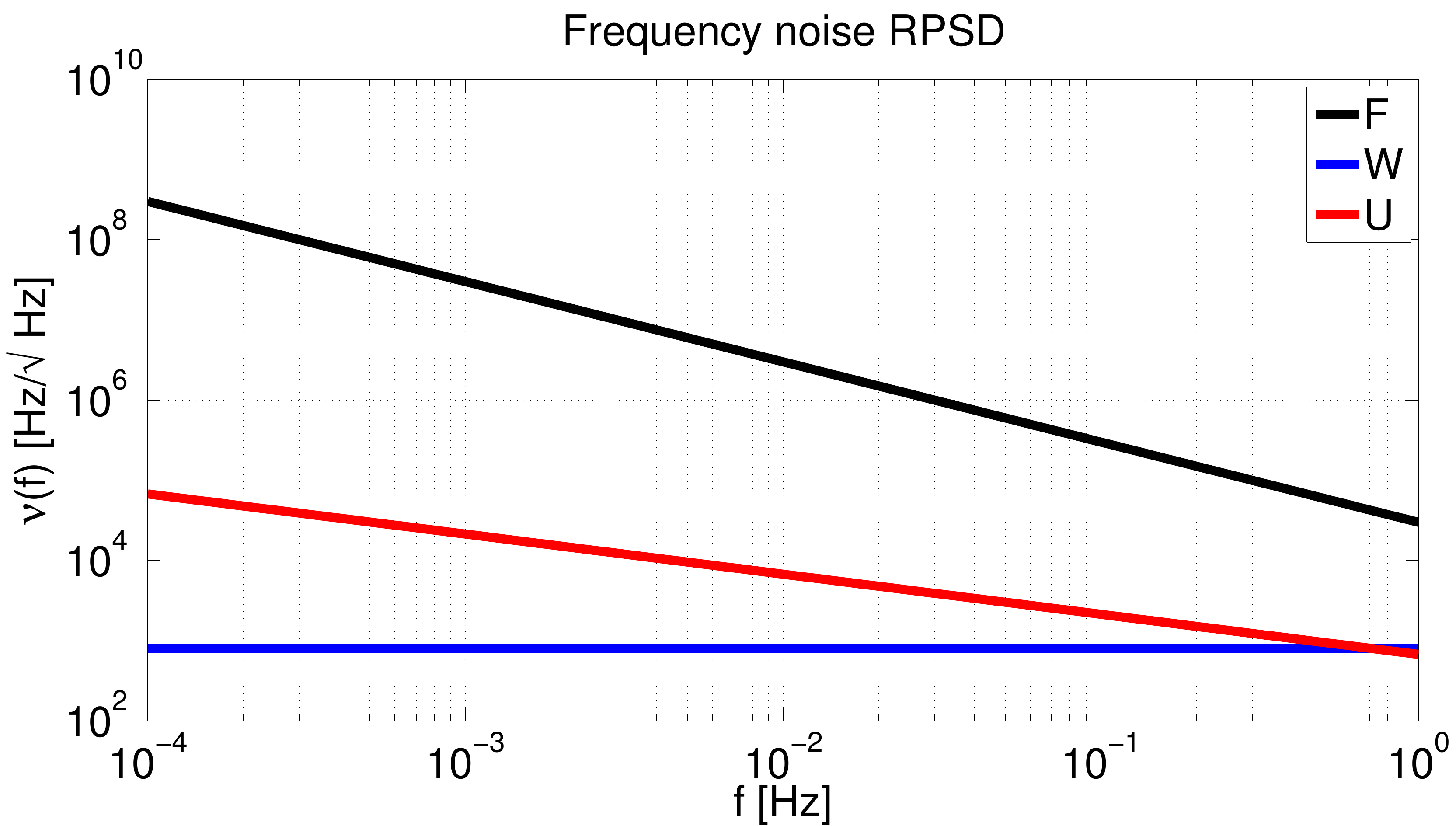}
\caption{Various assumptions of frequency noise  $\tilde{\nu}(f)$ as input to estimation of the initial   Doppler rate $\nu.$  The uppermost curve {\em F} represents the noise in a free-running laser, $\tilde{\nu}(f)=k_f/f,$ $k_f=\SI{1e4}{Hz^{3/2}}$;  the flat {\em W} curve represents white laser frequency noise, $\tilde{\nu}(f)=K_w = \SI{1e4}{Hz^{1/2}},$ and the {\em U} curve is a typical USO spectrum, $\tilde{\nu}_U(f)=k\nu_0/\sqrt{f};\ $ where $k= {2.4\times10^{-12}}$  and $\nu_0$ is the laser frequency, \SI{2.8e14}{Hz}.  \label{all-spectra}}
\end{center}
\end{figure}

The estimation based on USO noise is a new concept that requires some explanation.  The USO signals are imposed as $\approx 10\,$ GHz modulation sidebands on the carrier, for clock correction.  Just as the carrier is transponded by phase-locking the distant laser to the incoming laser, so too the USO sideband
can be transponded by feeding back the sideband/sideband beat to the phase-locking input of the USO.  Then at the master (local) spacecraft, the measured sideband/sideband beat is representative of the Doppler shift.  The measurement is made insensitive to laser frequency noise by subtracting the measured carrier/carrier phase, leaving noise only from the clock.  In this manner, the USO can substitute for an optical frequency reference for the purpose of measuring Doppler frequency.

The error in the average of the $k$'th time derivative of the doppler shift is a generalization of Equation~\ref{sigdef}:
\begin{equation}
\label{sigdef-k}
\sigma^2_k=\int_0^{\infty}{df\  (2\pi f)^{2k} \,\tilde{\nu}_M^2(f)}.
\end{equation}
Generalizing Equation~\ref{master} to allow for $j$ stages of averaging of the $k$'th time derivative,
\begin{equation}
\label{master-jk}
\sigma^2_{jk}=4\int_0^{\infty}{df\  (2\pi f)^{2k} \,\tilde{\nu}^2 {\rm sinc}^{2j}(fT)\sin^2(\pi f \tau)}.
\end{equation}
For $\tilde{\nu}=f^p,$ Equation~\ref{master-jk} gives finite $\sigma_{jk}$ only for $j\ge 1+p+k.$  Therefore, to estimate the first and second derivitaves of the Doppler shift ($k=1,2$, respectively) with white frequency noise ($p=0$) requires at least two and three stages of averaging ($j=2,3$) respectively.  Note that $j$ stages of averaging requires data of duration $jT.$  For the spectrum of free-running lasers ($p=-1$), the minimum number of averages to compute the first and second derivatives are $j=1,2,$ respectively.  We have computed  $\sigma_{jk}$ for $k=(0,1,2)$ and $p=(0,1)$ with up to $j=3$ stages of averaging, using the Symbolic Math package of Matlab~\cite{MATLAB}.  Sample numeric results are shown in Figure~\ref{free-run-deriv}.

\begin{figure}
\begin{center}
\includegraphics[width=.48\textwidth]{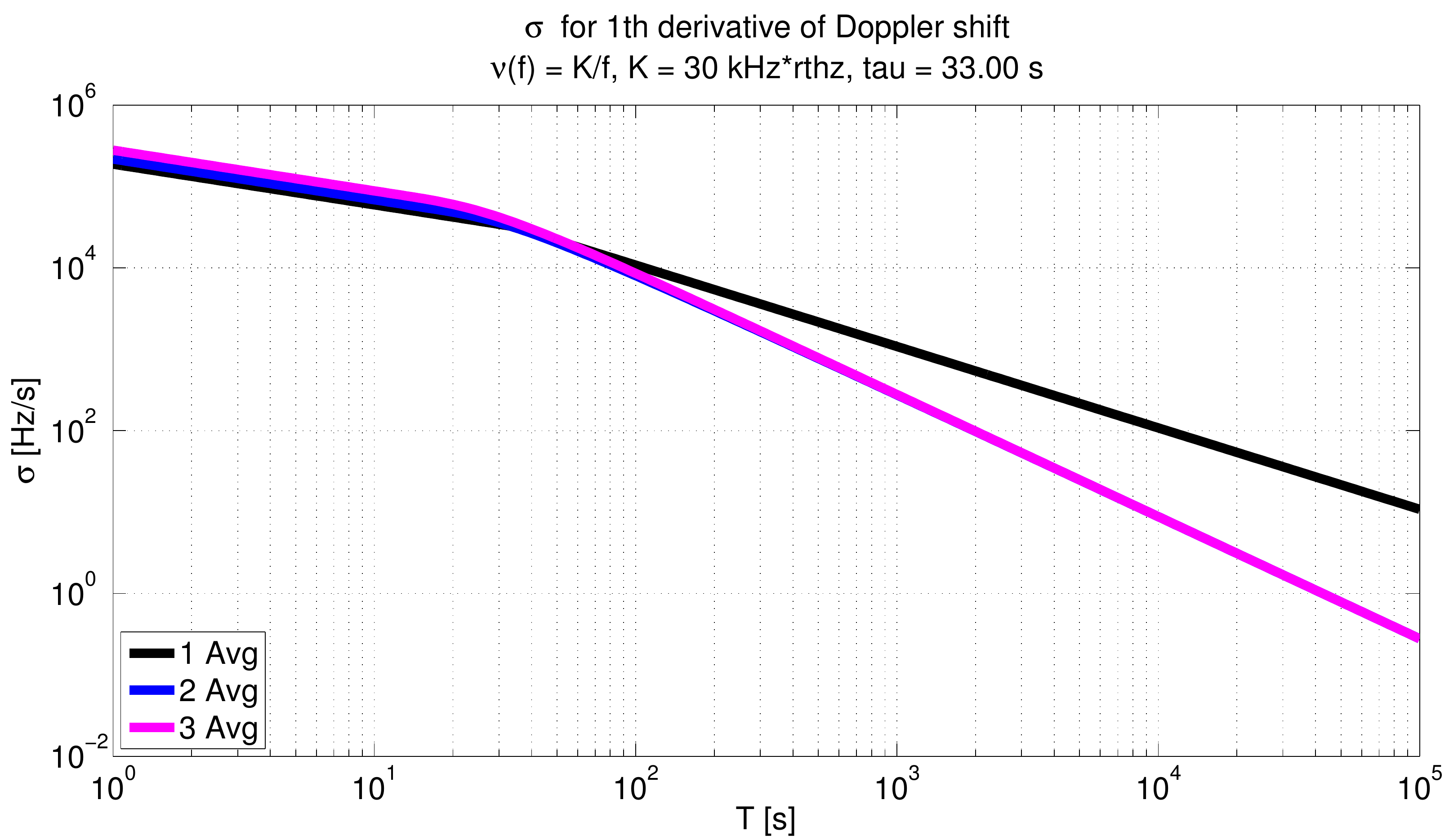}
\caption{Error in the estimate of the first time-derivative of the single-arm doppler shift, assuming the noise of free-running lasers, figure~\ref{all-spectra}.  
For averaging times $T<\tau=\SI{33}{s},$ there is no advantage to more than one stage of averaging.  
For $T>\SI{33}{s},$ two stages of averaging gives less error than one, but there is no advantage to three stages.  The estimation error from two stages, each of duration $T=\SI{100}{s},$ is $\SI{8}{kHz/s}.$
\label{free-run-deriv}}
\end{center}
\end{figure}

\section{Approximation of the sensors}
\label{appendix_sensors}
\begin{figure}[ht]
\begin{center}
\includegraphics[width=0.48\textwidth]{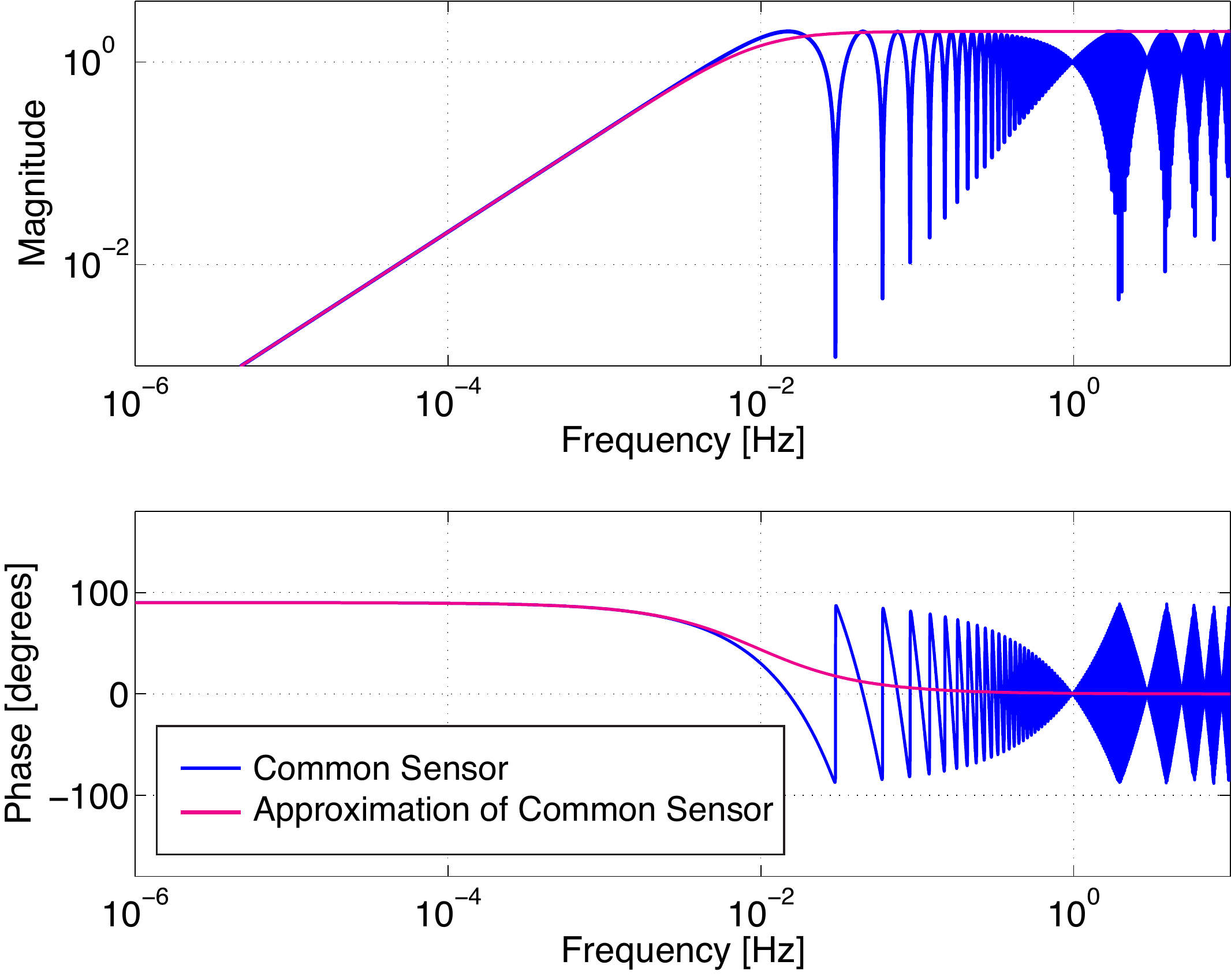}
\caption{Common arm locking sensor and it's approximation used in step responses.
\label{common_sensor_approx}}
\end{center}
\end{figure}

\begin{figure}[ht]
\begin{center}
\includegraphics[width=0.48\textwidth]{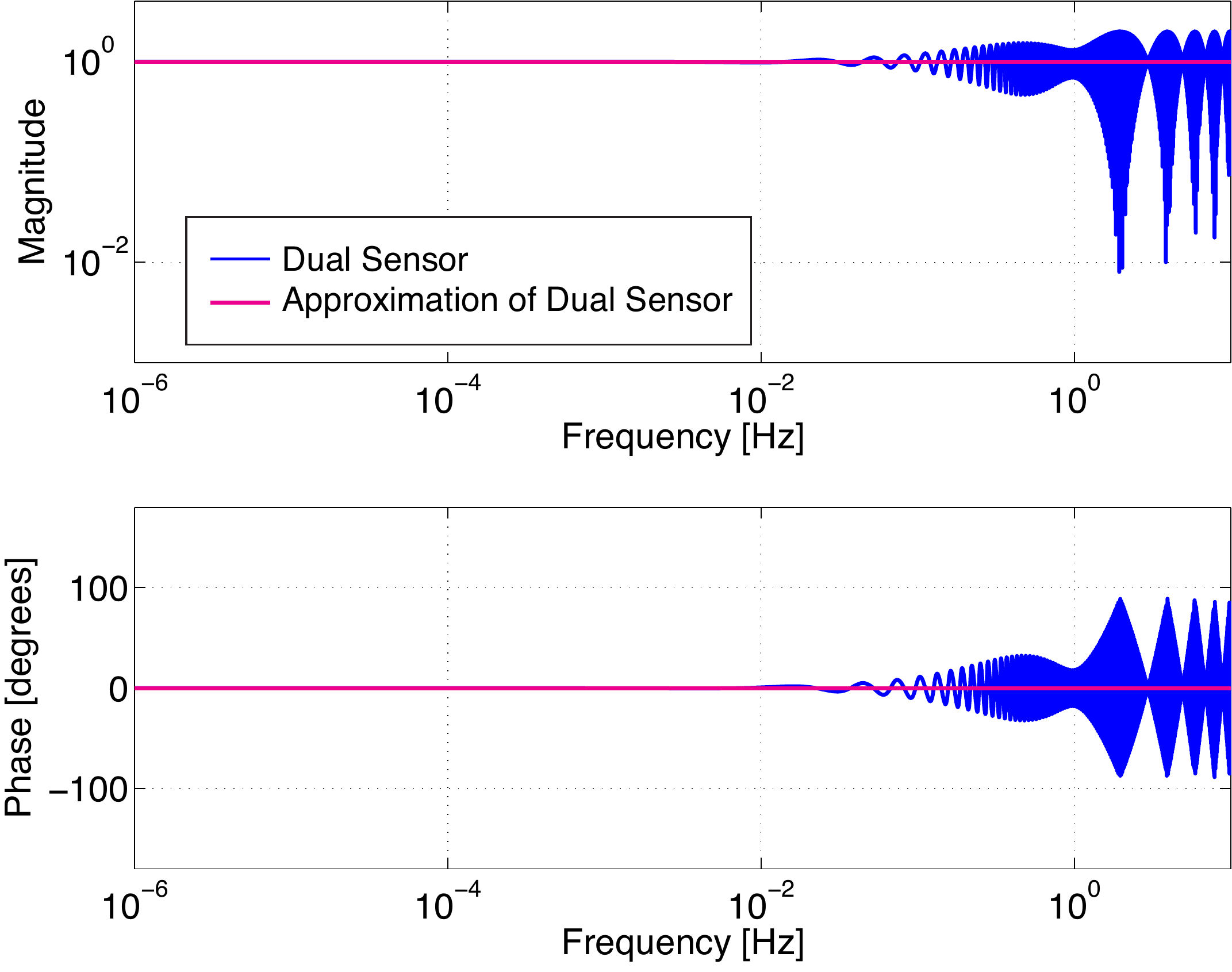}
\caption{Dual arm locking sensor and approximation used in step responses.
\label{dual_sensor_approx}}
\end{center}
\end{figure}
In section~\ref{section_doppler} we performed a study of the frequency pulling caused by the closed arm locking control loop. Due to the high computational time required to simulate the frequency pulling using the sensors in there exact form, the simulations spanning over a period of days after lock acquisition were performed using approximations of the sensors. These approximations are very good below the null frequencies, that is for $f<10$~mHz in common arm locking, and $f<2$~Hz for dual arm locking. This frequency band is responsible for the most significant pulling occurs.

In the study of pulling in common arm locking we approximate the frequency response of the sensor by
\begin{eqnarray}
P_+(\omega)|_{\rm approx} = \frac{4s}{s+2/\bar\tau},
\end{eqnarray}
which is plotted in figure~\ref{common_sensor_approx} along with the common arm locking sensor.

In the study of pulling in dual arm locking and modified dual arm locking we approximate the frequency response of the sensor by
\begin{eqnarray}
P_D(\omega)|_{\rm approx} = 2,
\end{eqnarray}
which is plotted in figure~\ref{dual_sensor_approx} along with the dual arm locking sensor.

The controllers used in calculating the frequency pulling in section~\ref{section_doppler} were based on the controller design presented in section~\ref{section_controller}. The shape of the common arm locking sensor differs significantly from that of the dual arm locking sensor (for which the controller was designed). To calculate the frequency pulling in common arm locking the controller response was modified to account for this. The controller used in common arm locking frequency pulling calculations was simply
\begin{eqnarray}
G_{1}(\omega) = \frac{G_{1}^*(\omega)}{P_+(\omega)|_{\rm approx}}.
\end{eqnarray}
where $G_1(\omega)^*$ is given by equation~\ref{equation_controller}.
\section{Orbits}
\label{appendix_orbits}
The orbit data used in this paper was provided by Peter Gath~\cite{PG}. Plots of the relevant parameters are shown in figures~\ref{mismatch},~\ref{comm_doppler}, and~\ref{diff_doppler}.

\begin{figure}
\begin{center}
\includegraphics[width=0.4\textwidth]{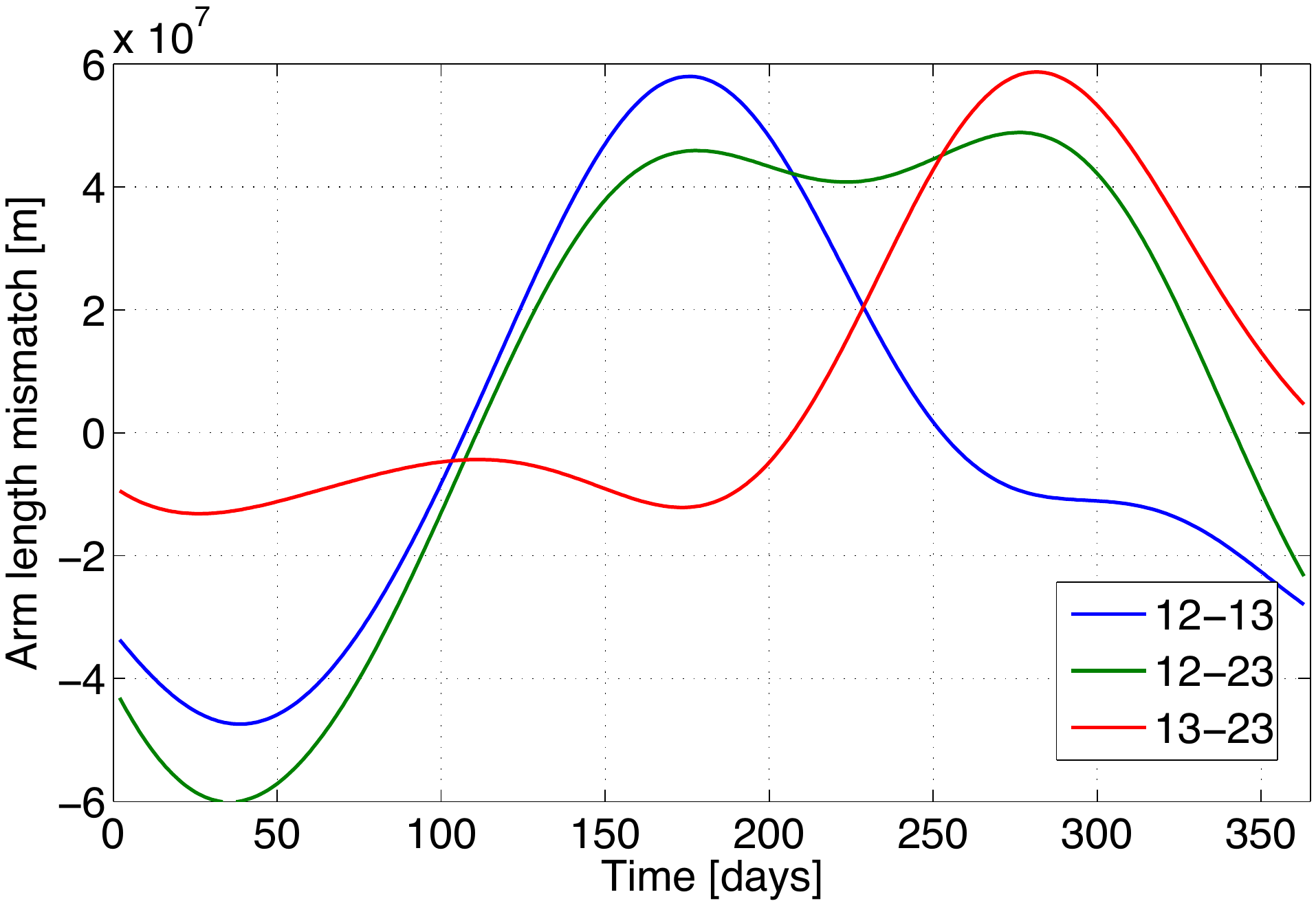}
\caption{Arm length mismatch of the sets of two arms. Calculations in this paper were performed using arm 12-13.
\label{mismatch}}
\end{center}
\end{figure}

\begin{figure}
\begin{center}
\includegraphics[width=0.4\textwidth]{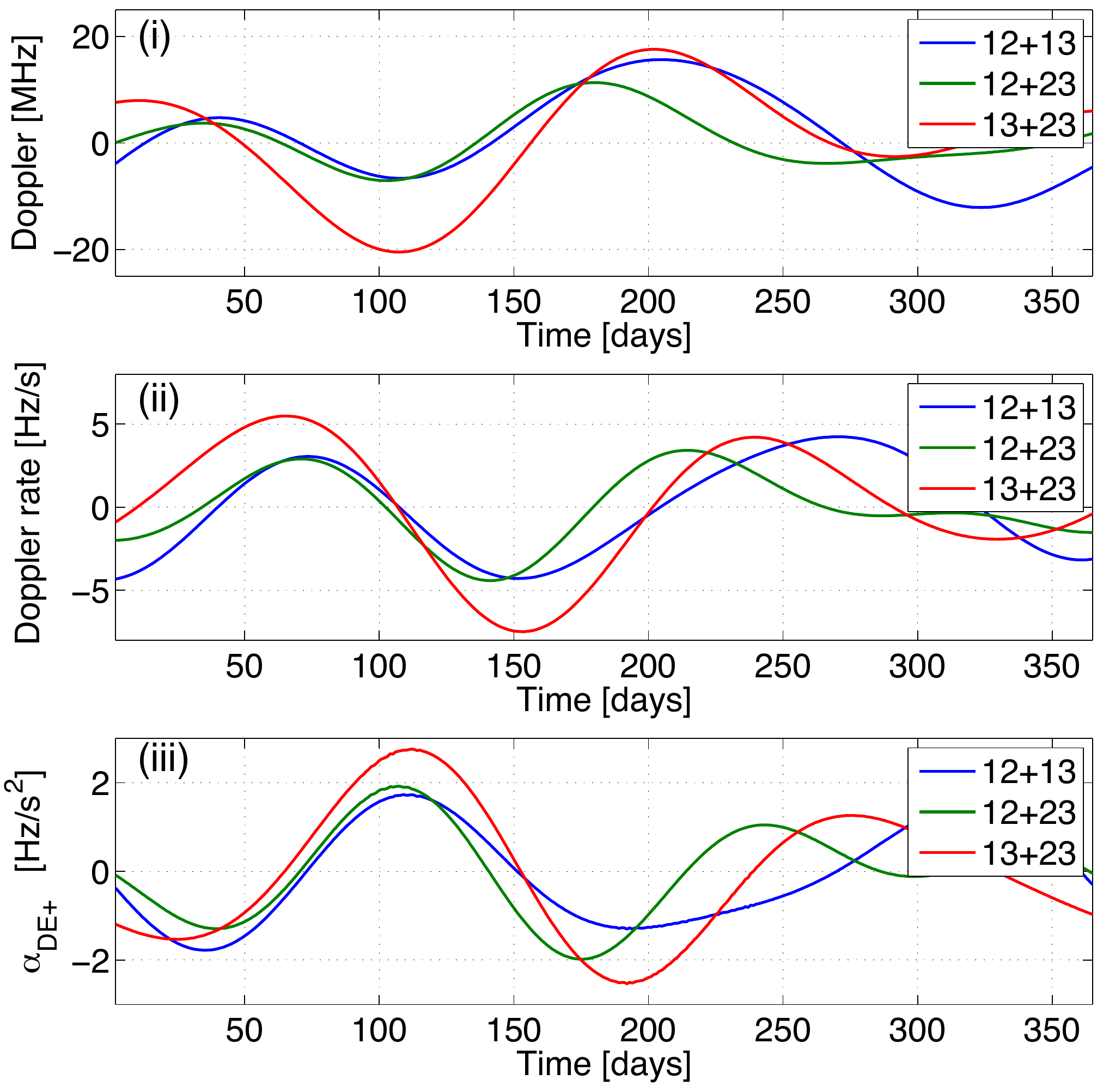}
\caption{(i) Common Doppler frequency, (ii) common Doppler rate, and (iii) 2nd time derivative of common Doppler frequency for the different arm pairs.
\label{comm_doppler}}
\end{center}
\end{figure}

\begin{figure}
\begin{center}
\includegraphics[width=0.4\textwidth]{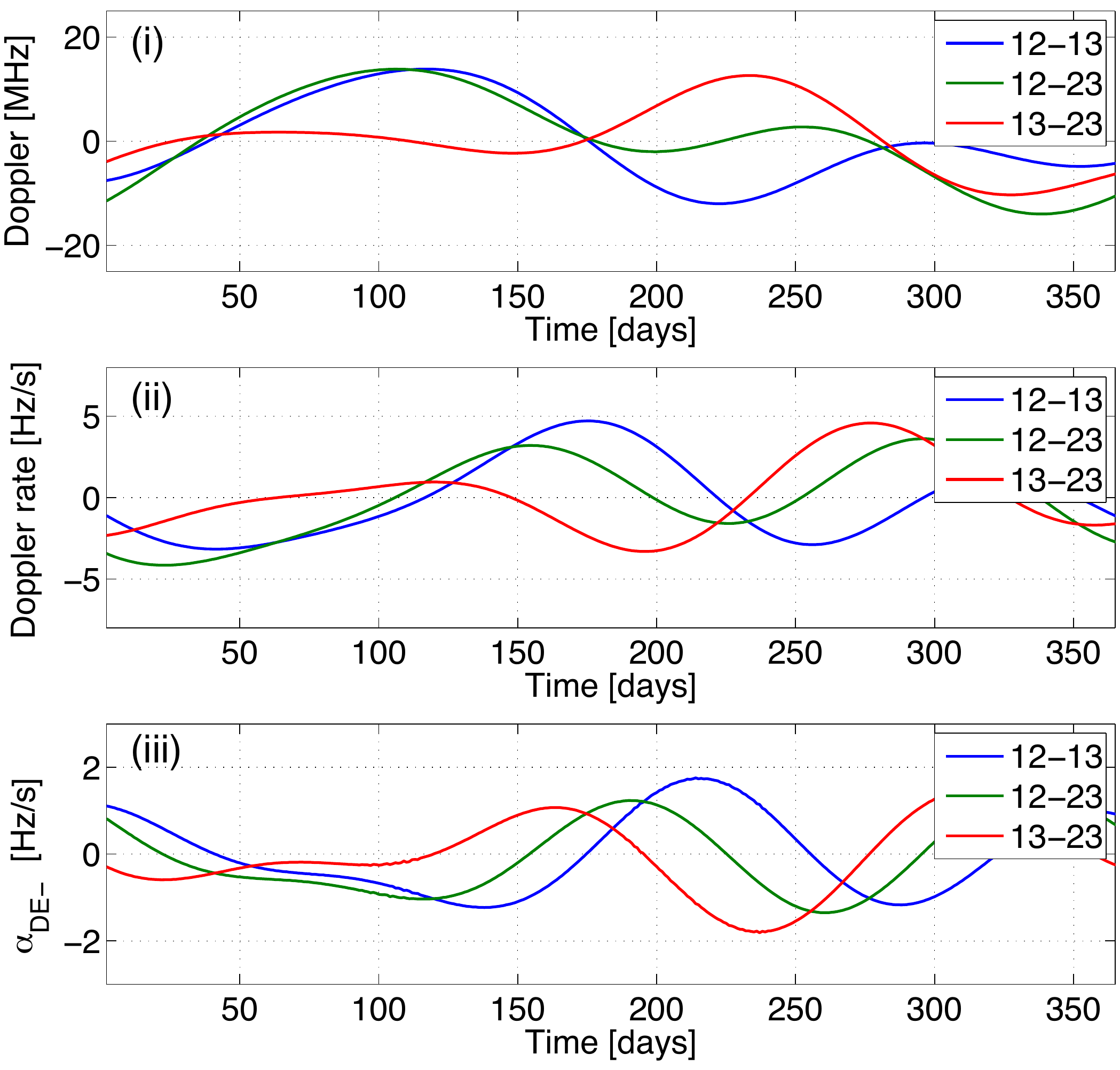}
\caption{ (i) Difference  Doppler frequency, (ii) difference Doppler rate, and (iii) 2nd time derivative of difference Doppler frequency for the different arm pairs.
\label{diff_doppler}}
\end{center}
\end{figure}

\end{document}